\documentclass[journal=jacsat,manuscript=article]{achemso}

\usepackage[version=3]{mhchem} 
\usepackage{graphicx} 
\usepackage{qcircuit}
\usepackage{braket}
\usepackage{amsmath}
\usepackage{amssymb}
\usepackage{amsfonts}
\usepackage{mathtools}
\usepackage{array}
\usepackage{pdflscape} 
\usepackage{tabularx} 
\usepackage{subcaption}
\usepackage[hidelinks]{hyperref}

\author{Anthony M. Smaldone}
\author{Victor S. Batista}
\email{{anthony.smaldone, victor.batista}@yale.edu}
\affiliation[Yale University]
{Department of Chemistry, Yale University, New Haven, 06511, CT, USA}

\title[An \textsf{achemso} demo]
  {Quantum to Classical Neural Network Transfer Learning Applied to Drug Toxicity Prediction}

\abbreviations{IR,NMR,UV}
\keywords{American Chemical Society, \LaTeX}

\begin{document}



\begin{abstract}
  Toxicity is a roadblock that prevents an inordinate number of drugs from being used in potentially life-saving applications. Deep learning provides a promising solution to finding ideal drug candidates; however, the vastness of chemical space coupled with the underlying $\mathcal{O}(n^3)$ matrix multiplication means these efforts quickly become computationally demanding. To remedy this, we present a hybrid quantum-classical neural network for predicting drug toxicity, utilizing a quantum circuit design that mimics classical neural behavior by explicitly calculating matrix products with complexity $\mathcal{O}(n^2)$. Leveraging the Hadamard test for efficient inner product estimation rather than the conventionally used swap test, we reduce the number qubits by half and remove the need for quantum phase estimation. Directly computing matrix products quantum mechanically allows for learnable weights to be transferred from a quantum to a classical device for further training. We apply our framework to the Tox21 dataset and show that it achieves commensurate predictive accuracy to the model's fully classical $\mathcal{O}(n^3)$ analog. Additionally, we demonstrate the model continues to learn, without disruption, once transferred to a fully classical architecture. We believe combining the quantum advantage of reduced complexity and the classical advantage of noise-free calculation will pave the way to more scalable machine learning models.

\end{abstract}

\section{Introduction}

The quest for novel pharmaceuticals is fraught with the challenge of ensuring drug safety. With 90\% of drug candidates failing in clinical trials and 30\% of these failures being attributed to toxicity\cite{sun_why_2022}, the cost to create a successful drug can exceed \$2 billion USD\cite{noauthor_research_2021}, making the early detection of toxicological properties in drug candidates crucial. 

Machine Learning (ML) has revolutionized the field of drug discovery\cite{mullowney_artificial_2023, sarkar_artificial_2023, volkamer_machine_2023}, offering a powerful tool to handle and interpret the complex datasets that are characteristic of pharmaceutical research. By leveraging algorithms that can learn from and make predictions on data, ML has enabled significant advancements in identifying new drug candidates \cite{bian_generative_2021,martinelli_generative_2022,tang_generative_2021} and predicting drug toxicity \cite{semenova_bayesian_2020,suzuki_predicting_2020,mayr_deeptox_2016, nguyen_graph_2022,peng_top_2020, jiang_ggl-tox_2021}. As ML continues to amaze, researchers are always searching for new computational tools to bolster the tractability and performance of their models.

Quantum computing, with its inherent capability to handle certain complex computations more efficiently than classical computing, emerges as a promising solution. Leveraging the principles of superposition and entanglement, quantum computing offers a new framework that potentially expands computational boundaries for ML. Recent work suggests that quantum ML (QML) models are more learnable and generalize better to unseen data than classical networks \cite{du_expressive_2020,du_quantum_2021,du_learnability_2021,caro_generalization_2022}. In pursuit of these potential advantages, researchers have built numerous QML models to address a range of chemical and biological problems \cite{gircha_hybrid_2023,batra_quantum_2021,sajjan_quantum_2022,mensa_quantum_2023,kao_exploring_2023,domingo_binding_2023,banerjee_hybrid_2023,dong_prediction_2023, vakili_quantum_2024, hong_quantum_2021}. More specifically, there have been several QML works that are trained to predict toxicity \cite{albrecht_quantum_2023,suzuki_predicting_2020,bhatia_quantum_2023}. Despite the success of these QML models, the limitations that plague noisy intermediate-scale quantum (NISQ) devices such as decoherence and gate errors are still a sobering reality.

While noise is not a concern for classical computing, it is not without its bottleneck either. Every year, as developers create deep learning models more powerful than the last, we are seeing a swift increase in the number of parameters \cite{villalobos_machine_2022}. This escalation in parameter count introduces a consequential challenge: the need for significantly more computing power and the development of more efficient algorithms. Matrix multiplication lies at the heart of neural networks. The standard method of matrix multiplication has complexity $\mathcal{O}(n^3)$, and while sub-cubic algorithms do exist such as the Strassen algorithm\cite{strassen_gaussian_1969} $\mathcal{O}(n^{2.8074})$ and the recording-setting algorithm by Williams \textit{et al.}\cite{williams_new_2023} $\mathcal{O}(n^{2.3716})$, these methods suffer from numerical instability\cite{miller_computational_1975}, high memory usage, or have a complexity prefactor that makes them galactic algorithms, rending them impractical for tractable computation.

In response to these computational challenges, this study introduces a hybrid quantum-classical neural network model designed to leverage the strengths of both quantum and classical computing paradigms and apply the framework to drug toxicity prediction. Utilizing the Hadamard test, our model employs quantum circuits that replicate the forward functionality of classical neural networks, allowing the model to efficiently calculate discrete inner products. Assuming efficient state preparation, the swap test has been shown to be the most computationally favorable algorithm \cite{shao_quantum_2018} for matrix multiplication with complexity $\mathcal{O}(n^2)$, and has been suggested for use in a quantum neuron \cite{stein_qucnn_2022,li_quantum_2020,zhao_building_2019,shao_quantum_2018-1} . Our use of the Hadamard test achieves the same complexity, one quantum circuit for each dot product, while using half the working qubits and without the need for quantum phase estimation. This QML approach allows for ``classical weights" to be easily derived from the quantum model, which can then be transferred to a classical machine to be fine-tuned without the presence of noise and errors which is characteristic of current NISQ devices.  All previous works of quantum transfer learning are either classical-to-quantum \cite{mari_transfer_2020,kim_classical--quantum_2023, schuman_towards_2023}, or simply use the initial quantum component as a feature extractor that is non-learnable after transferring \cite{mari_transfer_2020}. To our knowledge, this framework of fine-tuning weights classically that were initially embedded in a quantum circuit is a novel contribution to this growing field of quantum transfer learning.

In an effort to realize these potential quantum advantages, we investigate our model's performance on drug toxicity prediction using the Tox21 dataset \cite{tice_improving_2013,national_institute_of_health_tox21_nodate} with a convolutional neural network (CNN). We present the effectiveness of our quantum CNN (QCNN), which replaces the filter with a quantum circuit to compute the dot product. Additionally, we illustrate that a classical convolution filter can seamlessly continue the training process initially started by its quantum counterpart. Furthermore, we demonstrate that both the hybrid quantum-classical model and the fully classical equivalents exhibit comparable performance levels for drug toxicity prediction.

\section{Quadratic Matrix Multiplication}
\subsection{The Swap Test}
\begin{figure}[h]
    \centering
    \includegraphics[width=0.4\textwidth]{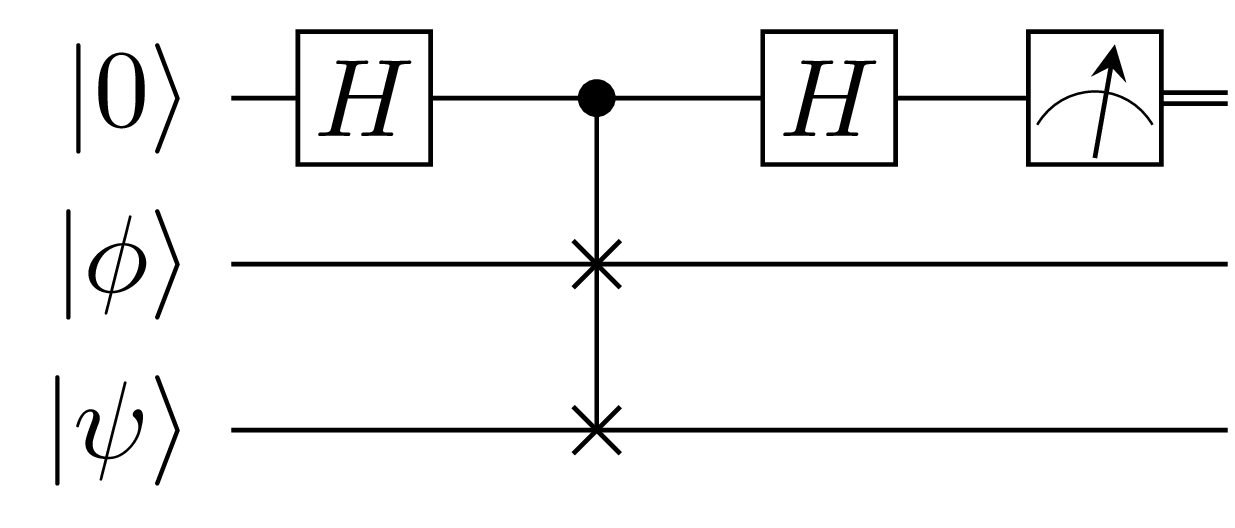}
    \caption{The Swap Test}
    \label{fig:swap_test}
\end{figure}

The swap test\cite{buhrman_quantum_2001}, shown in Figure \ref{fig:swap_test}, has been suggested for use in a quantum neuron and as a convolutional filter \cite{stein_qucnn_2022,li_quantum_2020,zhao_building_2019,shao_quantum_2018-1}. This idea is manifested by encoding the input data into one working register of qubits and by applying learnable quantum gates to another register. When the controlled swap between the two states of these registers is executed, the expectation value of the ancilla qubit will produce the squared modulus of the inner-product between the two states $\left|\braket{\phi|\psi}\right|^2$ with error $\epsilon$, in $\mathcal{O}(\frac{1}{\epsilon^2})$ repetitions. In this sense, the amplitudes of the parameterized qubit register can be thought of analogously as the ``weights". These ``weights" are being multiplied together with the input data and summed when the swap test is conducted and the ancilla is measured. However since the swap test only reveals the fidelity between two quantum states, it is directionless. The lack of the dot-product's sign is problematic for flexible machine learning, and places additional difficulty on the learning procedure. A remedy to this is quantum phase estimation (QPE), but the extra ancilla qubits required in QPE place an additional overhead on the required resources to run this algorithm.

\subsection{The Hadamard Test}
\begin{figure}[!h]
    \centering
    \includegraphics[width=0.4\textwidth]{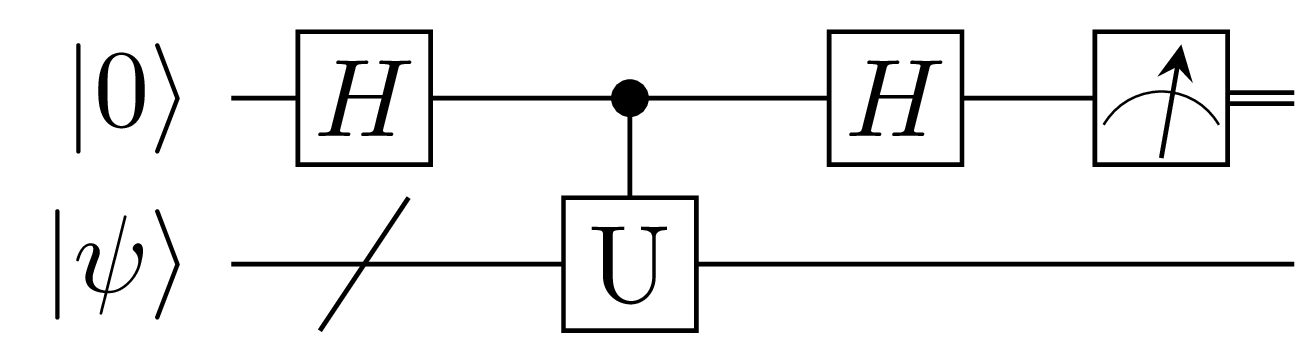}
    \caption{The Hadamard Test}
    \label{fig:hadamard_test}
\end{figure}

The Hadamard test, shown in Figure \ref{fig:hadamard_test}, is another cornerstone quantum algorithm \cite{cleve_quantum_1998} that has found use in calculating inner products for fields ranging from chemistry \cite{wang_electronic_2023}, to physics \cite{miquel_interpretation_2002}, to mathematics \cite{aharonov_polynomial_2009}. The test allows one to determine $\text{Re}\braket{\psi|U|\psi}$ with error $\epsilon$, in $\mathcal{O}(\frac{1}{\epsilon^2})$ repetitions. However, by choosing the correct $U$, we can use this algorithm to produce the inner product between two vectors of choice, $\text{Re}\braket{\psi|\phi}$. Since this gives a direction with the inner product unlike the swap test, we may dispense with the need for QPE as well as the need for separate quantum registers for the input data and parametric ``weights". Equations \ref{data_prep}-\ref{hadamard_test_equation} show how $U$ can be chosen, for input data, $\ket{\phi}$ and ``weights", $\ket{\psi}$.

\begin{equation}
    U_\psi \ket{0}^{\otimes n} = \ket{\psi}
    \label{data_prep}
\end{equation}
\begin{equation}
    U_\phi \ket{0}^{\otimes n} = \ket{\phi}
    \label{weight_prep}
\end{equation}
\begin{equation}
    U = U_\phi U^{\dagger}_\psi
\end{equation}
\begin{equation}
    \operatorname{Re}\bra{\psi}U\ket{\psi} = \operatorname{Re}\bra{\psi}U_\phi U^{\dagger}_\psi \ket{\psi} = \operatorname{Re}\bra{\psi}U_\phi U^{\dagger}_\psi U_\psi \ket{0}^{\otimes n} = \operatorname{Re}\braket{\psi|\phi}
    \label{hadamard_test_equation}
\end{equation}

\begin{figure}[!h]
    \centering
    \includegraphics[width=0.5\textwidth]{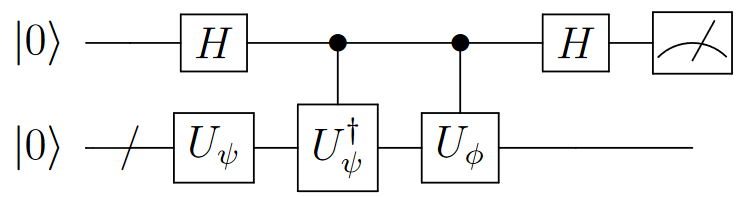}
    \caption{Quadratic Matrix Multiplication with the Hadamard Test}
    \label{fig:modified_hadamard_test}
\end{figure}

Equation \ref{hadamard_test_equation} shows that if $U$ is chosen to be the product of the unitary matrix that prepares the quantum state whose amplitudes act as the ``weights" and the adjoint of the unitary that prepares the input data, the real part of the desired inner-product is obtained when performing the Hadamard test. This circuit is visualized in Figure \ref{fig:modified_hadamard_test}. 

We note that the full complex-valued inner product $\braket{\psi|\phi}$ may easily be obtained. The Hadamard test may be slightly modified to include an $S^{\dagger}$ gate after the first Hadamard gate on the ancilla qubit to produce $\operatorname{Im}\braket{\psi|\phi}$. Summing the expectation values of these two different Hadamard tests yields the complex-valued inner product $\operatorname{Re}\braket{\psi|\phi} + i\operatorname{Im}\braket{\psi|\phi}=\braket{\psi|\phi}$.

As with the swap test, only one quantum circuit needs to be run for each resulting entry (or two circuits for complex-valued input data) when finding the product of two matrices, and thus scales $\mathcal{O}(n^2)$. While this method uses significantly fewer qubits than the swap test, it may come at the expense of deeper circuitry due to the controlled unitary gate $U^{\dagger}_\psi$ required to uncompute the input state. To this point, we note that both the swap test and the proposed Hadamard test method for multiplying arbitrary matrices with $\mathcal{O}(n^2)$ complexity rely on the assumption of efficiently prepared quantum states \cite{shao_quantum_2018}. As state preparation techniques become more refined, qubit control improves, and quantum devices become more coherent, the fulfillment of this assumption will also satisfy the concerns of deeper circuitry possessed by this Hadamard test method. Concurrently, the computational overhead of more than double the number of qubits and the requirement for QPE associated with the swap test will not be alleviated, as these are fundamental to the algorithm and are not dependent on quantum hardware.

\section{Architecture and Dataset}
\begin{figure}[h]
    \centering
    \includegraphics[width=1\textwidth]{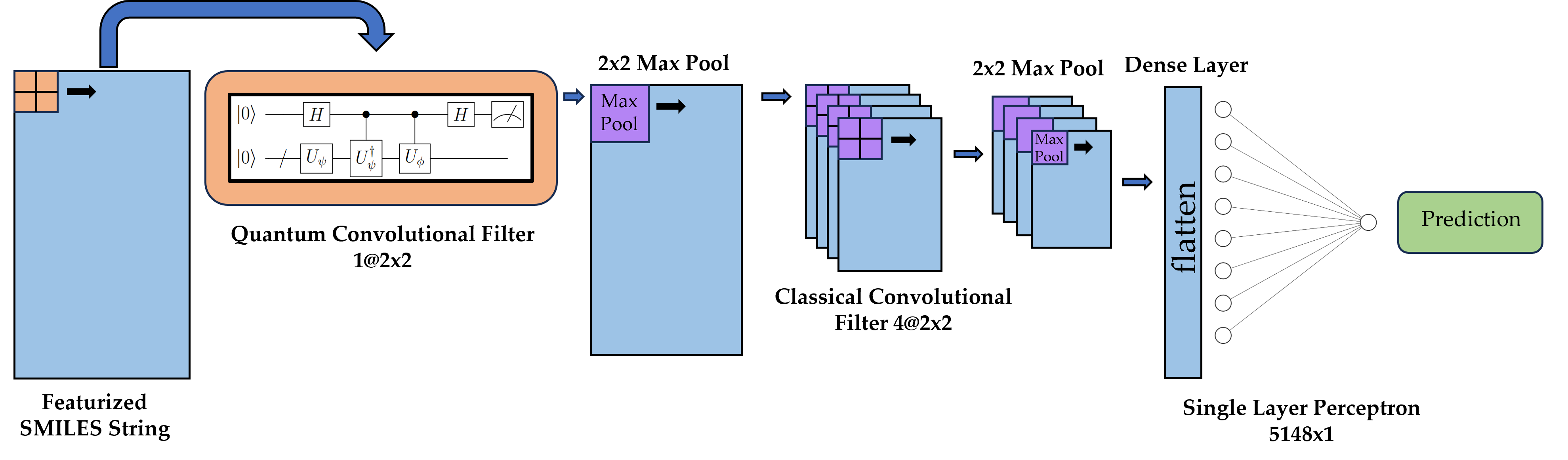}
    \caption{Quantum-Classical Neural Network Architecture}
    \label{fig:architecture}
\end{figure}

\subsection{Hybrid Architecture}

CNNs have been found to be successful for use in toxicity prediction tasks in the literature\cite{chen_bestox_2020, yuan_toxicity_2019, limbu_predicting_2022}, and are very popular QML analogs of classical architectures \cite{cong_quantum_2019,hur_quantum_2022,chen_quantum_2022, herrmann_realizing_2022, smaldone_quantum_2023}. The popularity of QCNNs partly stem from only needing to operate on small sections of input data at a time, rendering them feasible for NISQ devices. We employ a quantum-classical architecture where the first convolutional layer is replaced by a quantum circuit and is subsequently followed by classical pooling, a classical convolutional layer, and a fully connected layer as depicted in Figure \ref{fig:architecture}.

\subsection{Tox21 Dataset}

\begin{figure}[h]
    \centering
    \includegraphics[width=1\textwidth]{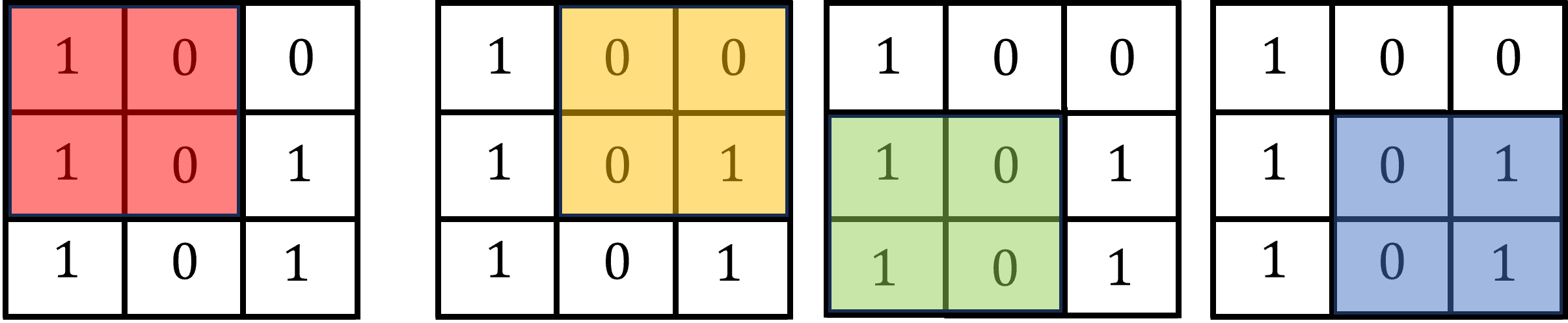}
    \caption{Binary data in a CNN. The red and green sections of this example input data will result in an identical output, and thus only one of the two dot products need to be computed. Since the number of unique dot products for binary data is $2^w$, where $w$, is the number of weights in a filter, for our input data of size 400 $\times$ 57 and a single 2 $\times$ 2 filter, there are only $2^{2 \times 2}$ different filters that will produce a unique dot product. Therefore, since the number of unique dot products is not dependent on the size of the input data, we only need to execute 15 circuits (the all-zero vector need not be computed), rather than the traditional 22,344 for each molecule. We note that the removal of non-unique calculations is applicable to both classical and quantum CNNs, however as quantum circuits are inherently more expensive than classical computations in the NISQ era, this process of removing duplicates becomes particularly valuable for QCNNs.}
    \label{fig:binary_filter}
\end{figure}

The Tox21 10k dataset \cite{tice_improving_2013,national_institute_of_health_tox21_nodate} is frequently used as a benchmark for evaluating ML models on their binary predictive ability for toxicity. This dataset comprises approximately 10,000 molecules represented as SMILES strings, each tagged with binary labels according to their response to 12 distinct assays. The Tox21 dataset suffers from significant imbalance, and although techniques such as SMOTE \cite{chawla_smote_2002} can enhance performance, the primary objective of this study is to showcase the synergy between quantum and classical ML frameworks, rather than attaining the highest level of predictive accuracy. We specifically choose to create a sparse, binary feature map from each SMILES string in order to reduce overhead on the quantum component of the model. The computational benefit of binary features in a CNN is visually depicted in Figure \ref{fig:binary_filter}. For each filter comprised of $w$ weights, there can exist a maximum of $2^w$ unique convolutions for binary data. Thus, for a single 2 $\times$ 2 filter there are only 16 unique dot-products that are required to fully stride across input data of any size. Additionally, the sparsity condition allows for the use of state-preparation techniques that do not scale exponentially \cite{de_veras_double_2022, gleinig_efficient_2021}, which is a necessity along the road to achieving implementable quadratic complexity for matrix multiplication.

\begin{table}[h!]
\centering
\begin{tabular}{llc}
\hline
\textbf{Feature type} & \textbf{Element} & \textbf{Bit(s)} \\
\hline
Symbol in SMILES & ( ) & 2 \\
 & [ ] & 2 \\
 &   & 1 \\
 & : & 1 \\
 & = & 1 \\
 & \# & 1 \\
 & \textbackslash & 1 \\
 & / & 1 \\
 & @ & 1 \\
 & + & 1 \\
 & - & 1 \\
 & . & 1 \\
 \hline
Number in SMILES & Atom Charge (2--7) & 6 \\
  & Ring being (yes/no) & 1 \\
  & Ring end (yes/no) & 1 \\
\hline
Atom type & C, H, O, N, others & 5 \\
\hline
Others & Surrounding Hydrogen Number (0--3) & 4 \\
 & Atom Formal Charge (-1, 0, 1) & 3 \\
 & Valence (1--5, other) & 6 \\
 & Ring Atom (yes/no) & 1 \\
 & Degree (1--4, other) & 5 \\
 & Aromaticity (yes/no) & 1 \\
 & Chirality (R/S/others) & 3 \\
 & Hybridization (sp, sp2, sp3, sp3d, sp3d2, unspecified, other) & 7 \\
\hline
Total & - & 57 \\
\hline
\end{tabular}
\caption{Feature Encoding Scheme for each SMILES Character}
\label{table:binary_encoded_smiles}
\end{table}

In this study, we utilize the Binary-Encoded-SMILES (BES) approach \cite{chen_bestox_2020}, whereby each character within the SMILES string is depicted as a bit vector. 27 bits are used to encode the SMILES symbols and alphabet, and 30 bits are used to encode atomic properties. The vector of each character is joined with all other character vectors forming a grid, and is padded to the longest string in the dataset. The descriptors used to create this length 57 vector for each SMILES character are detailed in Table \ref{table:binary_encoded_smiles}.

\subsection{Quantum Transfer Learning}
After training the quantum neural network (QNN), the real part of the first column of the unitary matrix $U_{\phi '}$ that prepares the learned state $\ket{\phi '}$ are the classical weights that may be used to further train the model on a classical device. This is the case since $U_{\phi '}$ always operates on $\ket{0}^{\otimes n}$, which yields only the first column of $U_{\phi '}$. Since the input data used in each quantum circuit must be L2 normalized to represent a valid quantum state, the corresponding classical component of the model must also work with normalized data to ensure a seamless transfer process. With regard to a CNN, this entails normalizing the data covered by a filter before taking the convolution with the weights, and is shown in Equation \ref{norm_CNN}. $\left|\bold{x}\right|_2$ is the normalized input data, $x_k$ is the $k$-th element of the $n$-dimensional input data vector, and $\bold{w}$ is the weight vector from the flattened convolutional filter.
\begin{equation}
    \text{output} = \left|\bold{x}\right|_2 \cdot \bold{w} = \left(\sqrt{\sum_{k=1}^n\left|x_k\right|^2}\right) \cdot \bold{w}
\label{norm_CNN}
\end{equation}

\begin{figure}[h]
    \centering
    \includegraphics[width=1\textwidth]{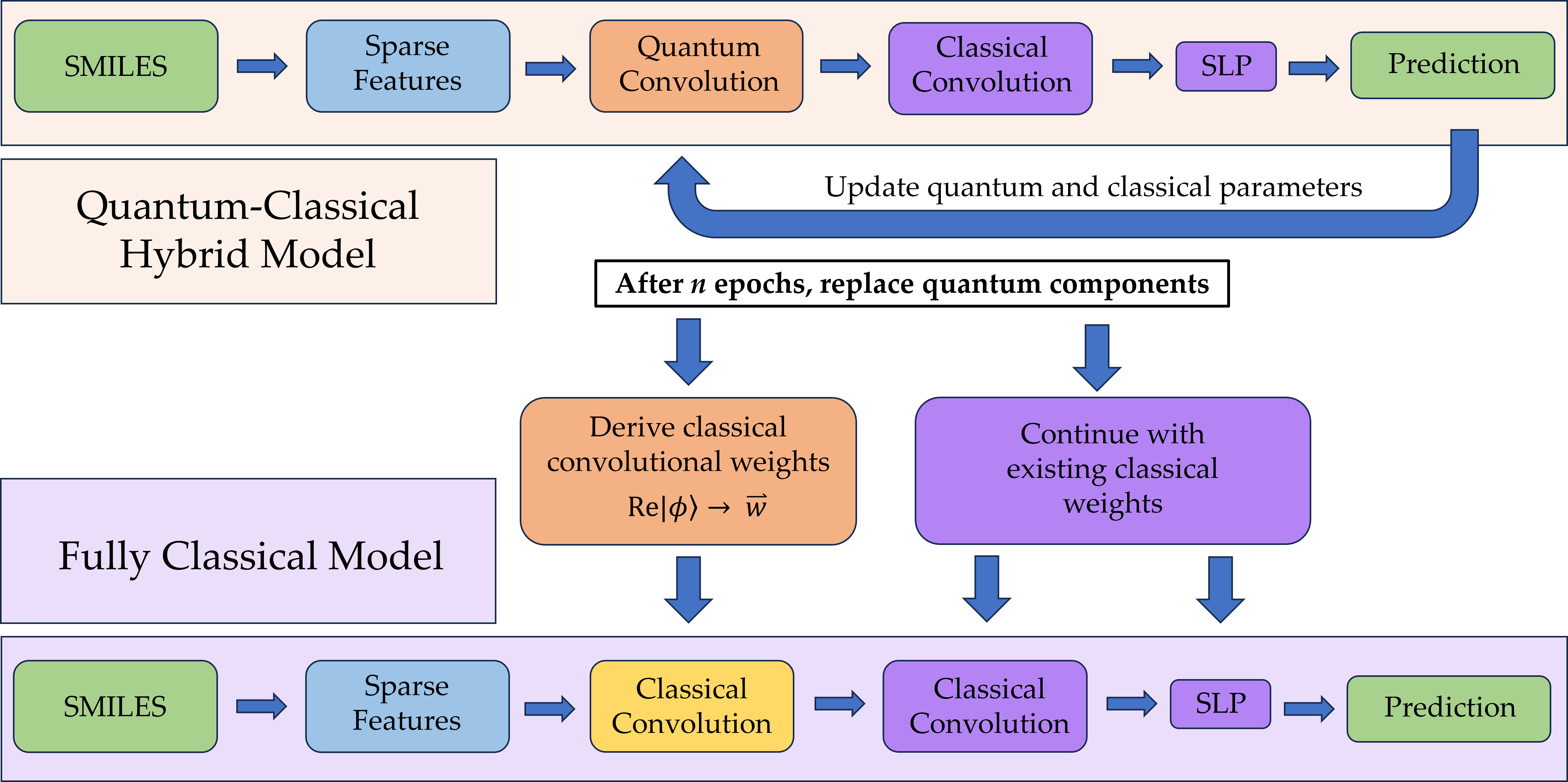}
    \caption{Quantum to classical neural network transfer pipeline. The convolutional weights are derived from the learned unitary matrix $U_\phi$ by operating this on the ground state as shown in Equation \ref{weight_prep}. The real part of the amplitudes of the evolved state $\ket{\phi}$ are identical to the normalized weights of a classical convolutional filter. The yellow classical convolutional block inherits the derived weights from the quantum state amplitudes. In order to maintain a mathematically exact learning transition between quantum and classical computations, this classical convolution L2-normalizes the input data covered by each filter. SLP indicates a Single Layer Perceptron. The exact architecture details are presented in Figure \ref{fig:architecture}.}
    \label{fig:pipeline}
\end{figure}

\begin{figure}[h]
    \centering
    \includegraphics[width=1\textwidth]{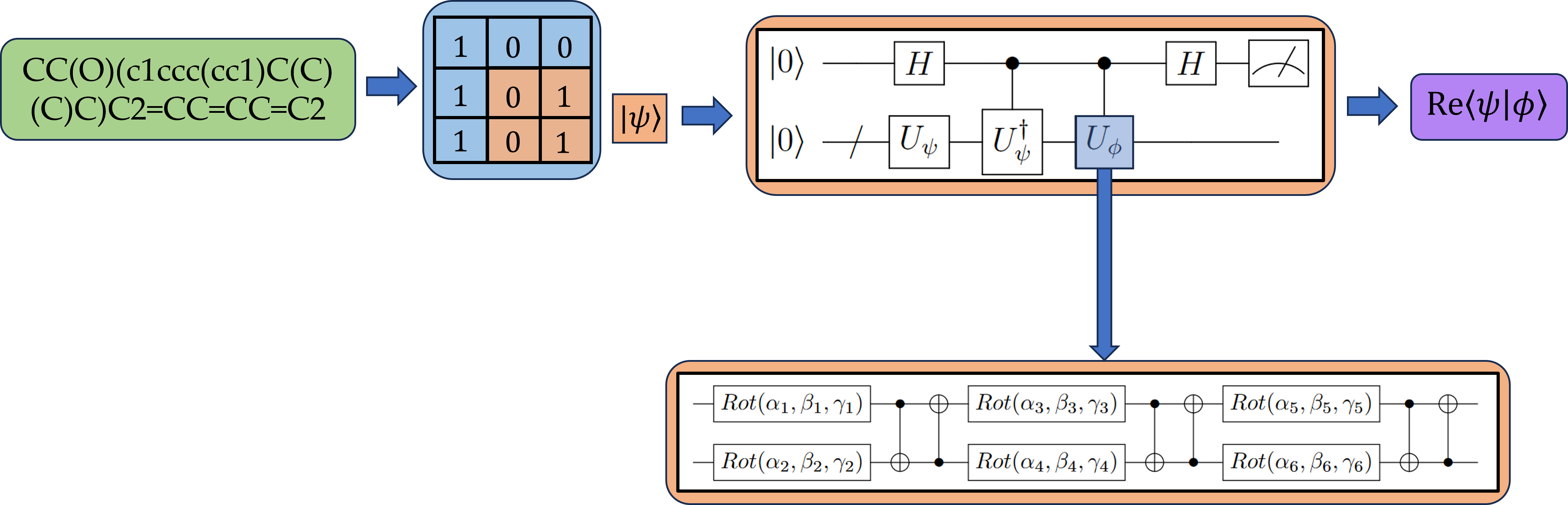}
    \caption{The variational quantum circuit used in the Hadamard test to calculate the real part of the inner product between an input data vector, $\ket{\psi}$ and a learnable weight vector $\ket{\phi}$. The unitary matrix of the \textit{Rot} gate is shown in Equation \ref{rot_gate}.}
    \label{fig:quantum_layer}
\end{figure}

The workflow of training and transferring our Hadamard test-based model from a hybrid QNN to a fully classical network is demonstrated in Figure \ref{fig:pipeline}. The details of the quantum component of the hybrid architecture are shown in Figure \ref{fig:quantum_layer}.

\section{Methods}

\subsection{Hardware}
All scripting was done in Python 3.9.18. The quantum simulation package Pennylane \cite{bergholm_pennylane_2022} version 0.35.0 was used, as well as PyTorch \cite{paszke_pytorch_2019} version 2.1.0  with CUDA Toolkit \cite{noauthor_cuda_nodate} 11.8 . An Intel i7-13700KF CPU, 12GB Nvidia GeForce RTX 3080Ti GPU, and 64GB of 3600MHz CL18 RAM were used for all computations. 
\subsection{Dataset}
The Tox21 dataset was downloaded from the National Institute of Health \cite{national_institute_of_health_tox21_nodate}. RDKit version 2023.09.1 was used to load the SMILES string and calculate the atomic features shown in Table \ref{table:binary_encoded_smiles}. SMILES strings with invalid valence shells and deuterium isotopes were discarded.  80\% of each assay dataset was used for training and 20\% for the test set. To ensure all binary encoded SMILES were the same dimension, they were padded with zeros for a final shape of 400 $\times$ 57.

\subsection{Machine Learning}
Binary cross entropy was used for the loss function and weights were updated with the Adam optimizer \cite{kingma_adam_2017}. A learning rate of 0.001 was used. Quantum weights were initialized from a uniform distribution between $0$ and $2\pi$. To ensure fair comparisons between the quantum and classical models, all other classical parameters were randomly initialized and copied to be identical between the models. The exponential decay rate for the first and second moment estimations were set to 0.9 and 0.999, respectively for all training. Epsilon was set to 1.0 $\times$ $10^{-8}$. The bias component of the quantum convolutional layer was trained classically. The output of each hidden layer in the model architecture was activated with ReLU. Given the imbalanced nature of the binary dataset, we find it appropriate to measure performance with the area under the curve of the receiver operating characteristic (ROC-AUC) \cite{hanley_method_1983}. This metric calculates the area under the curve of the plot of the true positive rate (Equation \ref{TPR}) versus the false positive rate (Equation \ref{FPR}). An ROC-AUC score of $1.0$ indicates the model predicts all samples correctly, while a score of $0.5$ indicates the model is no better than random chance.

\begin{equation}
    TPR = \frac{\text{True Positive}}{\text{True Positive} + \text{False Negative}}
    \label{TPR}
\end{equation}

\begin{equation}
    FPR = \frac{\text{False Positive}}{\text{False Positive} + \text{True Negative}}
    \label{FPR}
\end{equation}

\subsection{Quantum Computing}
Quantum states were prepared with the algorithm presented by Mottonen \textit{et al.} \cite{mottonen_transformation_2004} available in Pennylane, however in practice with this feature map, the algorithms by Veras \textit{et al.} \cite{de_veras_double_2022}  and Gleinig \textit{et al.}  \cite{gleinig_efficient_2021} would be better suited for use with NISQ devices. The variational component of the quantum circuit used three layers from Pennylane's \textit{StronglyEntanglingLayers}, inspired by Schuld \textit{et al.} \cite{schuld_circuit-centric_2020}. The unitary matrices comprising the general rotation gates are show in Equations \ref{rot_gate} and \ref{ry_rz_gate}.

\begin{equation}
    Rot\left(\alpha, \beta, \gamma\right) = RZ\left(\gamma\right)RY\left(\beta\right)RZ\left(\alpha\right)
\label{rot_gate}
\end{equation}

\begin{equation}
    RZ\left(\alpha\right) = \begin{bmatrix}
                                e^{-i\frac{\alpha}{2}} & 0 \\
                                0 & e^{i\frac{\alpha}{2}} \\
                            \end{bmatrix},
    RY\left(\beta\right) = \begin{bmatrix}
                                \cos{\left(\frac{\beta}{2}\right)} & -\sin{\left(\frac{\beta}{2}\right)} \\
                                \sin{\left(\frac{\beta}{2}\right)} & \cos{\left(\frac{\beta}{2}\right)} \\
                            \end{bmatrix}                        
\label{ry_rz_gate}
\end{equation}
The expectation value of the ancilla qubit in the computational basis was used as the input to the subsequent convolutional layer.

\section{Results and Discussion}

We compare our results to a fully classical model with identical architecture, except where the quantum convolutional filter is replaced by a classical filter of the same dimensions. It is shown in Table \ref{table:tox21_results} that the hybrid quantum-classical neural network (QNN) and fully classical neural network (CNN) perform similarly for each assay. On the overall dataset, the quantum and classical models achieved an ROC-AUC of 0.692 and 0.689, respectively. These results demonstrate that the same accuracy as the fully classical $\mathcal{O}(n^3)$ neural network can be achieved by the more efficient QNN with the $\mathcal{O}(n^2)$ quantum component.

\begin{table}[h]
\centering
\caption{Tox21 Test ROC-AUC}
\label{table:tox21_results}
\begin{tabular}{|>{\centering\arraybackslash}m{3.0cm}|>{\centering\arraybackslash}m{2.5cm}|>{\centering\arraybackslash}m{2.5cm}|>{\centering\arraybackslash}m{5.5cm}|}
\hline
\textbf{Assay} & \textbf{QNN} & \textbf{CNN} & \textbf{QNN \textit{transferred to} CNN} \\ \hline
nr-ahr & 0.619 & 0.632 & 0.619\\ 
nr-ar & 0.725 & 0.764 & 0.727\\ 
nr-ar-lbd & 0.800 & 0.847 & 0.800\\ 
nr-aromatase & 0.724 & 0.722 & 0.726\\ 
nr-er & 0.677 & 0.674 & 0.676\\
nr-er-lbd & 0.641 & 0.623 & 0.640\\
nr-ppar-gamma & 0.636 & 0.586 & 0.626\\
sr-are & 0.701 & 0.705 & 0.701\\
sr-atad5 & 0.682 & 0.644 & 0.679\\
sr-hse & 0.677 & 0.668 & 0.675\\
sr-mmp & 0.722 & 0.698 & 0.721\\
sr-p53 & 0.692 & 0.689 & 0.691\\ \hline
\textbf{Average} & \textbf{0.692} & \textbf{0.689} & \textbf{0.691}\\ \hline
\end{tabular}
\end{table}

\begin{figure}[!h]
    \centering
    \begin{subfigure}[b]{0.47\textwidth}
        \includegraphics[width=\textwidth]{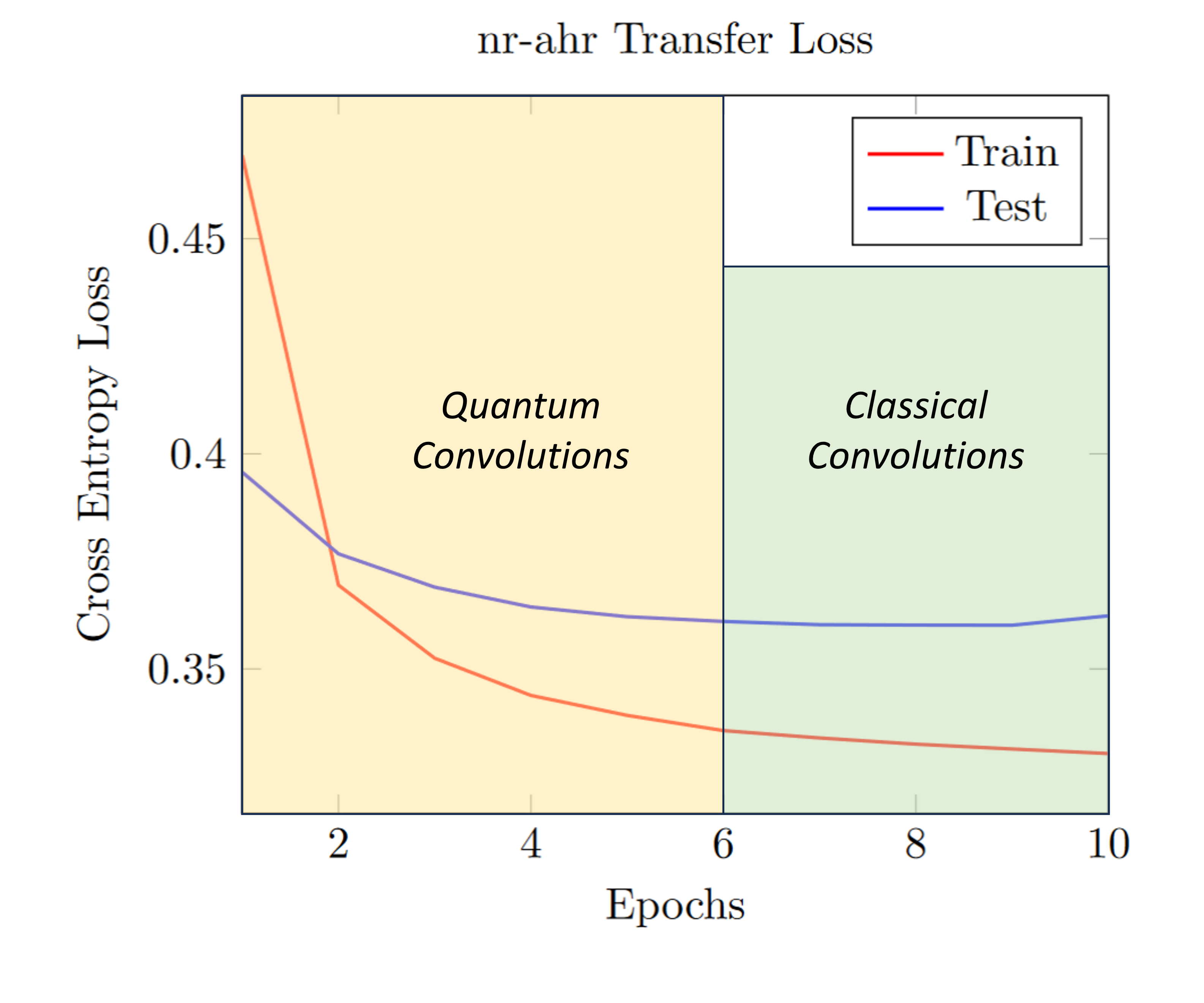}
        \caption{}
        \label{fig:nr_ahr_transfer_loss}
    \end{subfigure}
    \hfill
    \begin{subfigure}[b]{0.47\textwidth}
        \includegraphics[width=\textwidth]{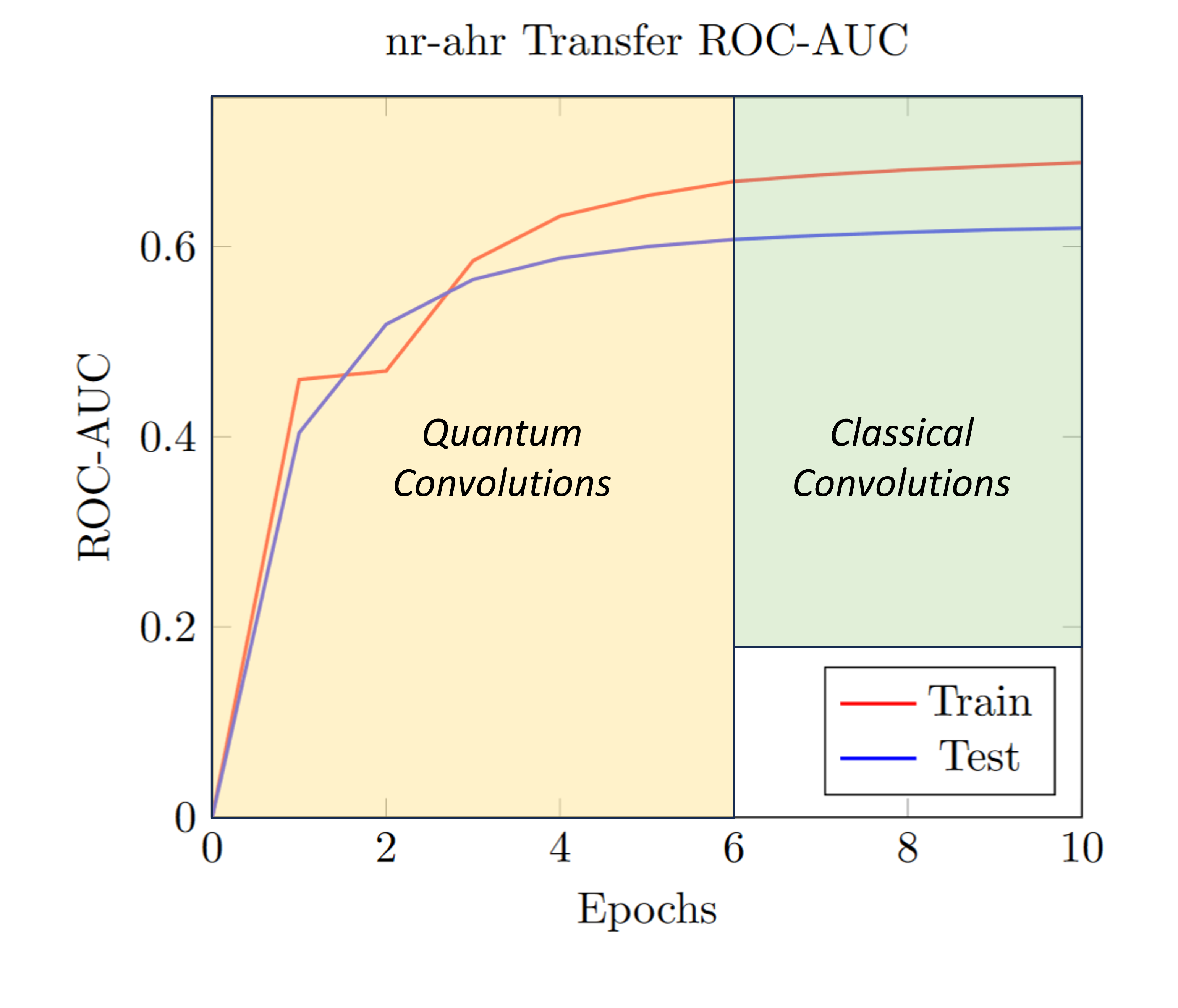}
        \caption{}
        \label{fig:nr_ahr_transfer_roc_auc}
    \end{subfigure}
    \begin{subfigure}[b]{0.47\textwidth}
        \includegraphics[width=\textwidth]{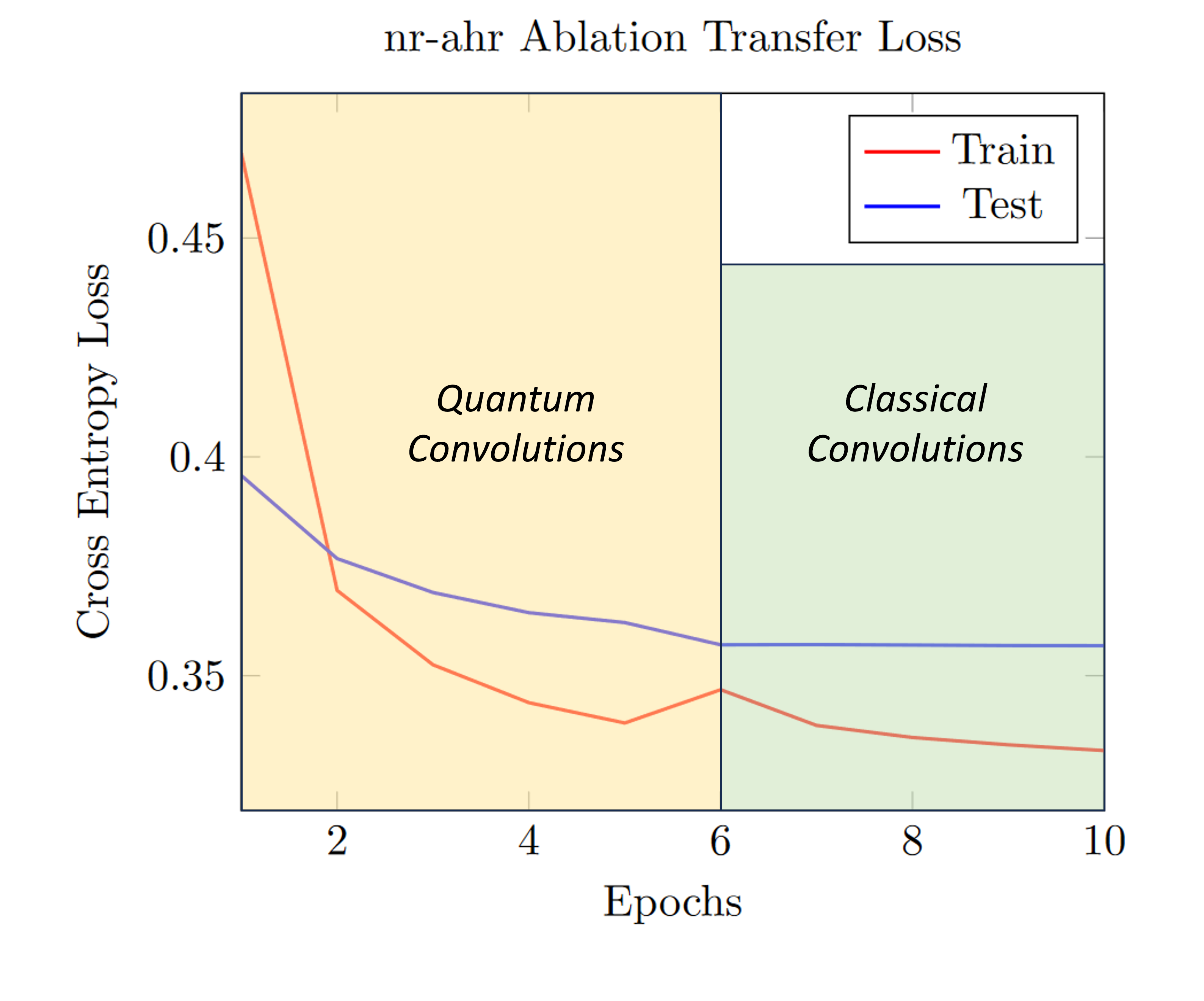}
        \caption{}
        \label{fig:nr_ahr_ablation_loss}
    \end{subfigure}
    \hfill
    \begin{subfigure}[b]{0.47\textwidth}
        \includegraphics[width=\textwidth]{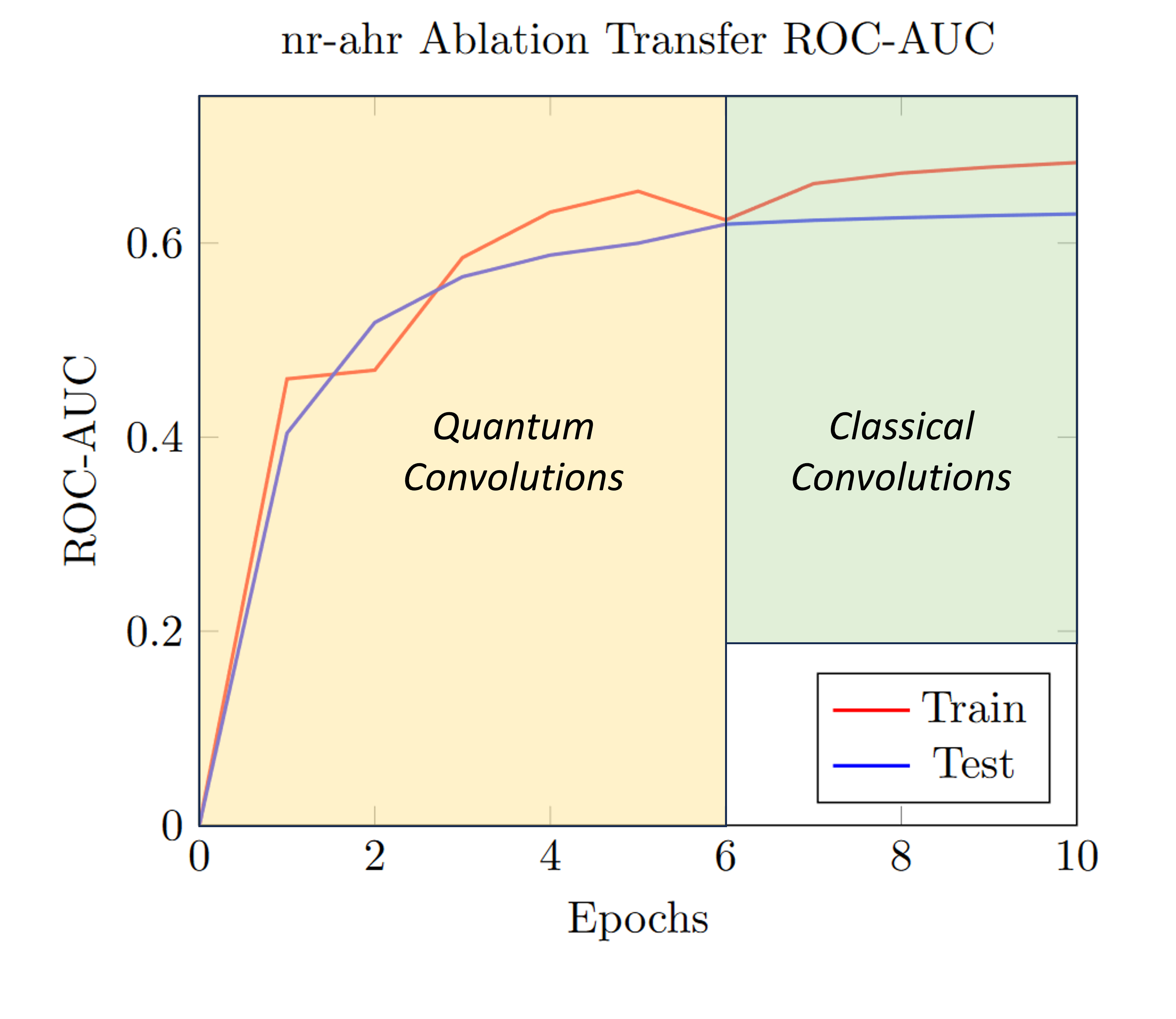}
        \caption{}
        \label{fig:nr_ahr_ablation_roc_auc}
    \end{subfigure}
    \caption{Model performance before and after weight transfer and ablation study results. (a) Transfer learning loss, (b) Transfer learning ROC-AUC, (c) Ablation study loss, (d) Ablation study ROC-AUC. The yellow shading (Epochs 1-5) indicates where the model was trained with quantum convolutions. The green shading (Epochs 6-10) indicates where the model was trained with classical convolutions. The ablation study (c and d) shows the model's performance where the weights from the quantum layer were initialized to random values instead of deriving them from $U_\phi$.}
    \label{fig:nr_ahr_combined}
\end{figure}

Additionally, we demonstrate that the fully classical model is able to resume the training process started by the quantum model. As shown in Figures \ref{fig:nr_ahr_transfer_loss} and \ref{fig:nr_ahr_transfer_roc_auc}, the training continues seamlessly after 5 epochs, where weights for the classical convolutional filter were derived from Equation \ref{weight_prep}. The training curves for all assays are made available in the Supplemental Information.

The first convolutional layer transferred from quantum training to classical training contains only a small portion of the model's total parameters. Despite this, the first layer of feature extraction plays a crucial role in the model's overall performance, and improperly computing the weights during the transfer procedure will have a demonstrable impact. To validate our framework of properly transferring these weights, we show the resulting model's performance after randomly initializing the weights from the first convolutional layer, while copying the remaining learned weights. Epoch 6, the first epoch after transferring to a fully classical architecture, shows a spike in cross entropy loss and a decrease in ROC-AUC in Figures \ref{fig:nr_ahr_ablation_loss} and \ref{fig:nr_ahr_ablation_roc_auc}. Despite being so few parameters, this ablation study is indicative that our framework correctly transfers these weights to be fine-tuned classically.

\section{Conclusions}
In summary, we present the first quantum-to-classical transfer learning framework where weights initially learned from a quantum circuit can be fine-tuned classically. Assuming efficient state preparation, we show a theoretical quantum advantage by reducing the complexity of calculating matrix products from $\mathcal{O}(n^3)$ to $\mathcal{O}(n^2)$ using a modified version of the Hadamard test, which uses half the qubits of the previously proposed swap test for a quantum neuron and without the need for quantum phase estimation. We apply these methods to toxicity prediction using the Tox21 dataset, and show that our frameworks perform equivalently to a fully classical model with approximately the same number of parameters. As quantum devices improve and the assumption of efficient state preparation becomes fulfilled, we hope this framework will provide an implementable polynomial speedup for machine learning networks, and consequently more efficiently navigate the chemical space to find life-saving pharmaceuticals.

\section{Data and Software Availability}
The data and code to reproduce the results presented here are available as open-source at \href{https://github.com/anthonysmaldone/Quantum-to-Classical-Transfer-Learning}{https://github.com/anthonysmaldone/Quantum-to-Classical-Transfer-Learning}.

\section{Author Contributions}
AMS designed the idea, AMS developed the software, AMS conducted the research, AMS and VSB analyzed the data, AMS wrote the article.

\section{Acknowledgements}
We acknowledge financial support from the National Institute of Health Biophysical Training Grant (1T32GM149438-01) [AS], as well as the financial support provided by the CCI Phase I: National Science Foundation Center for Quantum Dynamics on Modular Quantum Devices (CQD-MQD) under Award Number 2124511 [VSB]. Furthermore, we acknowledge seed funding from Yale University through the National Science Foundation Engines Development Award: Advancing Quantum Technologies (CT) under Award Number 2302908 and high-performance computer time from the National Energy Research Scientific Computing Center and from the Yale University Faculty of Arts and Sciences High Performance Computing Center.

\bibliography{references}

\newpage
\section{Supplemental Information}

\setcounter{table}{0} 
\renewcommand{\thetable}{S\arabic{table}}

\begin{table}[h]
\centering
\caption{Model Parameter Count}
\label{table:parameters}
\begin{tabular}{|>{\centering\arraybackslash}m{3.0cm}|>{\centering\arraybackslash}m{2.5cm}|}
\hline
\textbf{Model} & \textbf{Parameters} \\ \hline
QNN & 5188 \\ 
CNN & 5174 \\
QNN to CNN & 5188 to 5174 \\\hline
\end{tabular}
\end{table}

\setcounter{figure}{0} 
\renewcommand{\thefigure}{S\arabic{figure}} 

\begin{figure}[!h]
    \centering
    \begin{subfigure}[b]{0.5\textwidth}
        \includegraphics[width=\textwidth]{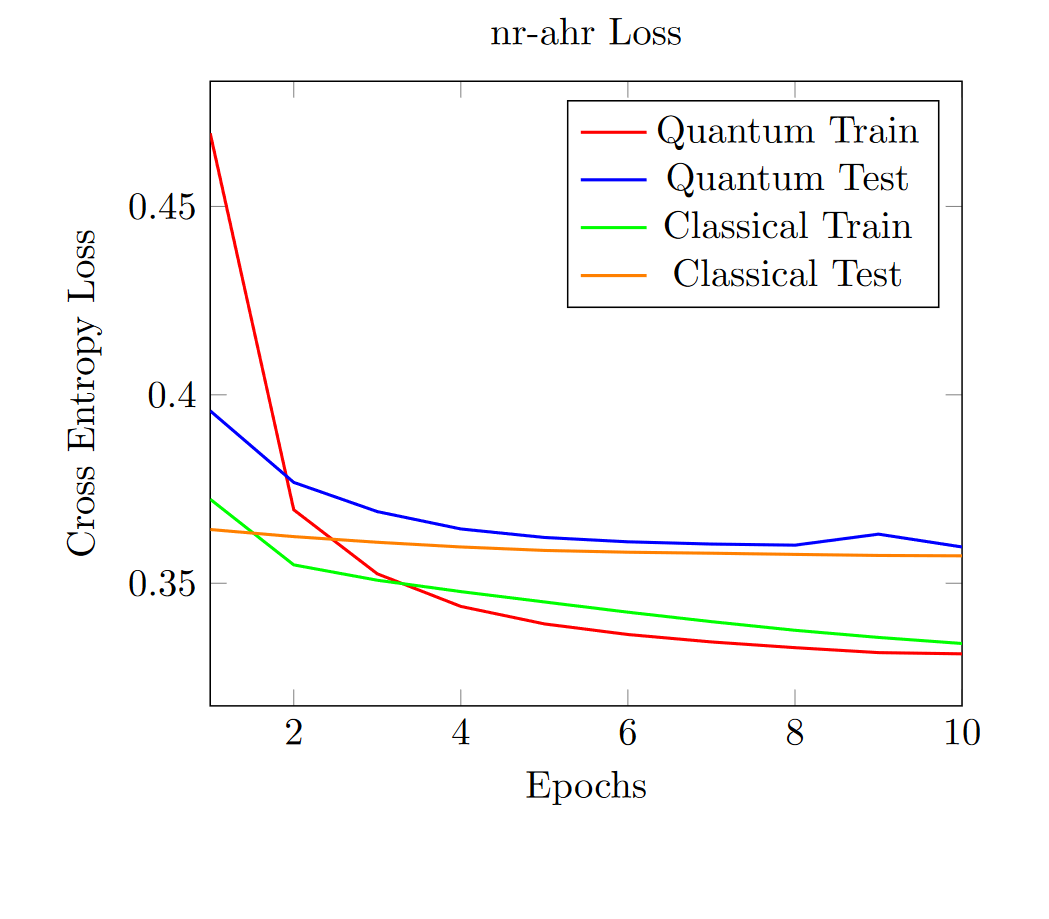}
        \caption{}
    \end{subfigure}
    \hfill
    \begin{subfigure}[b]{0.5\textwidth}
        \includegraphics[width=\textwidth]{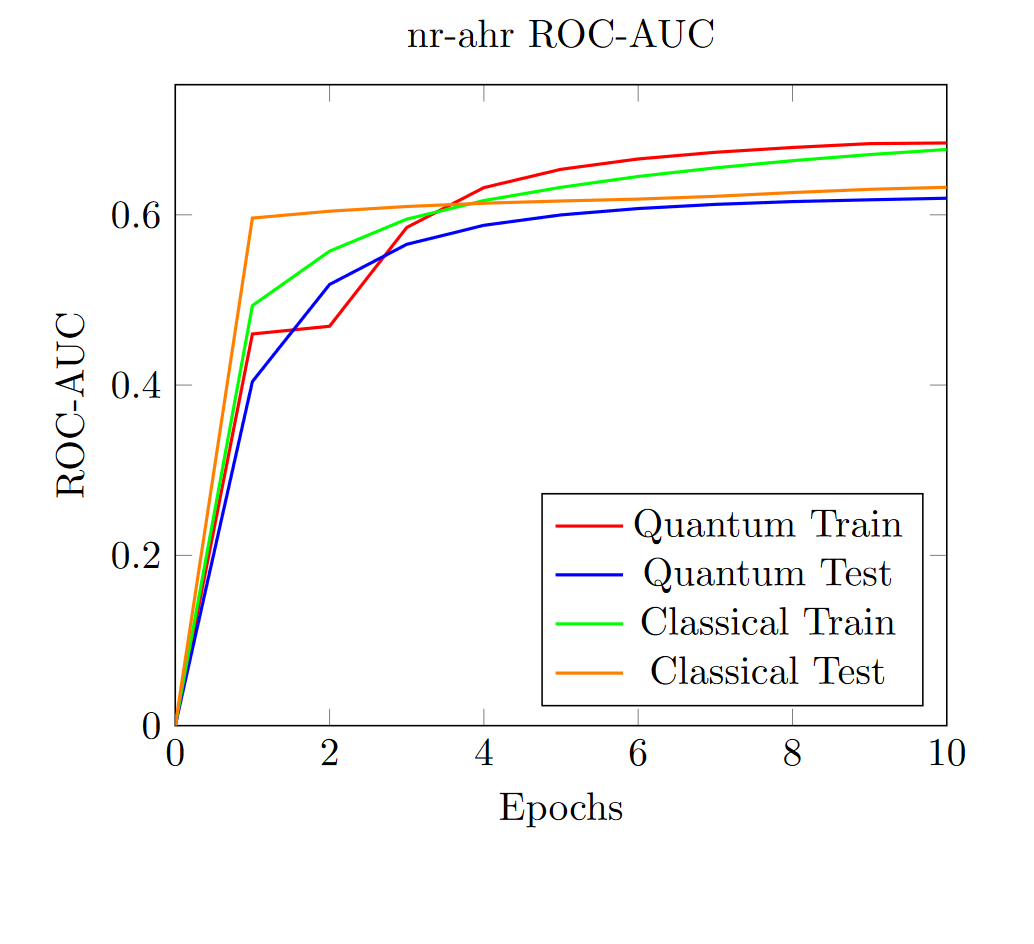}
        \caption{}
    \end{subfigure}
    \caption{Training curves for the nr-ahr assay}
\end{figure}

\begin{figure}[!h]
    \centering
    \begin{subfigure}[b]{0.47\textwidth}
        \includegraphics[width=\textwidth]{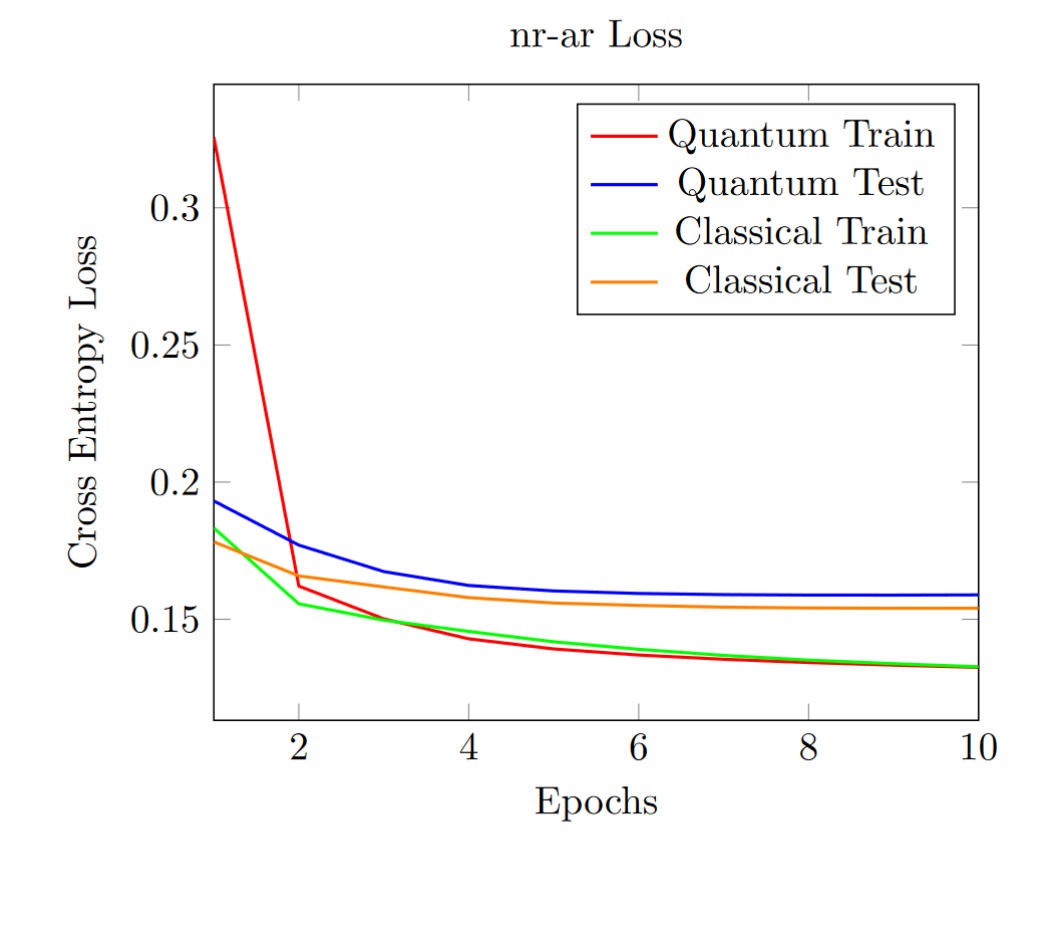}
        \caption{}
    \end{subfigure}
    \hfill
    \begin{subfigure}[b]{0.47\textwidth}
        \includegraphics[width=\textwidth]{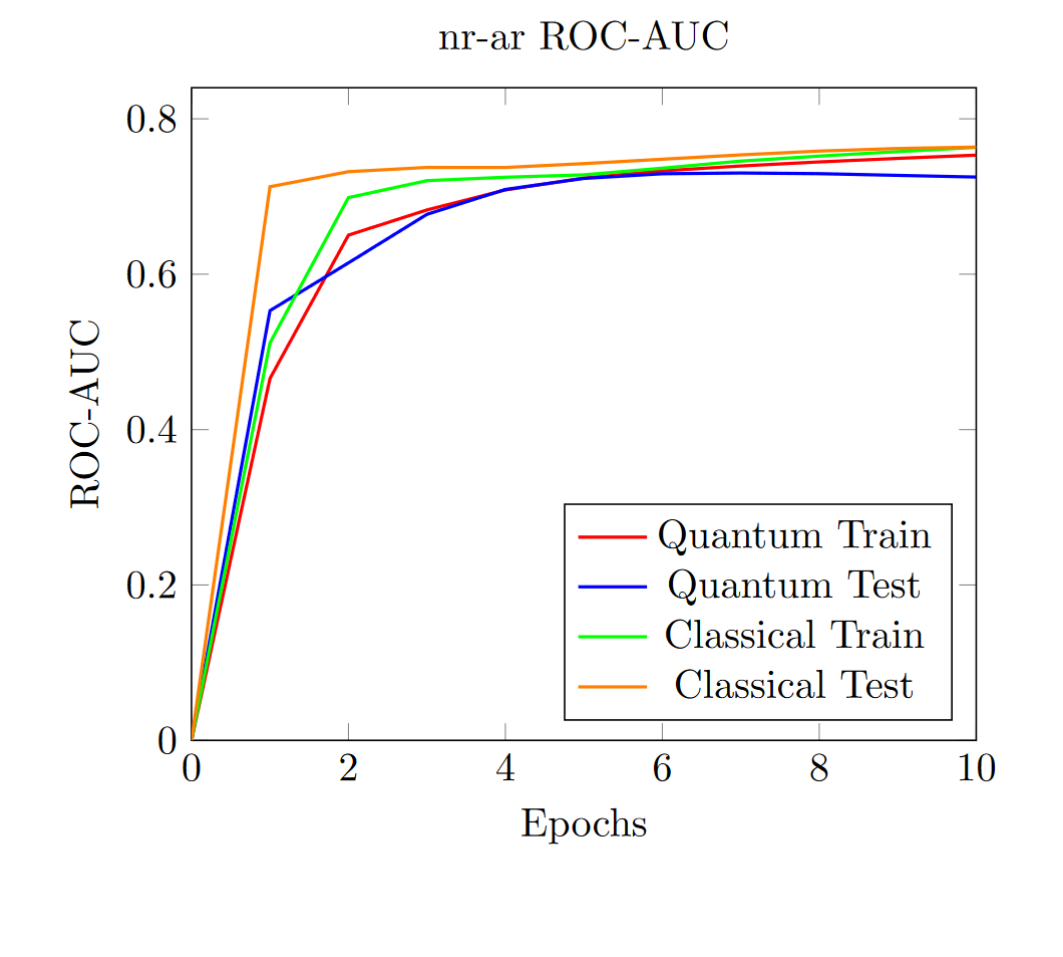}
        \caption{}
    \end{subfigure}
    \begin{subfigure}[b]{0.47\textwidth}
        \includegraphics[width=\textwidth]{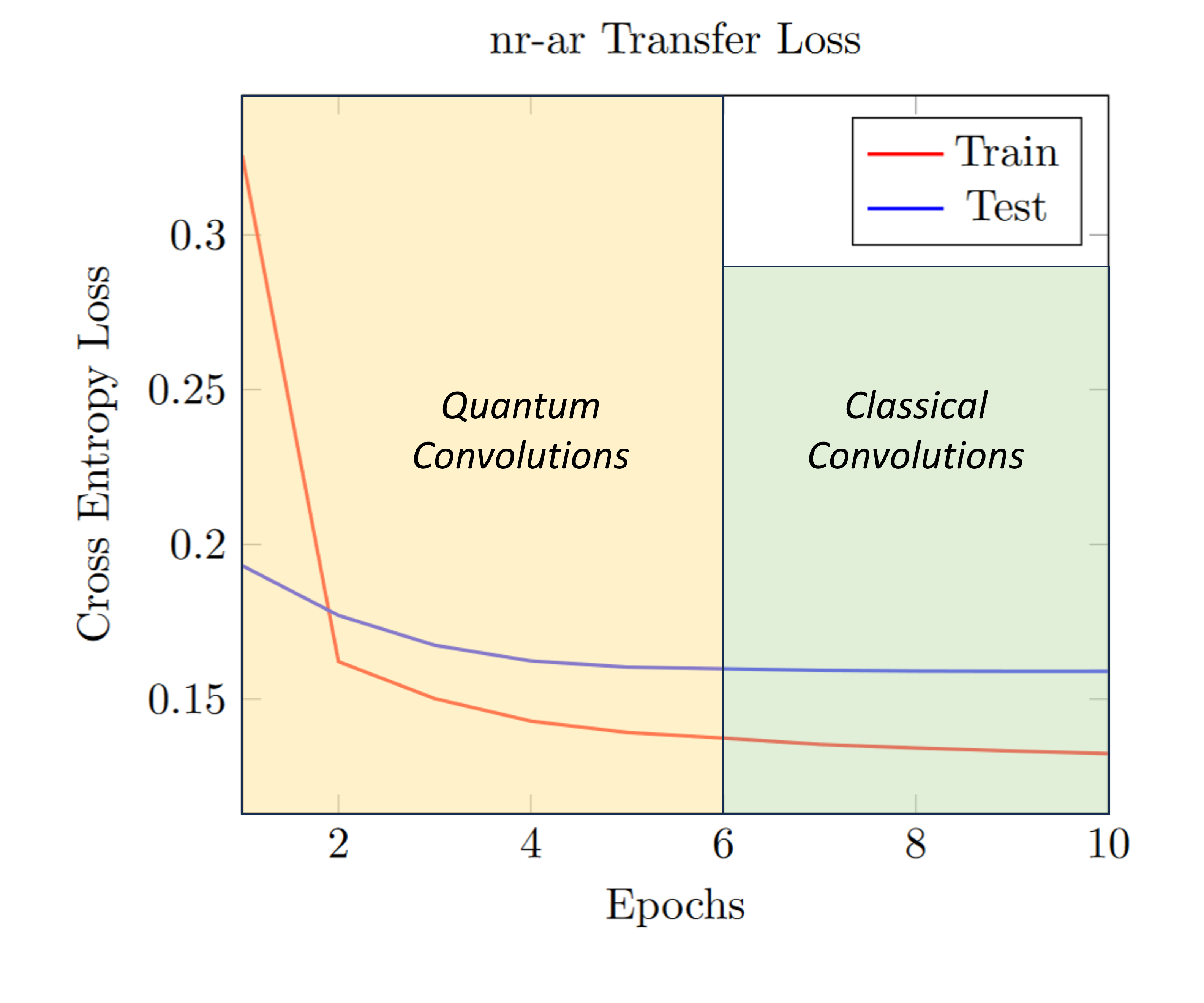}
        \caption{}
    \end{subfigure}
    \hfill
    \begin{subfigure}[b]{0.47\textwidth}
        \includegraphics[width=\textwidth]{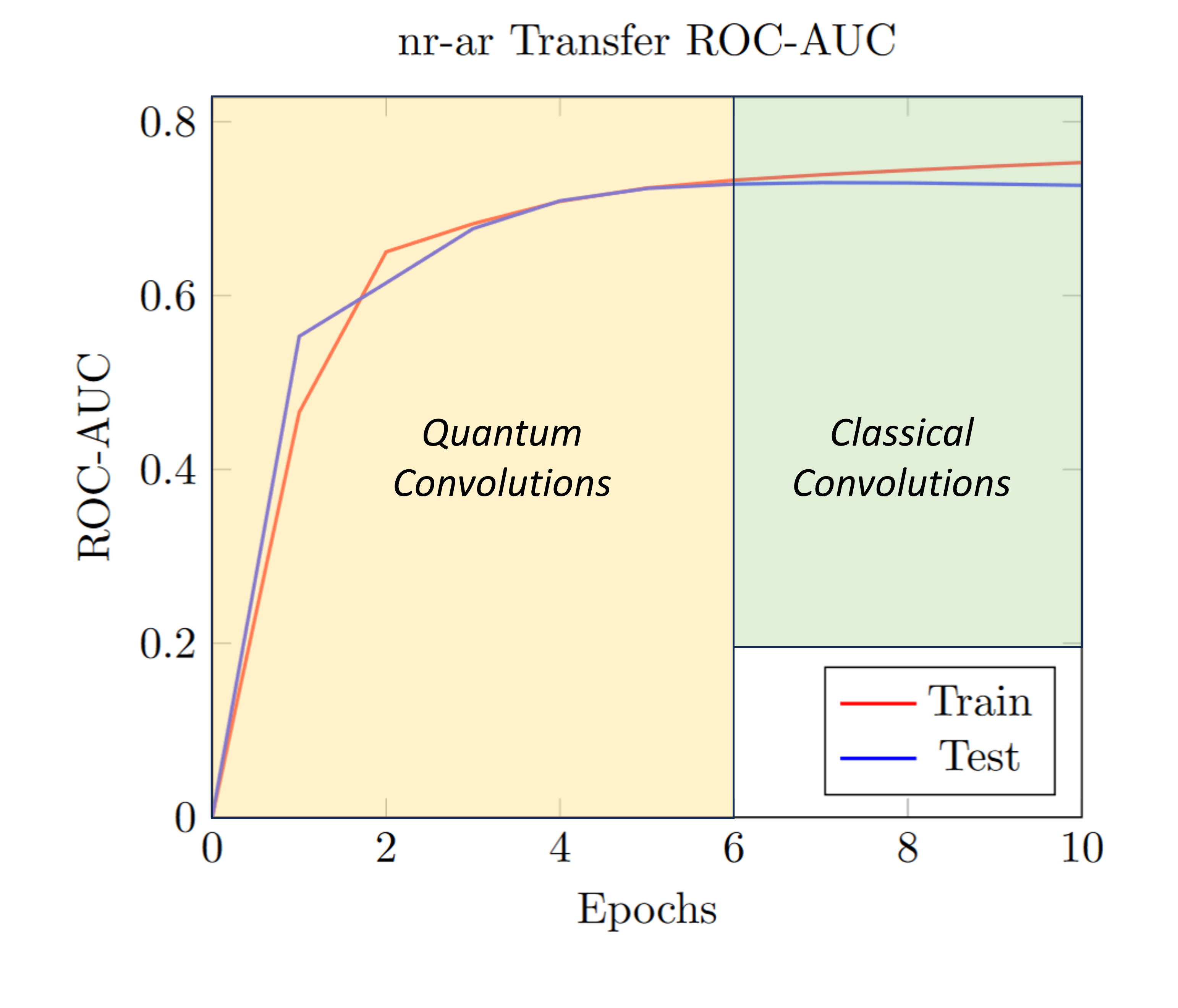}
        \caption{}
    \end{subfigure}
    \caption{Training curves for the nr-ar assay}
\end{figure}

\begin{figure}[!h]
    \centering
    \begin{subfigure}[b]{0.47\textwidth}
        \includegraphics[width=\textwidth]{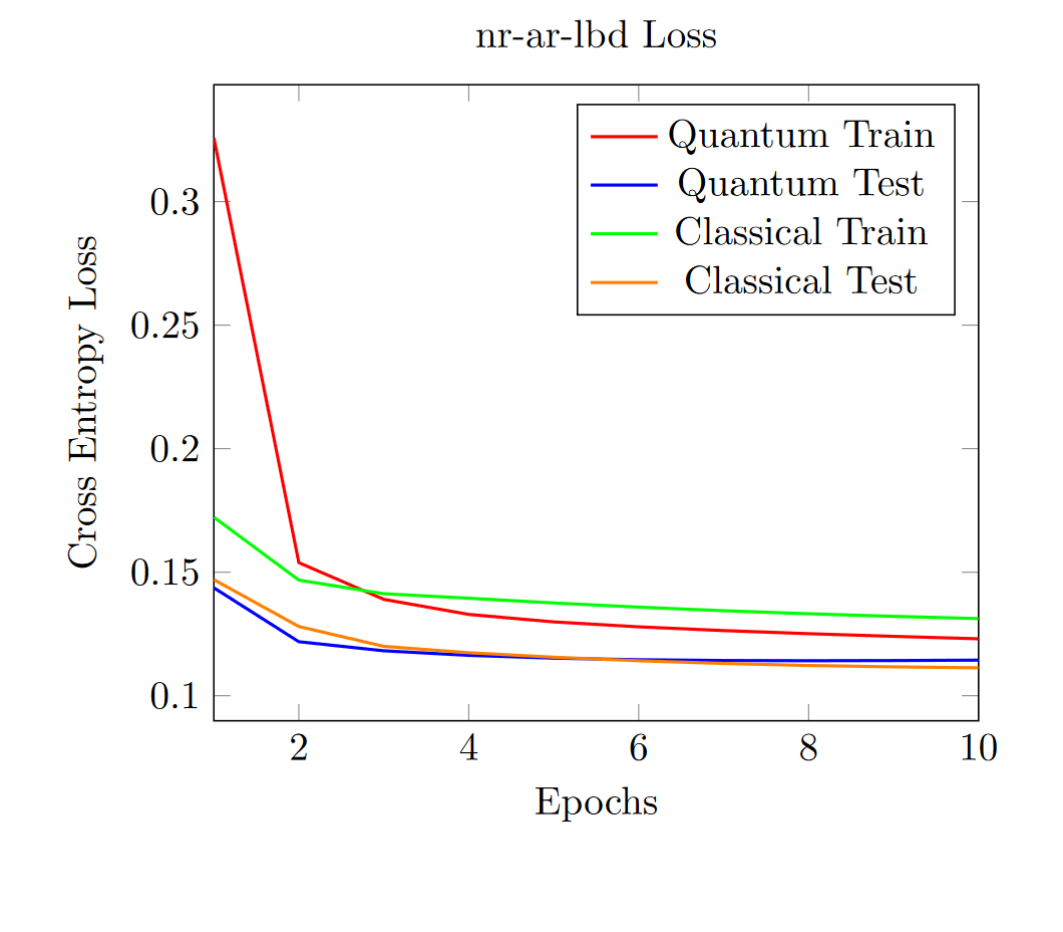}
        \caption{}
    \end{subfigure}
    \hfill
    \begin{subfigure}[b]{0.47\textwidth}
        \includegraphics[width=\textwidth]{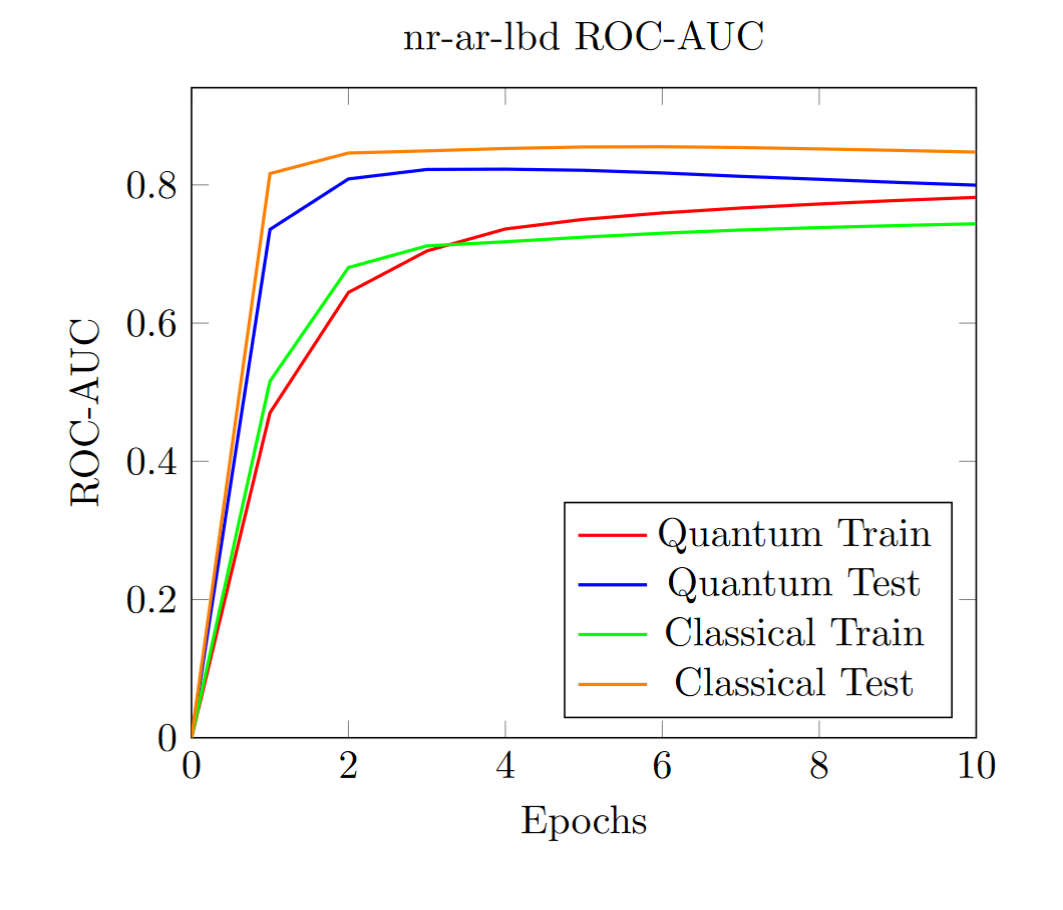}
        \caption{}
    \end{subfigure}
    \begin{subfigure}[b]{0.47\textwidth}
        \includegraphics[width=\textwidth]{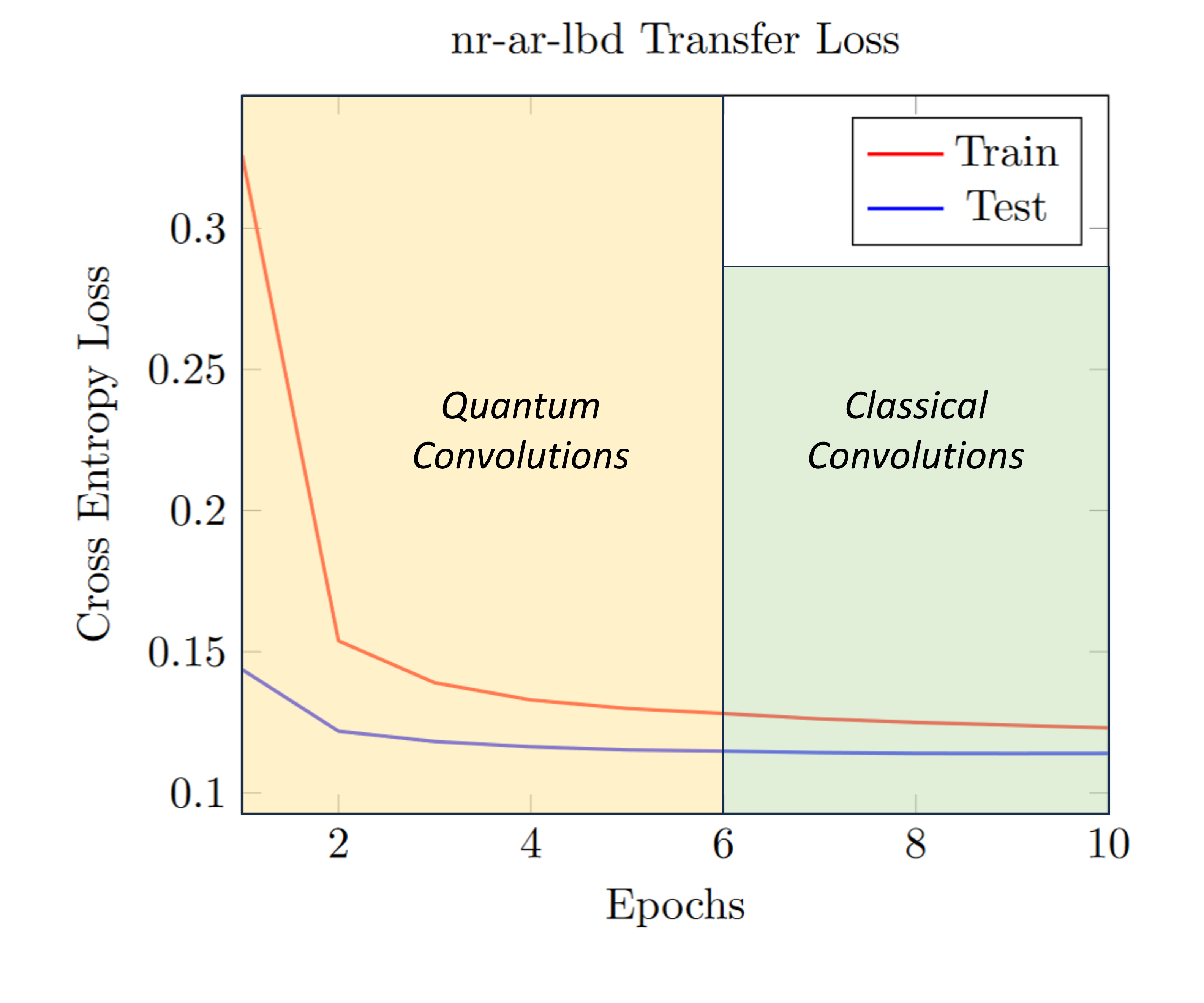}
        \caption{}
    \end{subfigure}
    \hfill
    \begin{subfigure}[b]{0.47\textwidth}
        \includegraphics[width=\textwidth]{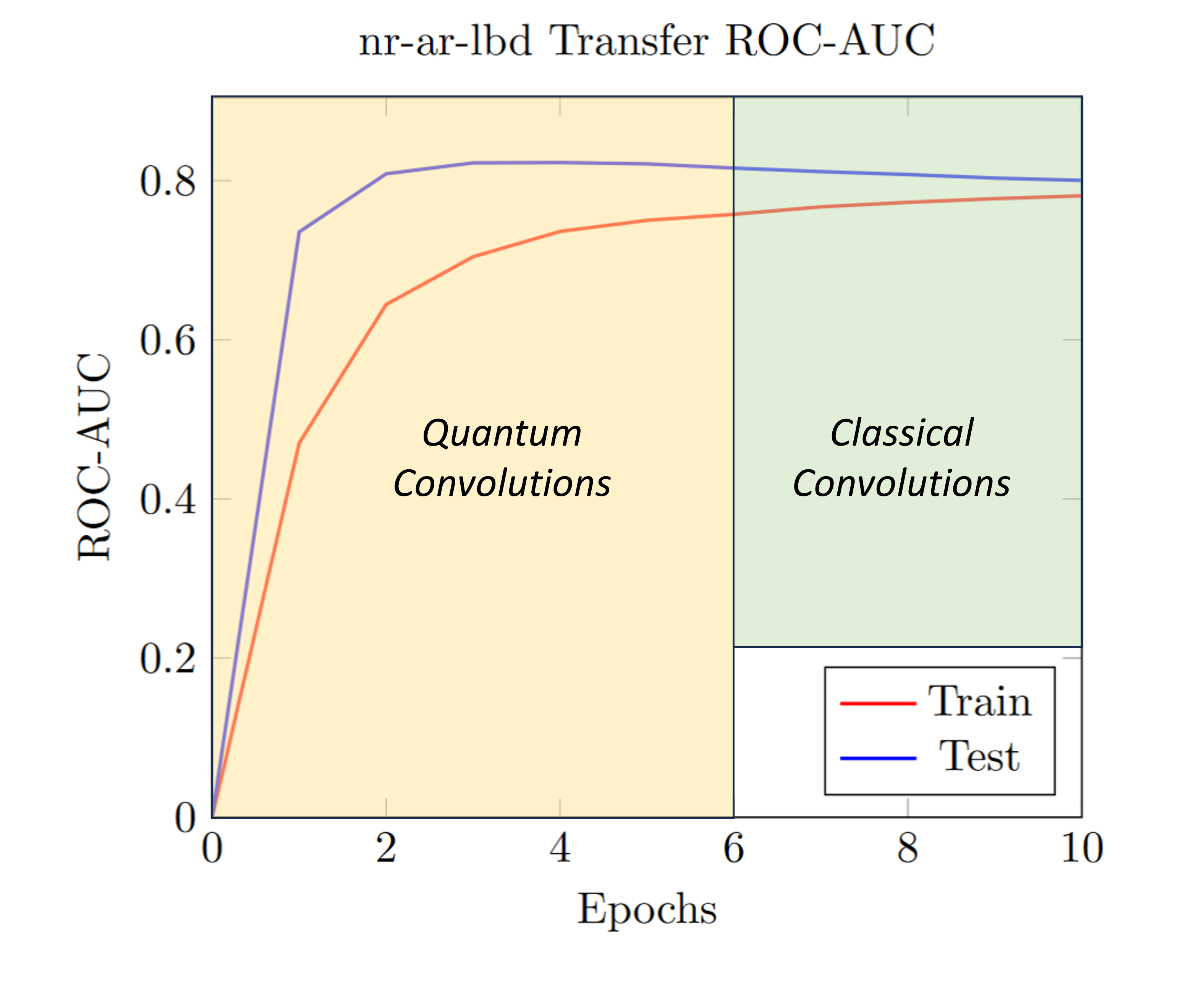}
        \caption{}
    \end{subfigure}
    \caption{Training curves for the nr-ar-lbd assay}
\end{figure}

\begin{figure}[!h]
    \centering
    \begin{subfigure}[b]{0.47\textwidth}
        \includegraphics[width=\textwidth]{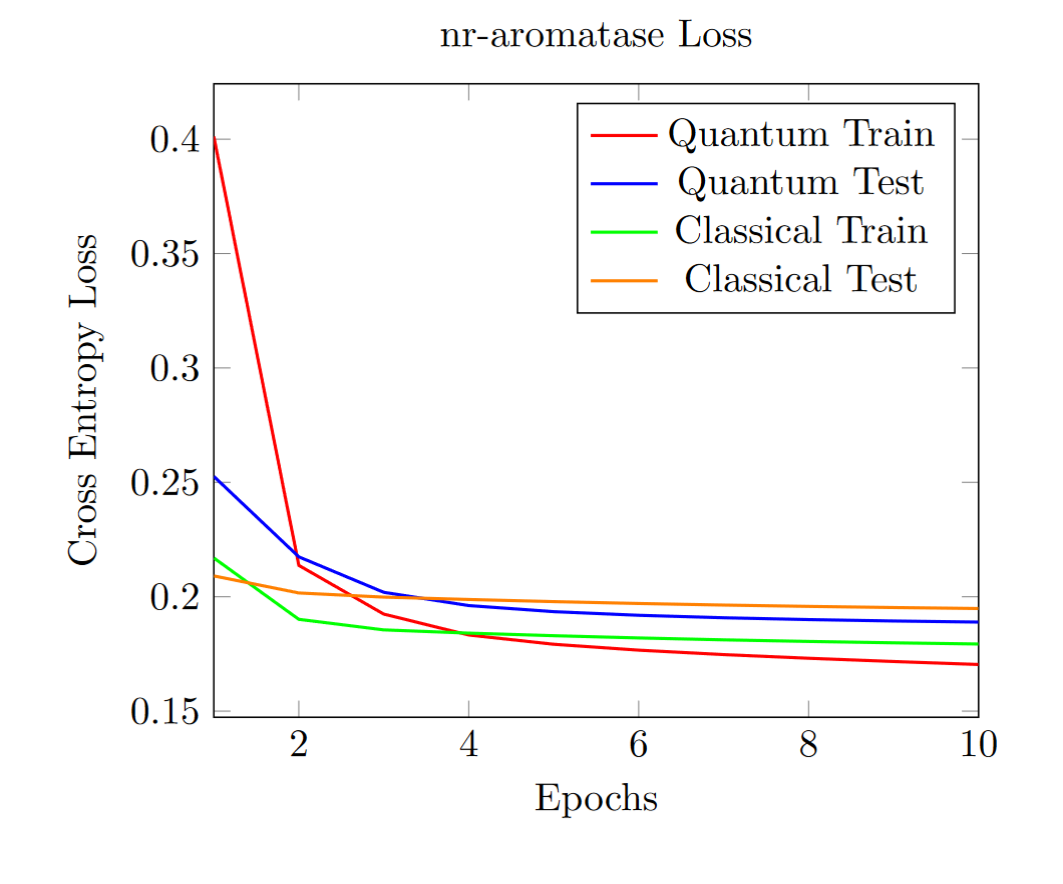}
        \caption{}
    \end{subfigure}
    \hfill
    \begin{subfigure}[b]{0.47\textwidth}
        \includegraphics[width=\textwidth]{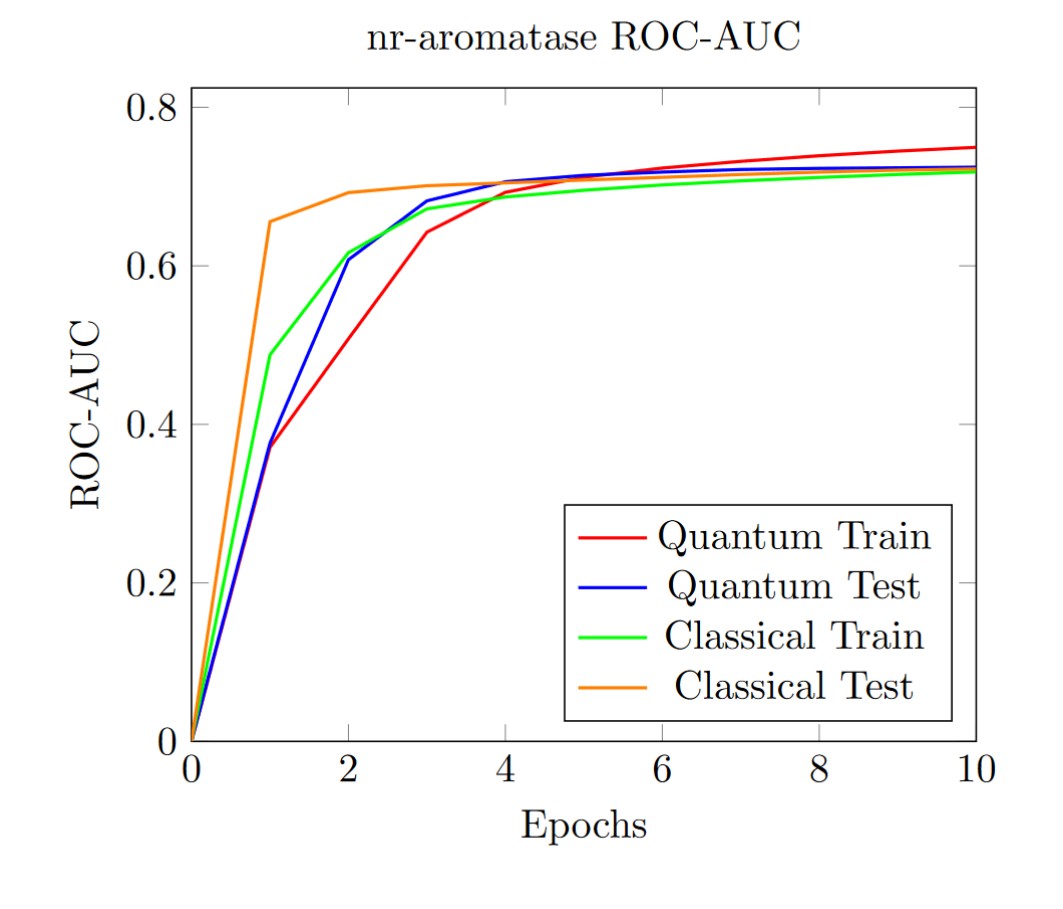}
        \caption{}
    \end{subfigure}
    \begin{subfigure}[b]{0.47\textwidth}
        \includegraphics[width=\textwidth]{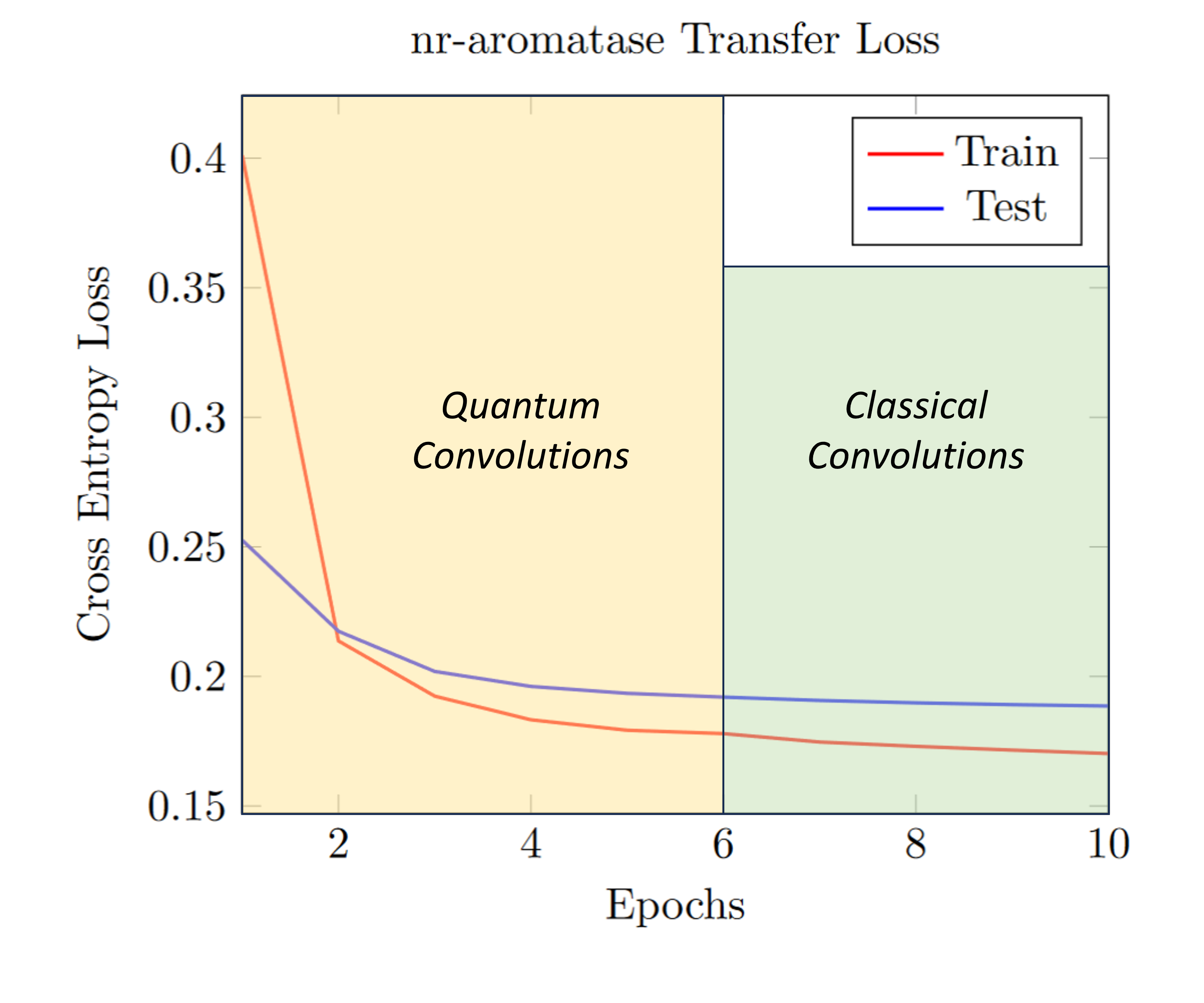}
        \caption{}
    \end{subfigure}
    \hfill
    \begin{subfigure}[b]{0.47\textwidth}
        \includegraphics[width=\textwidth]{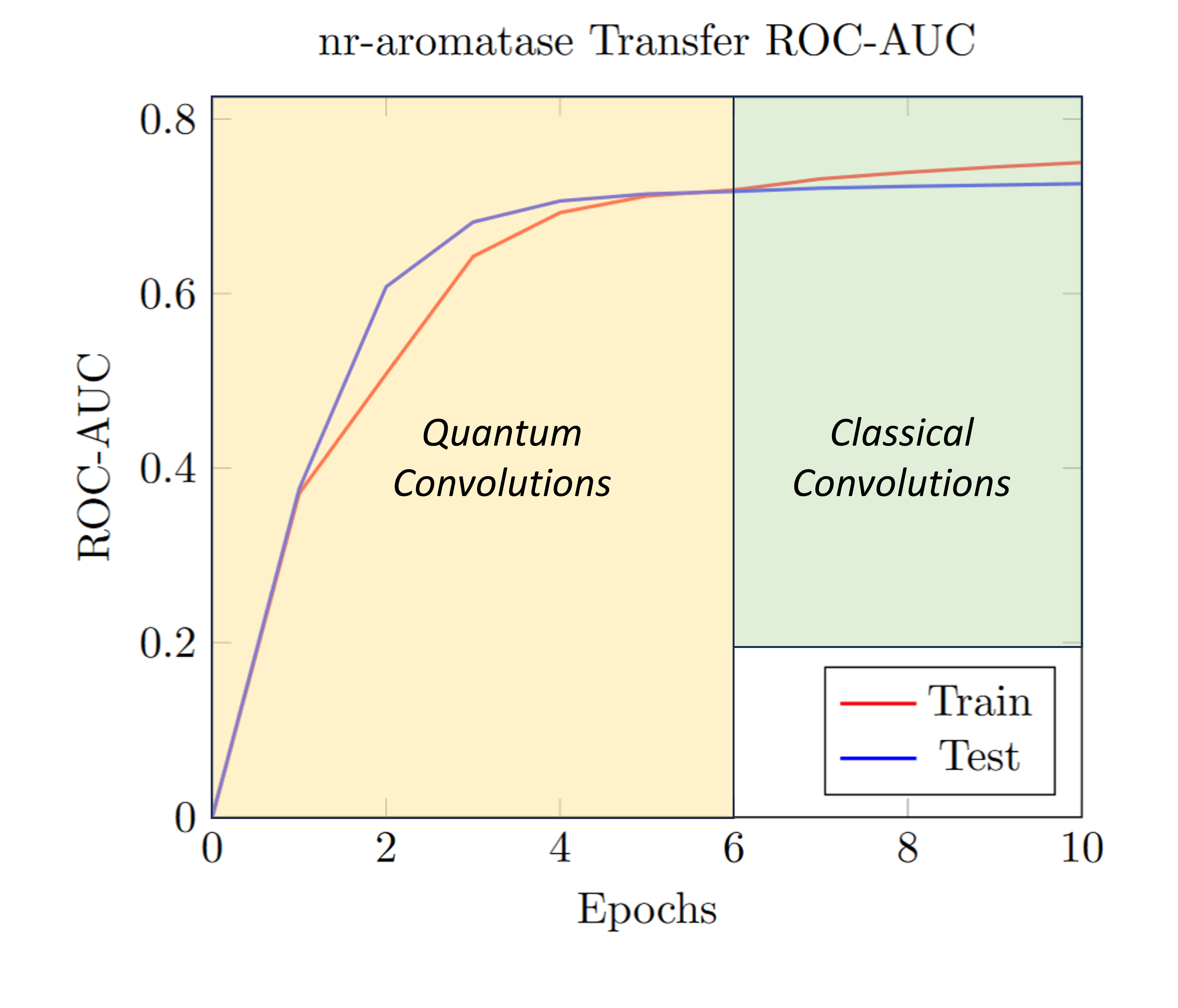}
        \caption{}
    \end{subfigure}
    \caption{Training curves for the nr-aromatase assay}
\end{figure}

\begin{figure}[!h]
    \centering
    \begin{subfigure}[b]{0.48\textwidth}
        \includegraphics[width=\textwidth]{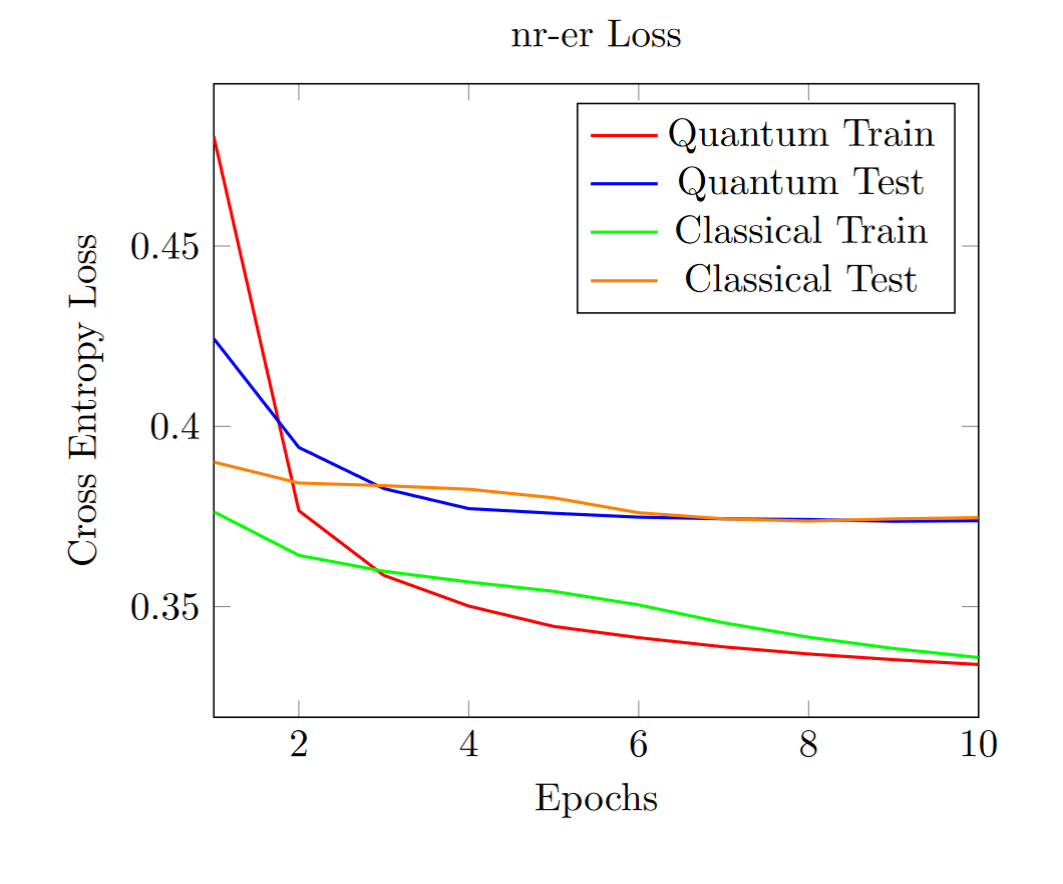}
        \caption{}
    \end{subfigure}
    \hfill
    \begin{subfigure}[b]{0.46\textwidth}
        \includegraphics[width=\textwidth]{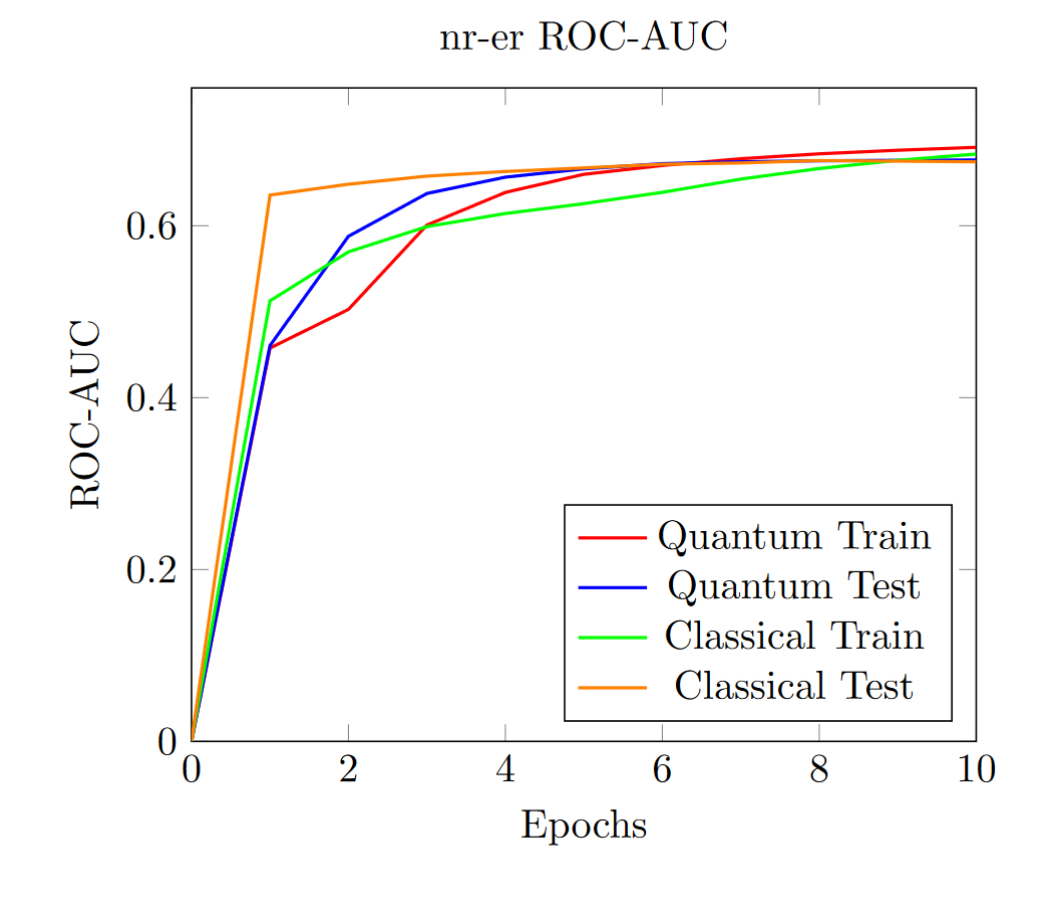}
        \caption{}
    \end{subfigure}
    \begin{subfigure}[b]{0.47\textwidth}
        \includegraphics[width=\textwidth]{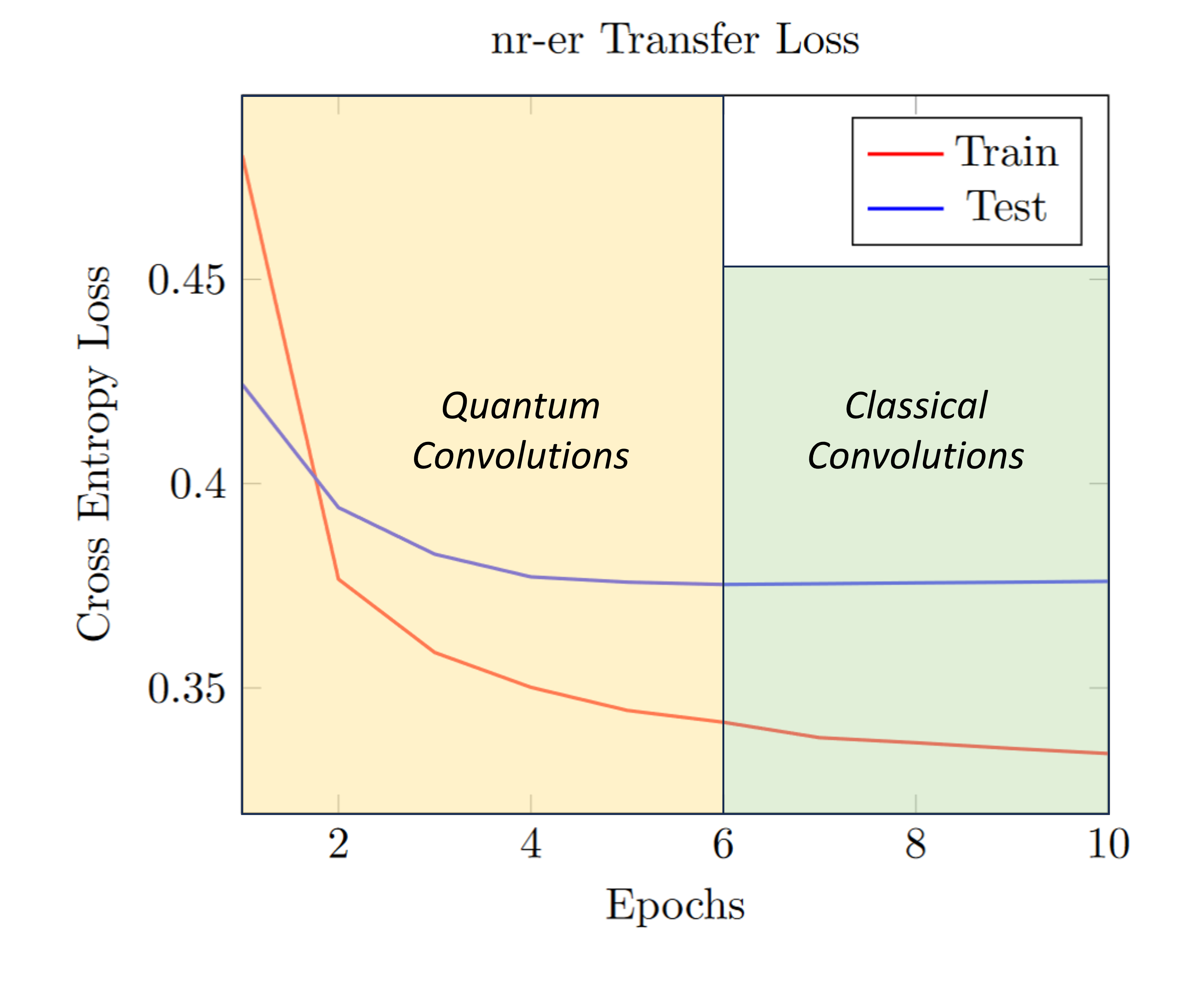}
        \caption{}
    \end{subfigure}
    \hfill
    \begin{subfigure}[b]{0.47\textwidth}
        \includegraphics[width=\textwidth]{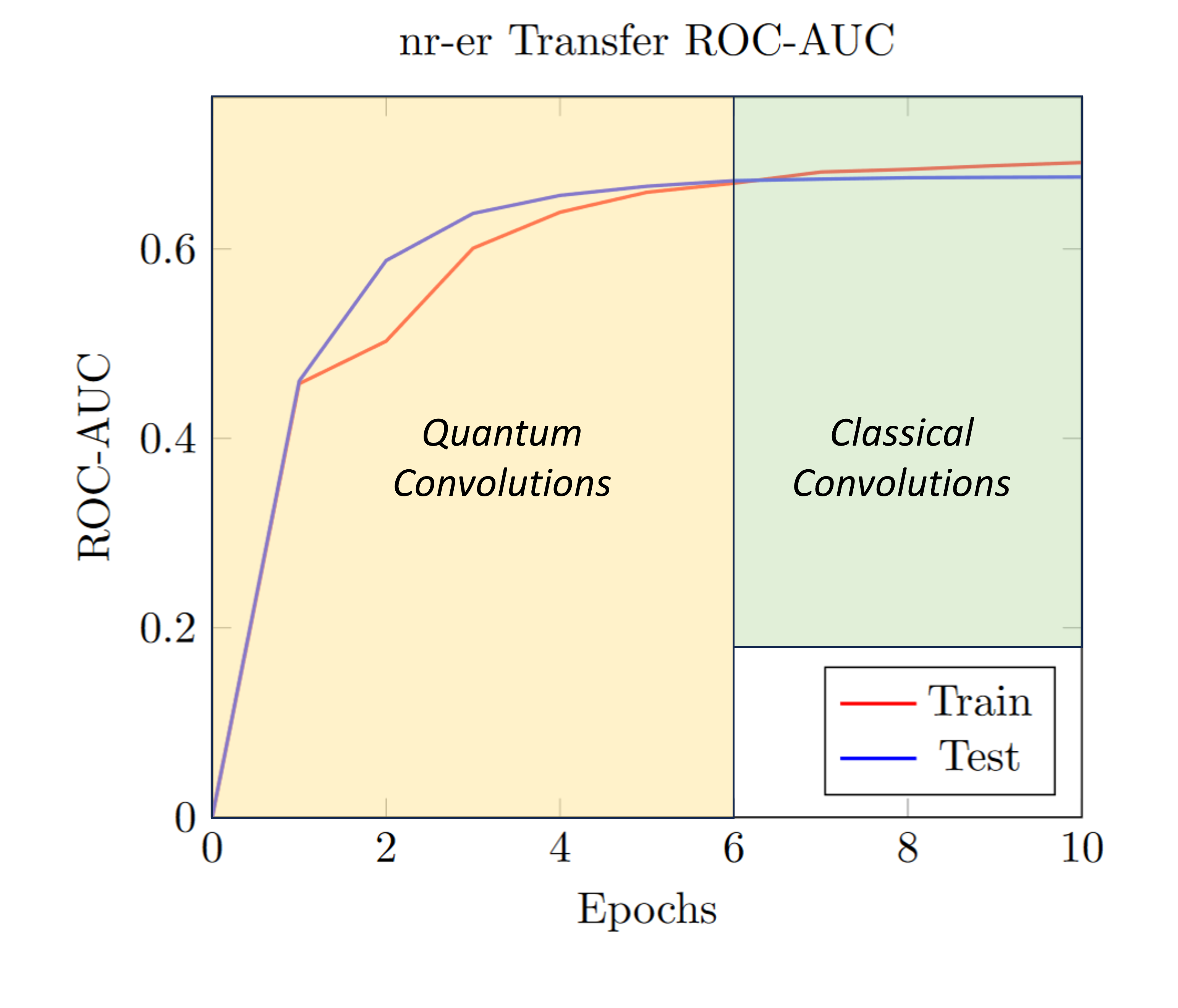}
        \caption{}
    \end{subfigure}
    \caption{Training curves for the nr-er assay}
\end{figure}

\begin{figure}[!h]
    \centering
    \begin{subfigure}[b]{0.47\textwidth}
        \includegraphics[width=\textwidth]{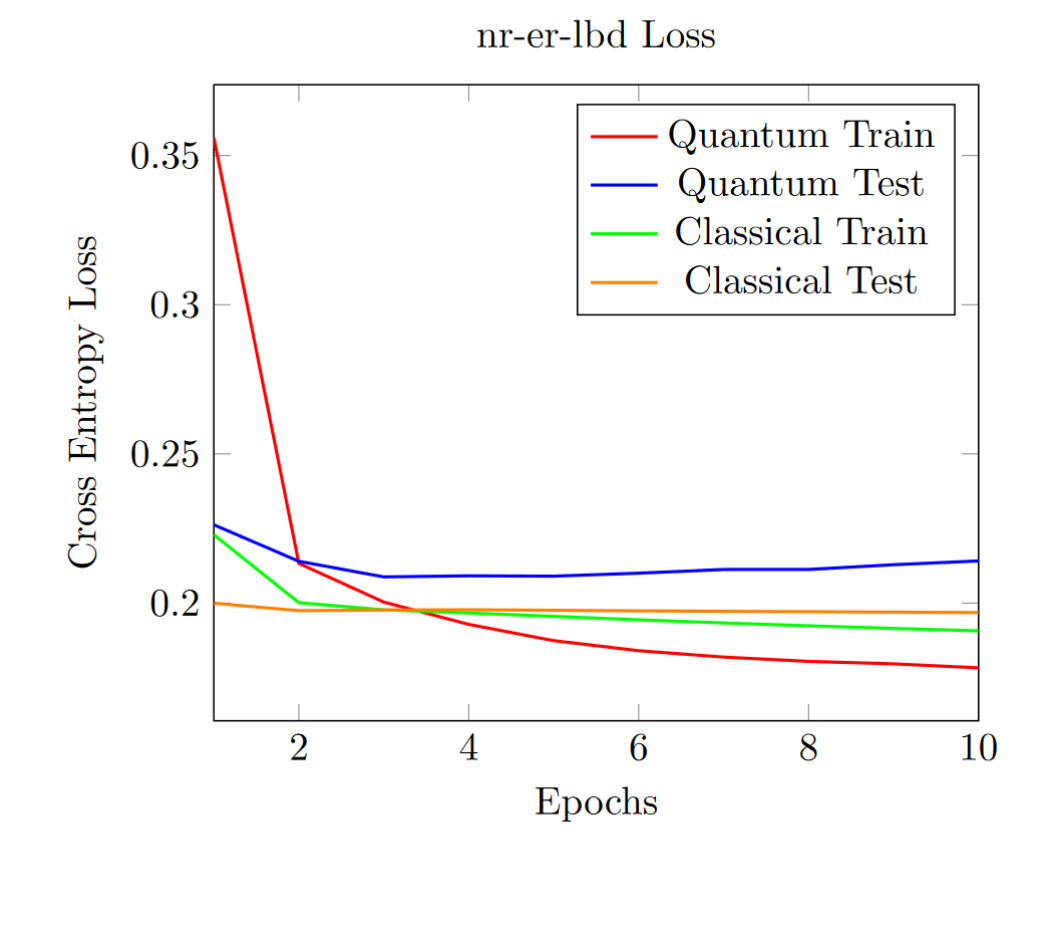}
        \caption{}
    \end{subfigure}
    \hfill
    \begin{subfigure}[b]{0.47\textwidth}
        \includegraphics[width=\textwidth]{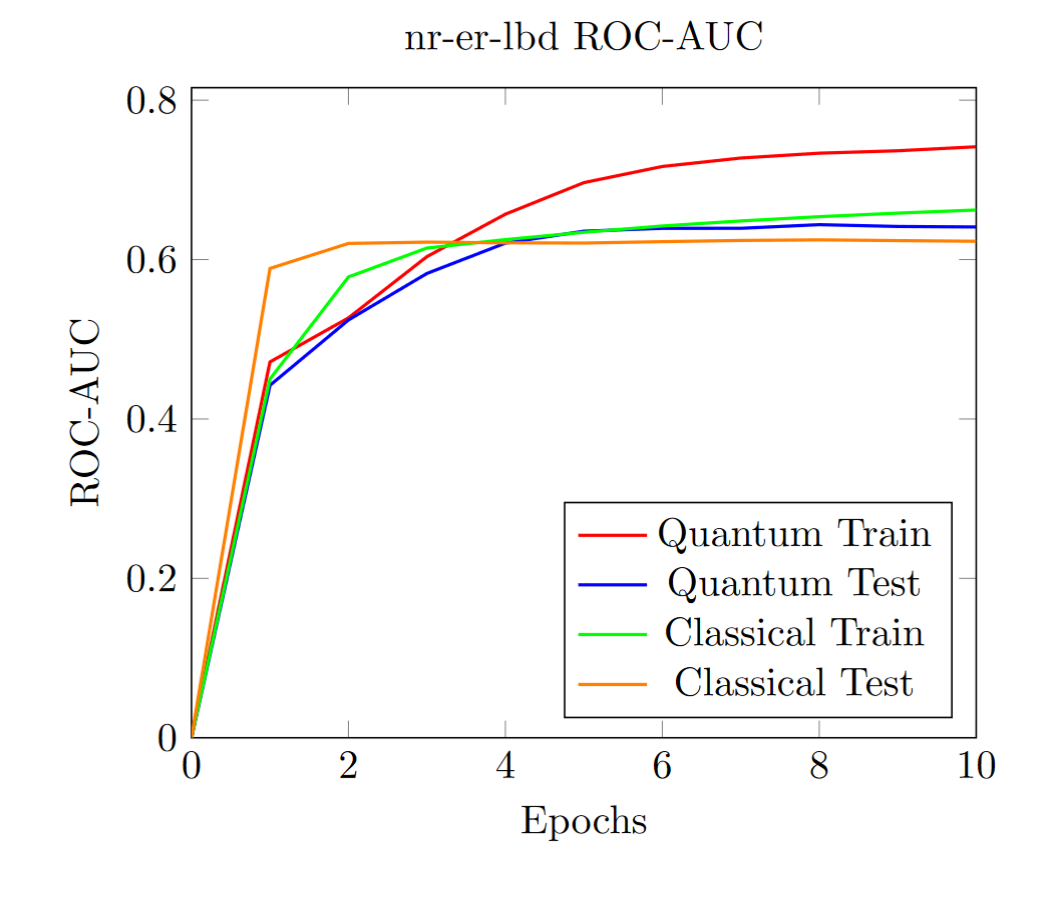}
        \caption{}
    \end{subfigure}
    \begin{subfigure}[b]{0.47\textwidth}
        \includegraphics[width=\textwidth]{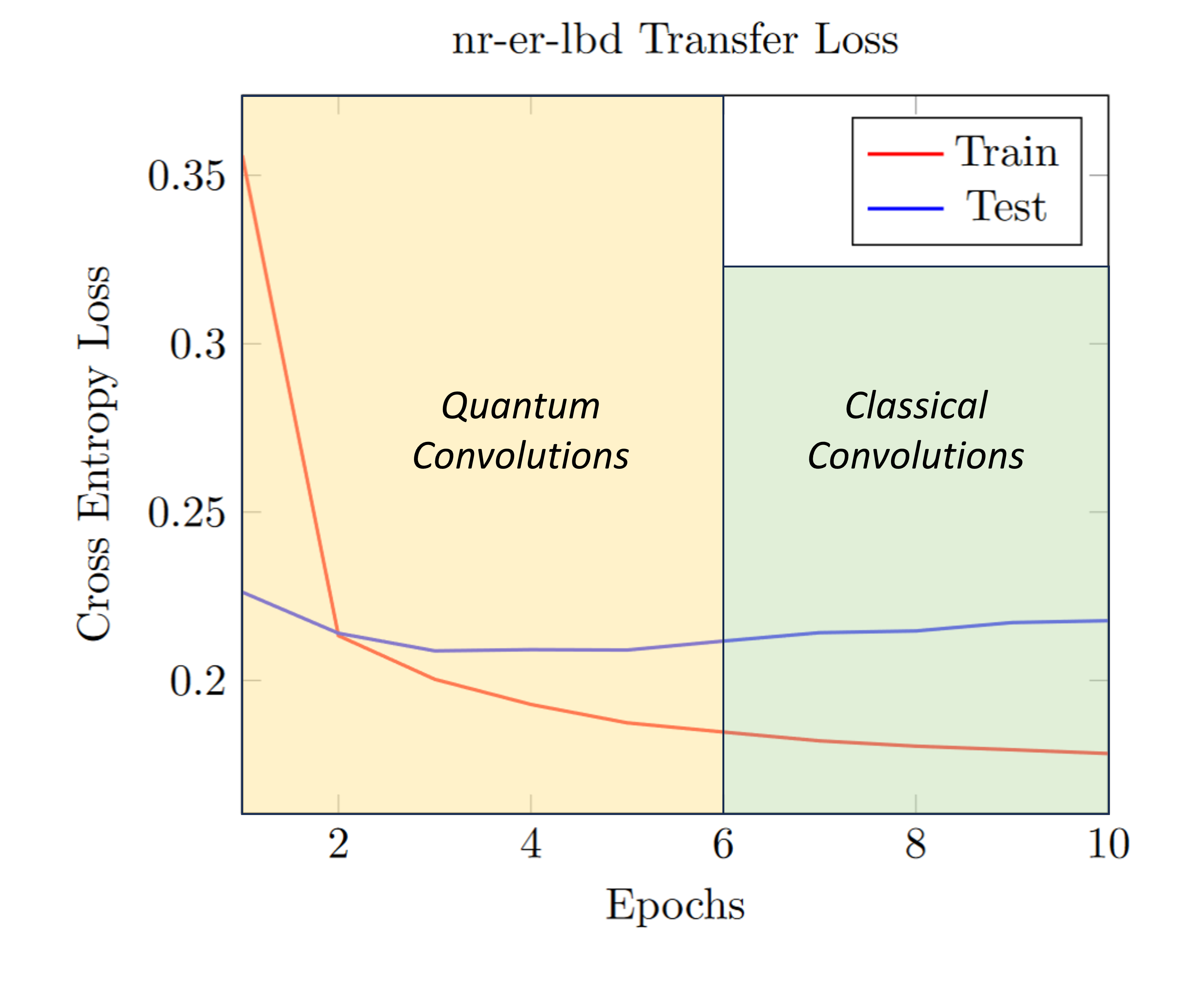}
        \caption{}
    \end{subfigure}
    \hfill
    \begin{subfigure}[b]{0.47\textwidth}
        \includegraphics[width=\textwidth]{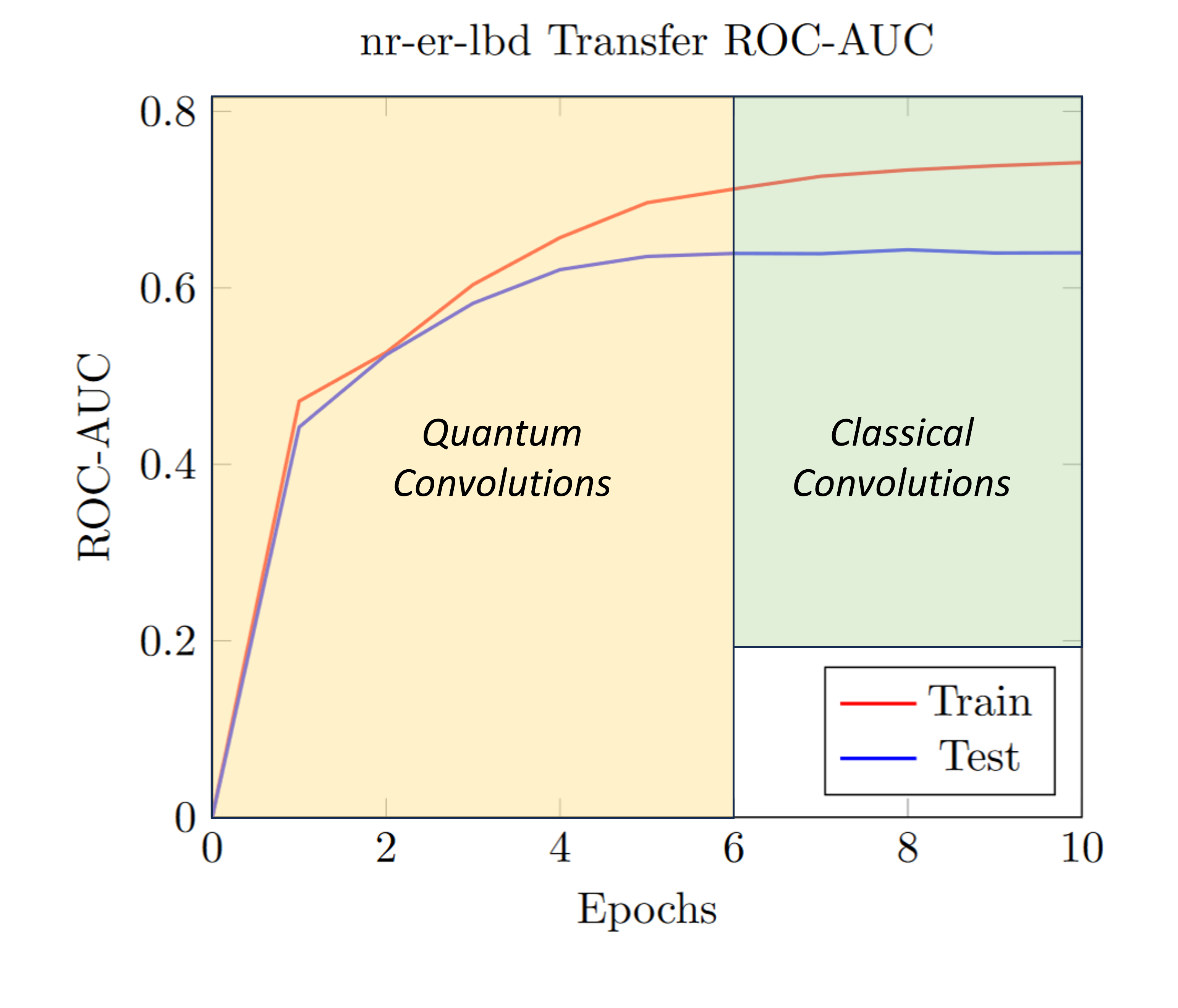}
        \caption{}
    \end{subfigure}
    \caption{Training curves for the nr-er-lbd assay}
\end{figure}

\begin{figure}[!h]
    \centering
    \begin{subfigure}[b]{0.47\textwidth}
        \includegraphics[width=\textwidth]{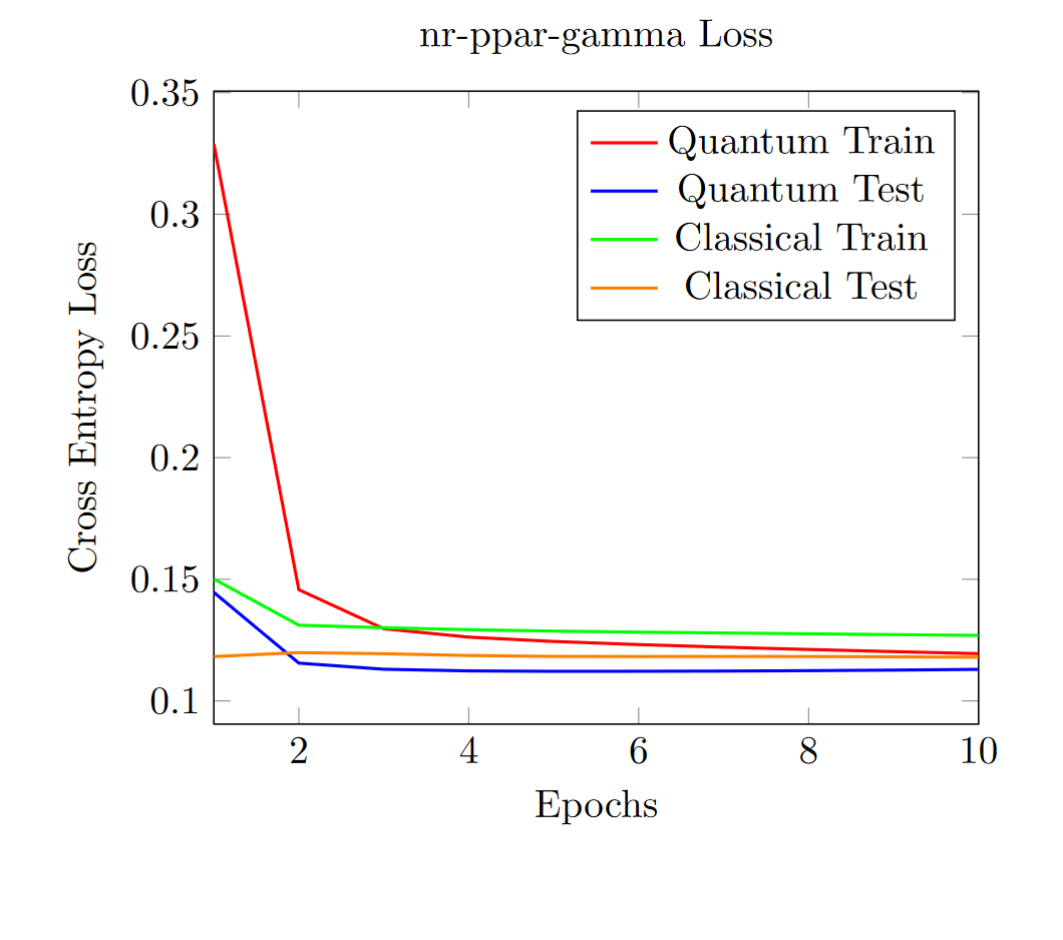}
        \caption{}
    \end{subfigure}
    \hfill
    \begin{subfigure}[b]{0.47\textwidth}
        \includegraphics[width=\textwidth]{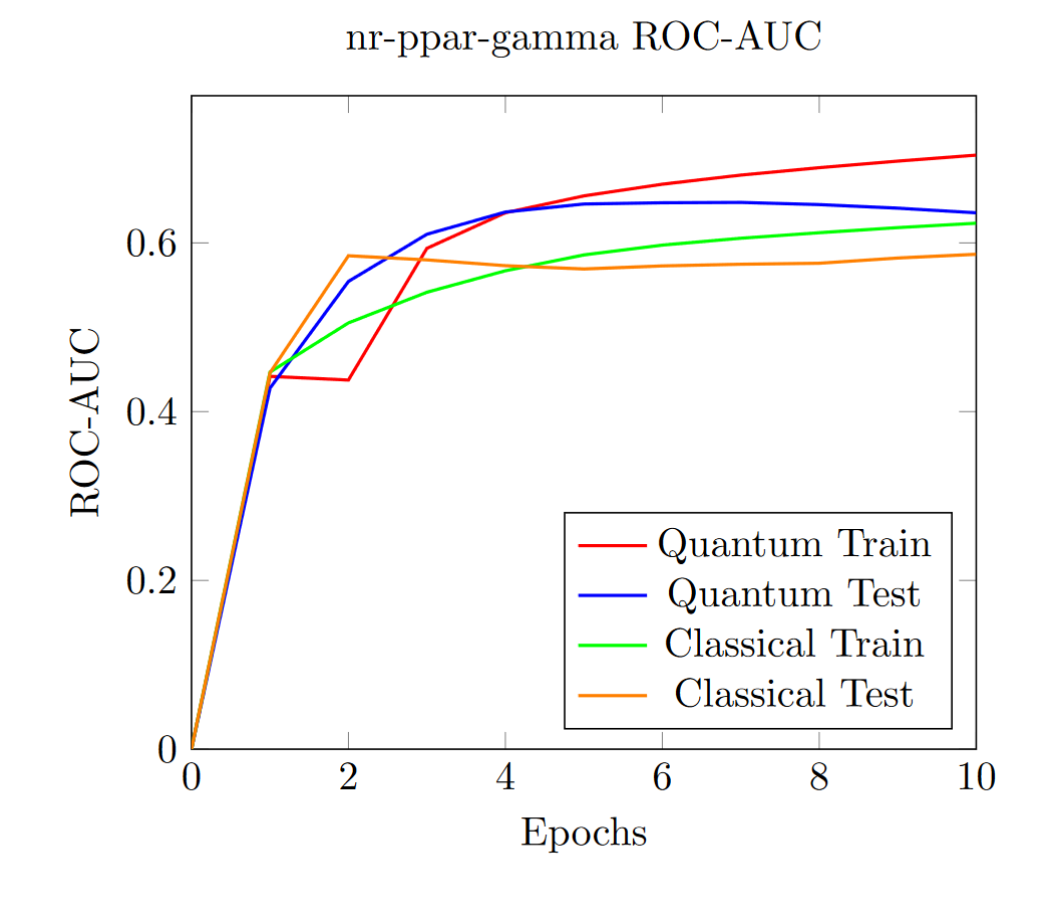}
        \caption{}
    \end{subfigure}
    \begin{subfigure}[b]{0.47\textwidth}
        \includegraphics[width=\textwidth]{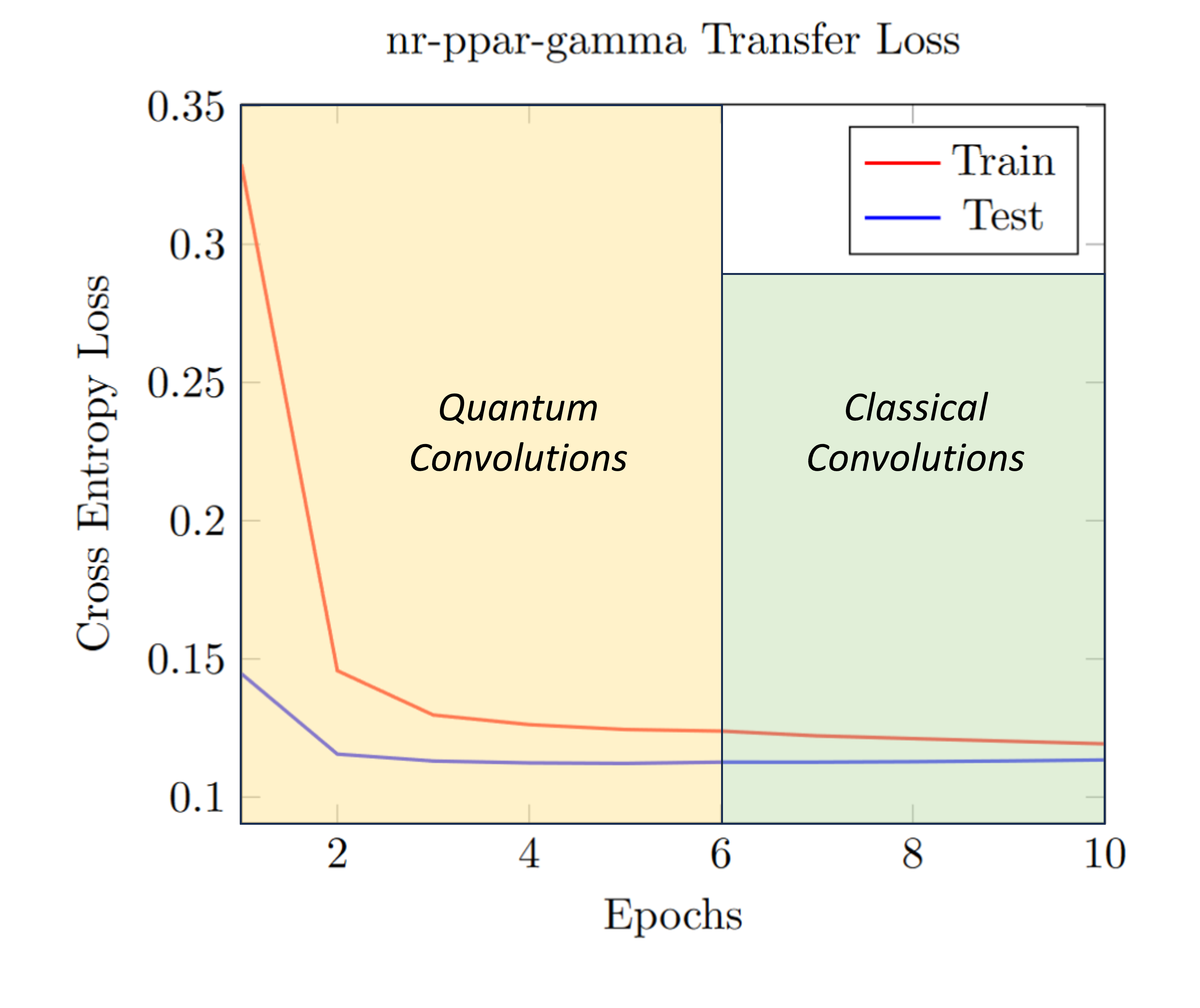}
        \caption{}
    \end{subfigure}
    \hfill
    \begin{subfigure}[b]{0.47\textwidth}
        \includegraphics[width=\textwidth]{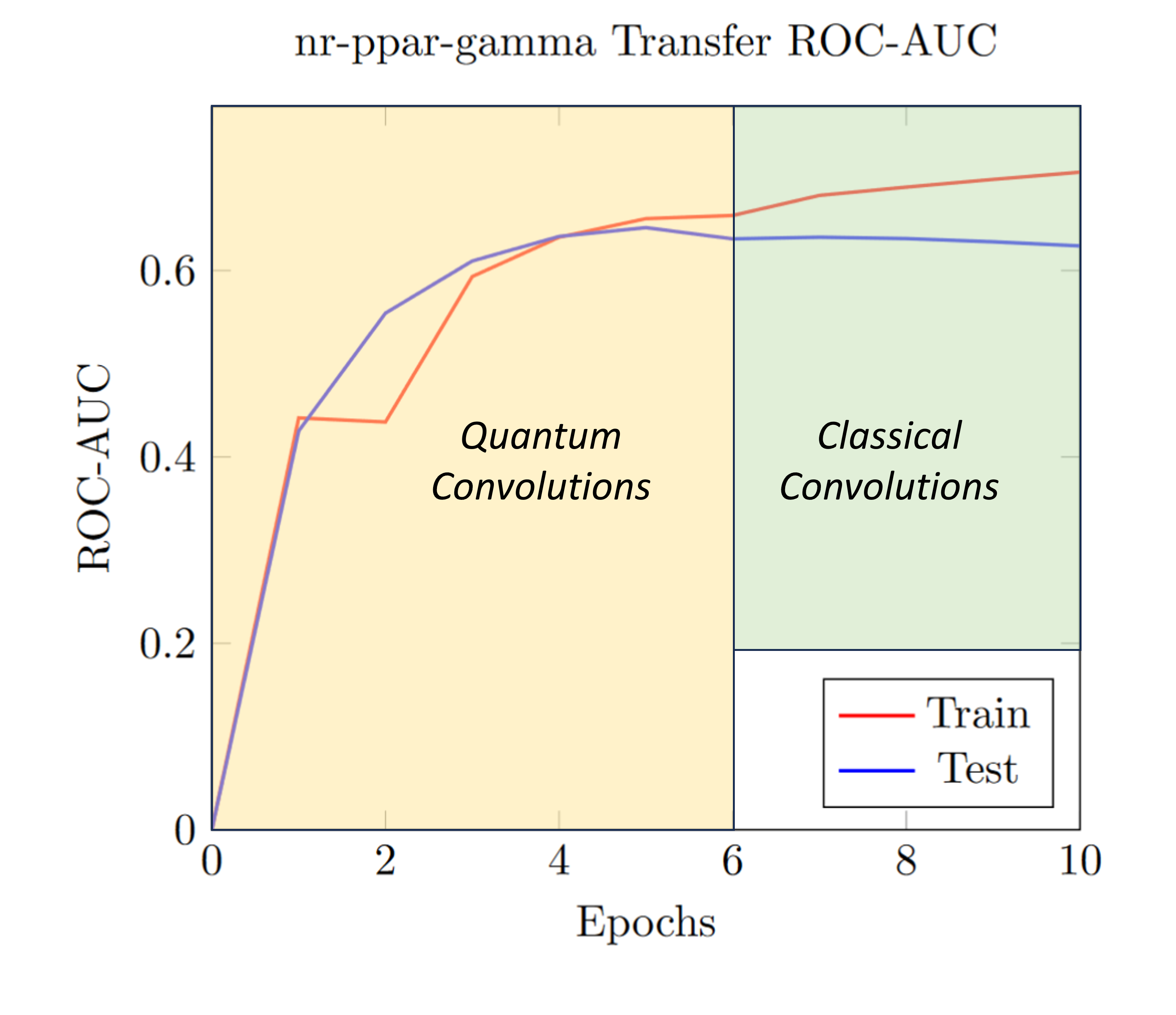}
        \caption{}
    \end{subfigure}
    \caption{Training curves for the nr-ppar-gamma assay}
\end{figure}

\begin{figure}[!h]
    \centering
    \begin{subfigure}[b]{0.47\textwidth}
        \includegraphics[width=\textwidth]{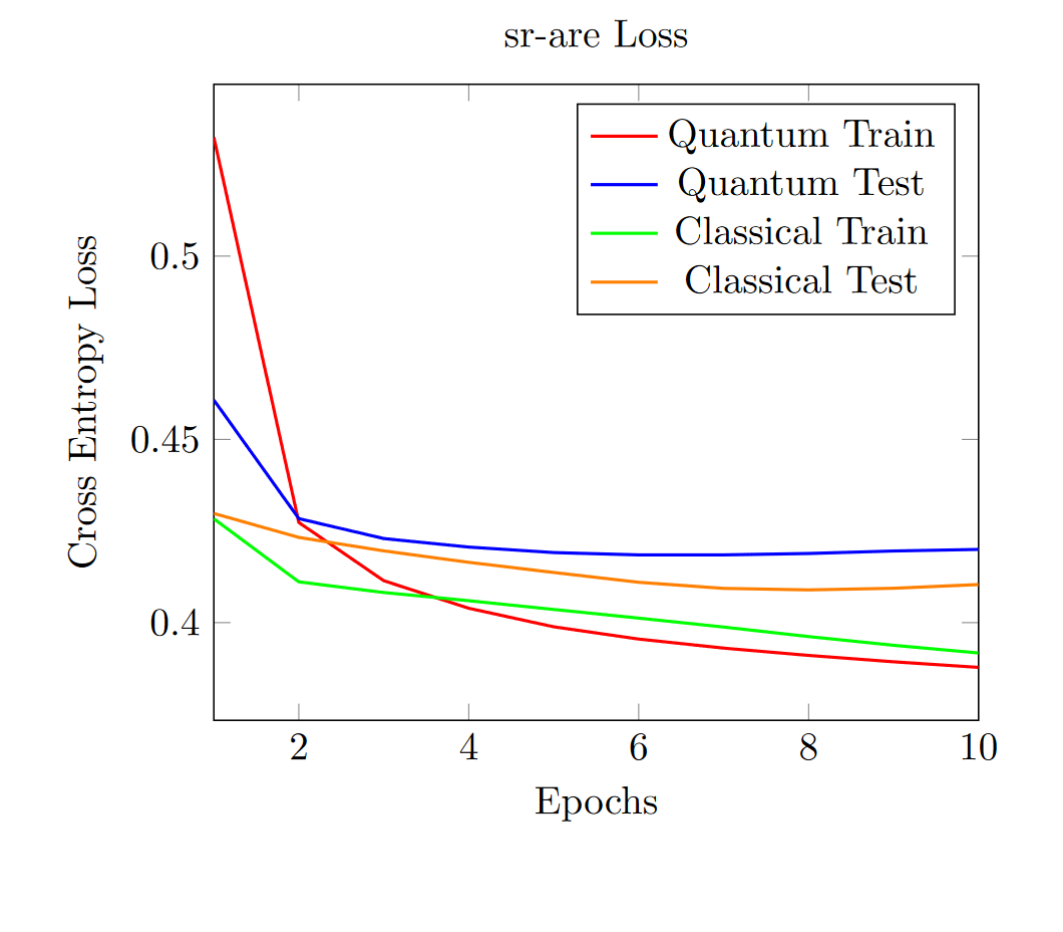}
        \caption{}
    \end{subfigure}
    \hfill
    \begin{subfigure}[b]{0.47\textwidth}
        \includegraphics[width=\textwidth]{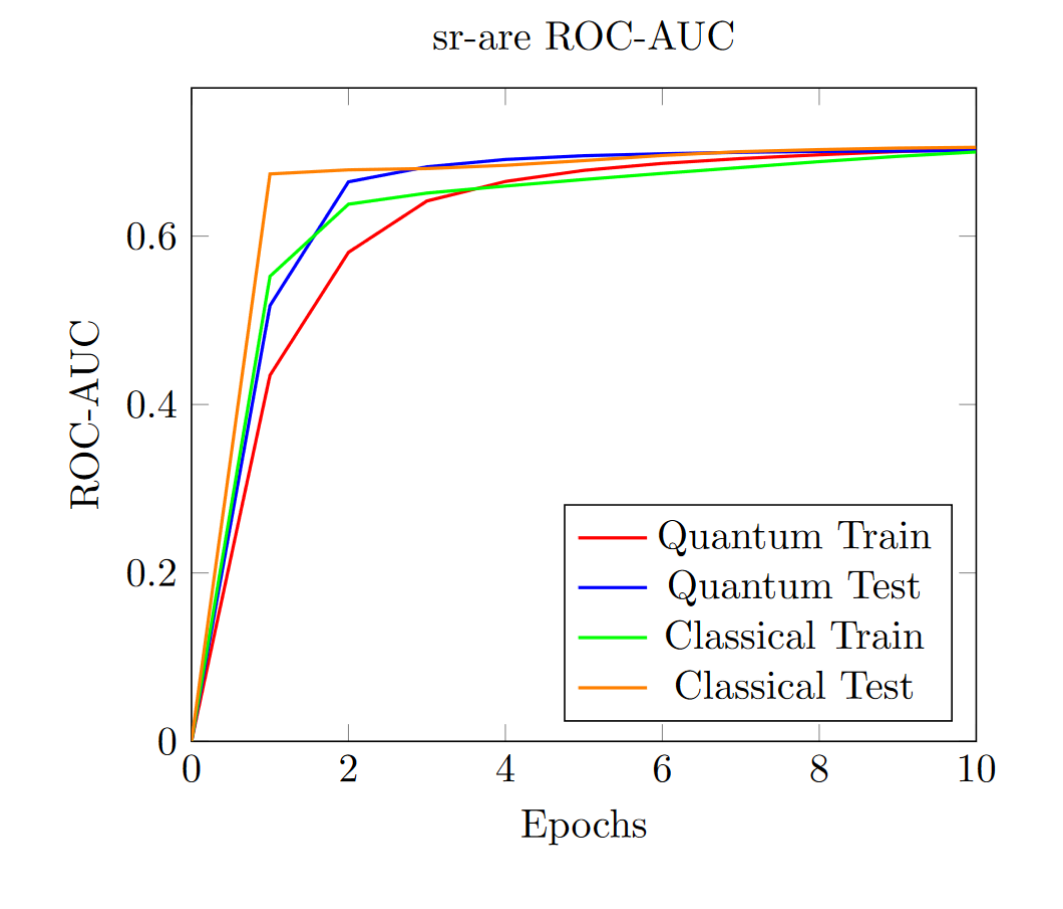}
        \caption{}
    \end{subfigure}
    \begin{subfigure}[b]{0.47\textwidth}
        \includegraphics[width=\textwidth]{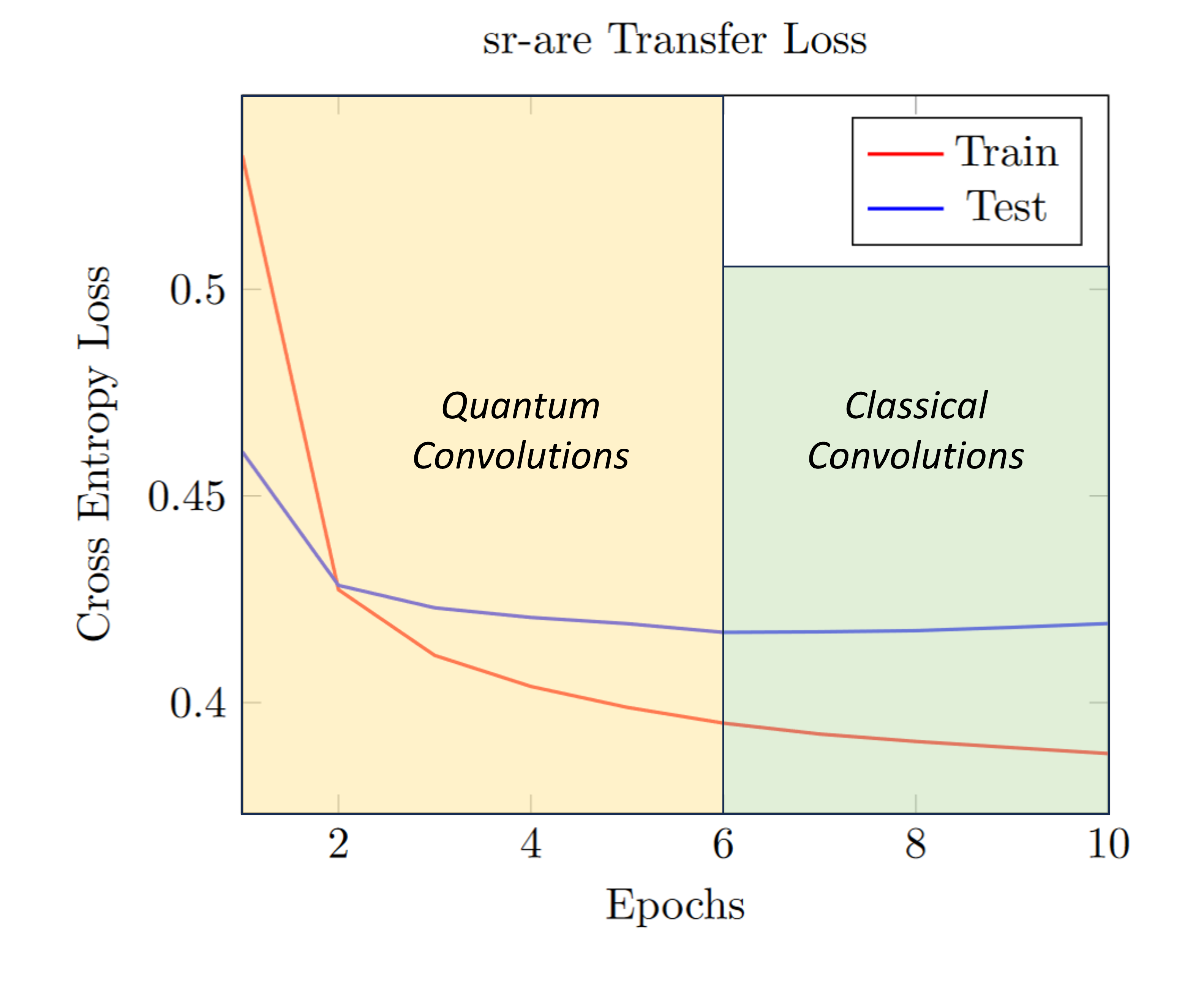}
        \caption{}
    \end{subfigure}
    \hfill
    \begin{subfigure}[b]{0.47\textwidth}
        \includegraphics[width=\textwidth]{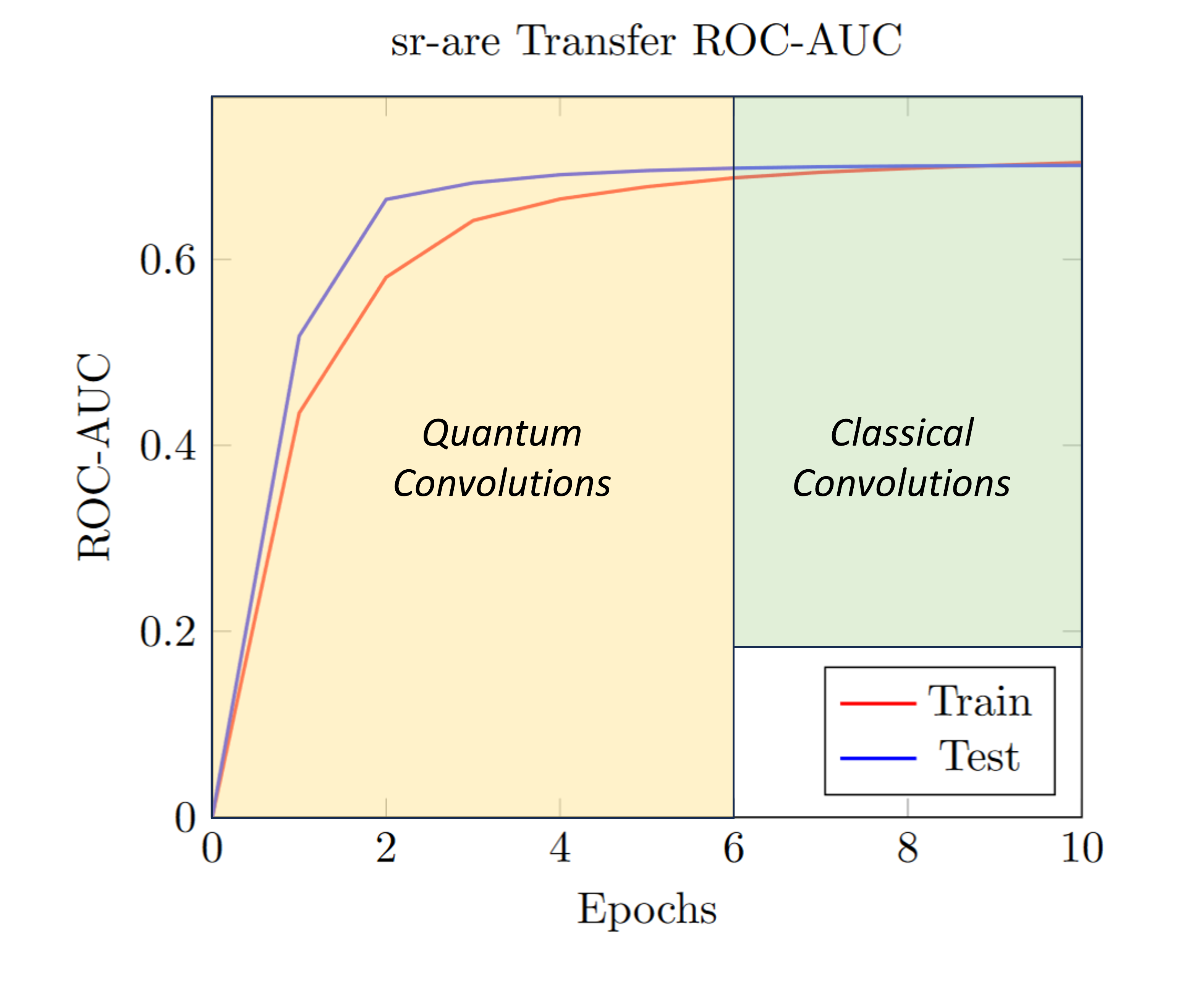}
        \caption{}
    \end{subfigure}
    \caption{Training curves for the sr-are assay}
\end{figure}

\begin{figure}[!h]
    \centering
    \begin{subfigure}[b]{0.47\textwidth}
        \includegraphics[width=\textwidth]{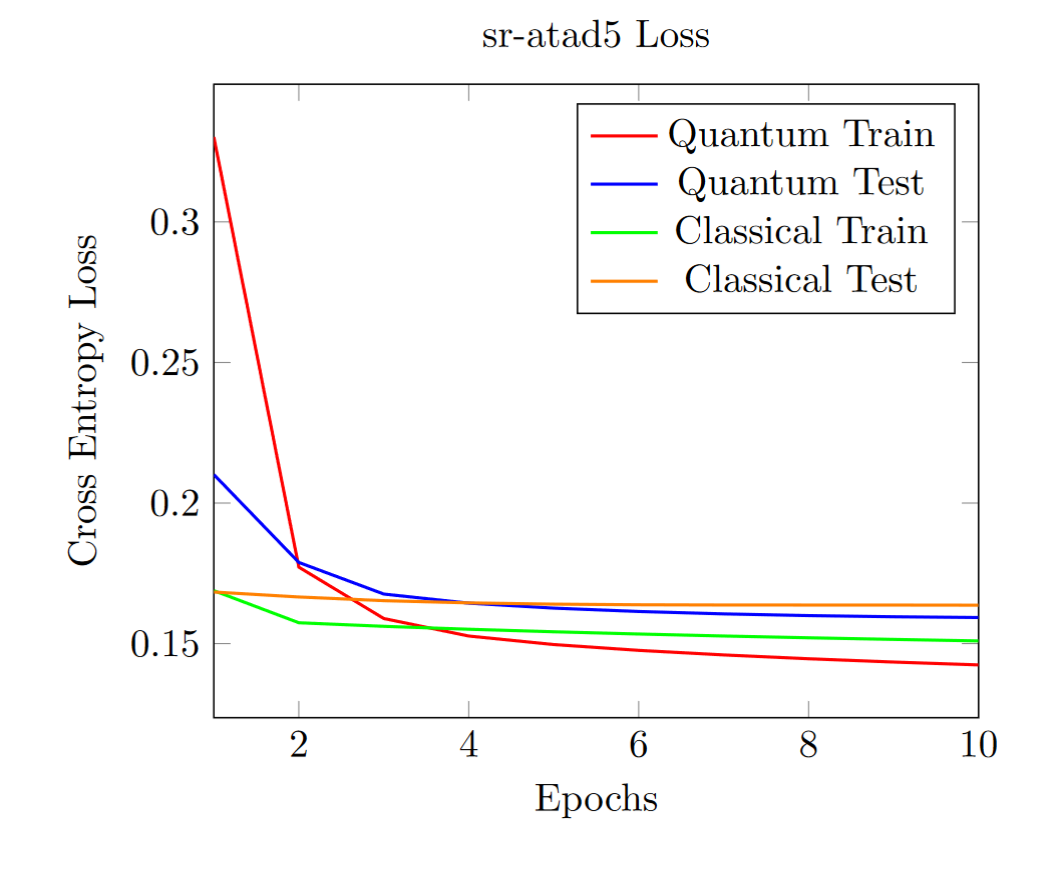}
        \caption{}
    \end{subfigure}
    \hfill
    \begin{subfigure}[b]{0.47\textwidth}
        \includegraphics[width=\textwidth]{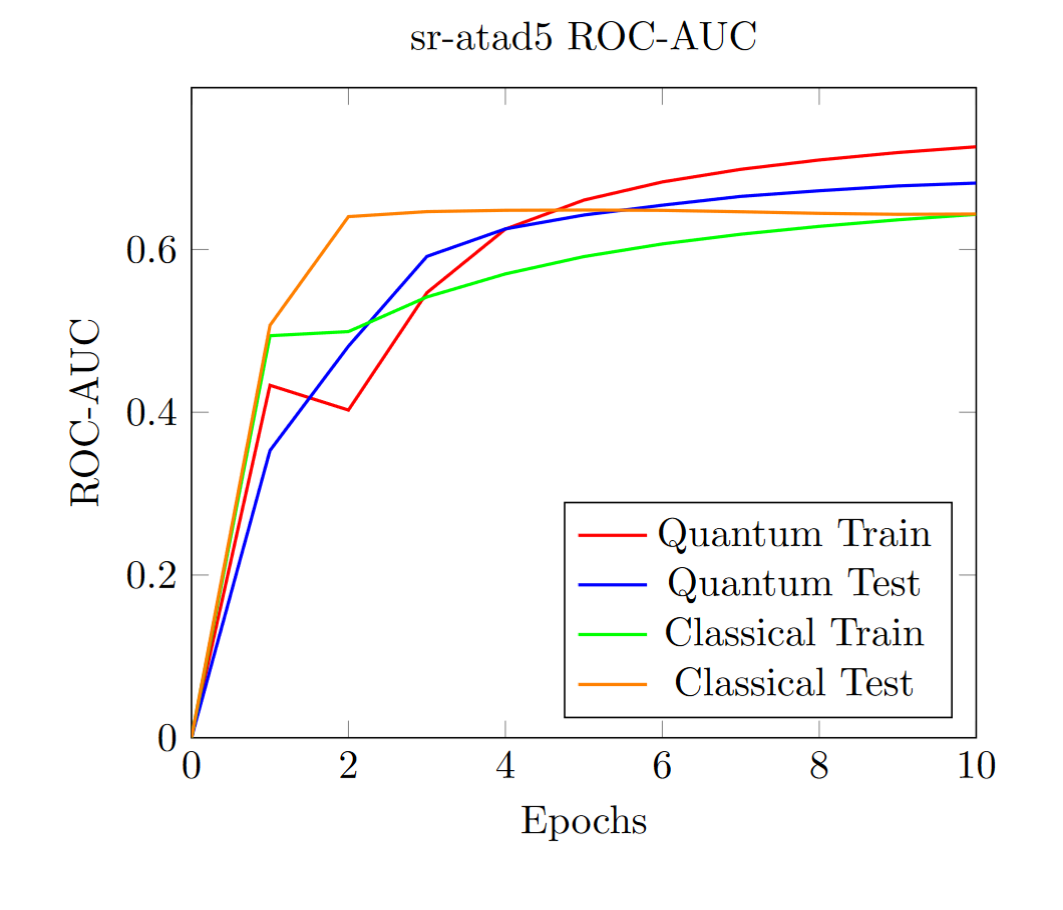}
        \caption{}
    \end{subfigure}
    \begin{subfigure}[b]{0.47\textwidth}
        \includegraphics[width=\textwidth]{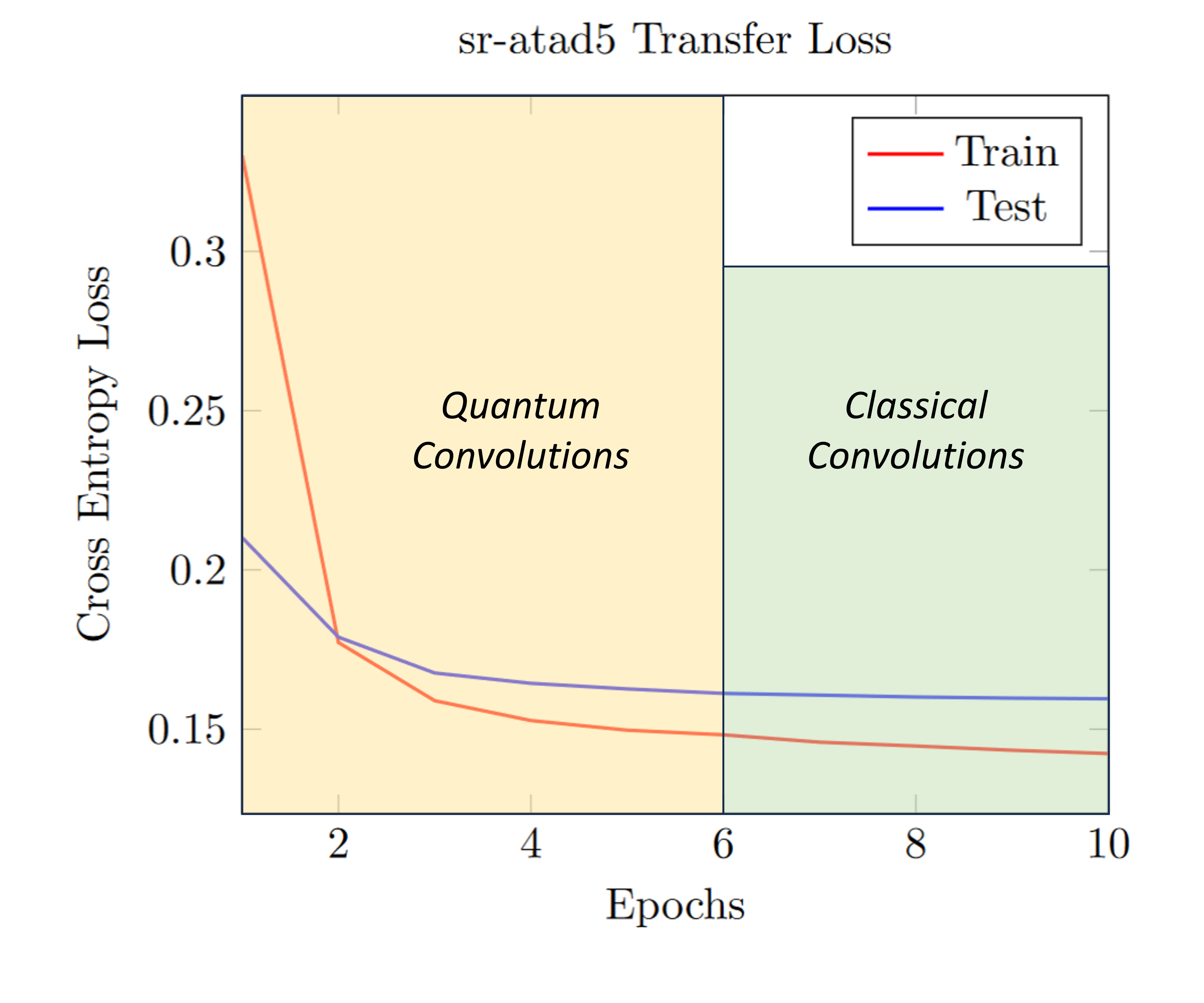}
        \caption{}
    \end{subfigure}
    \hfill
    \begin{subfigure}[b]{0.47\textwidth}
        \includegraphics[width=\textwidth]{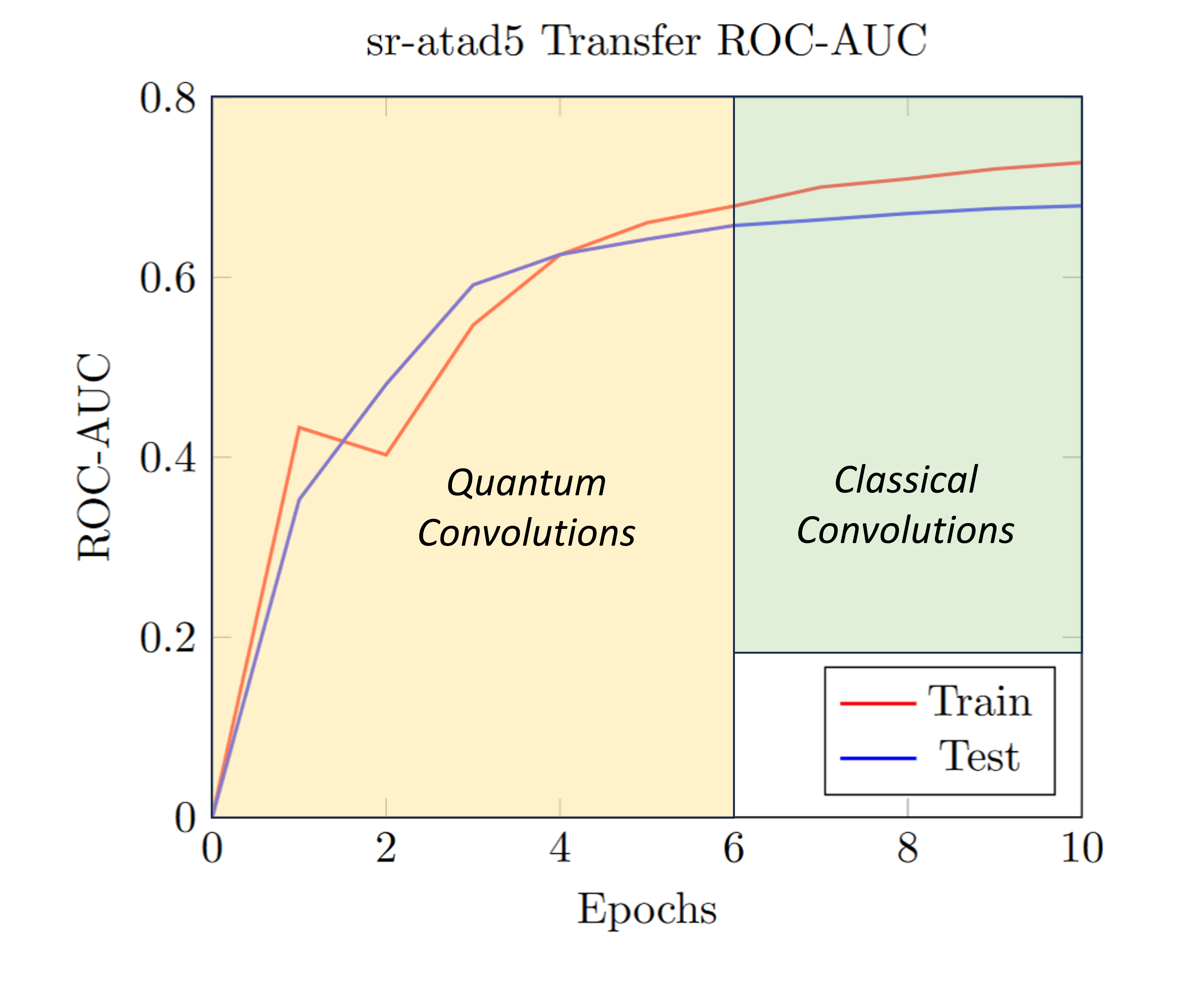}
        \caption{}
    \end{subfigure}
    \caption{Training curves for the sr-atad5 assay}
\end{figure}

\begin{figure}[!h]
    \centering
    \begin{subfigure}[b]{0.47\textwidth}
        \includegraphics[width=\textwidth]{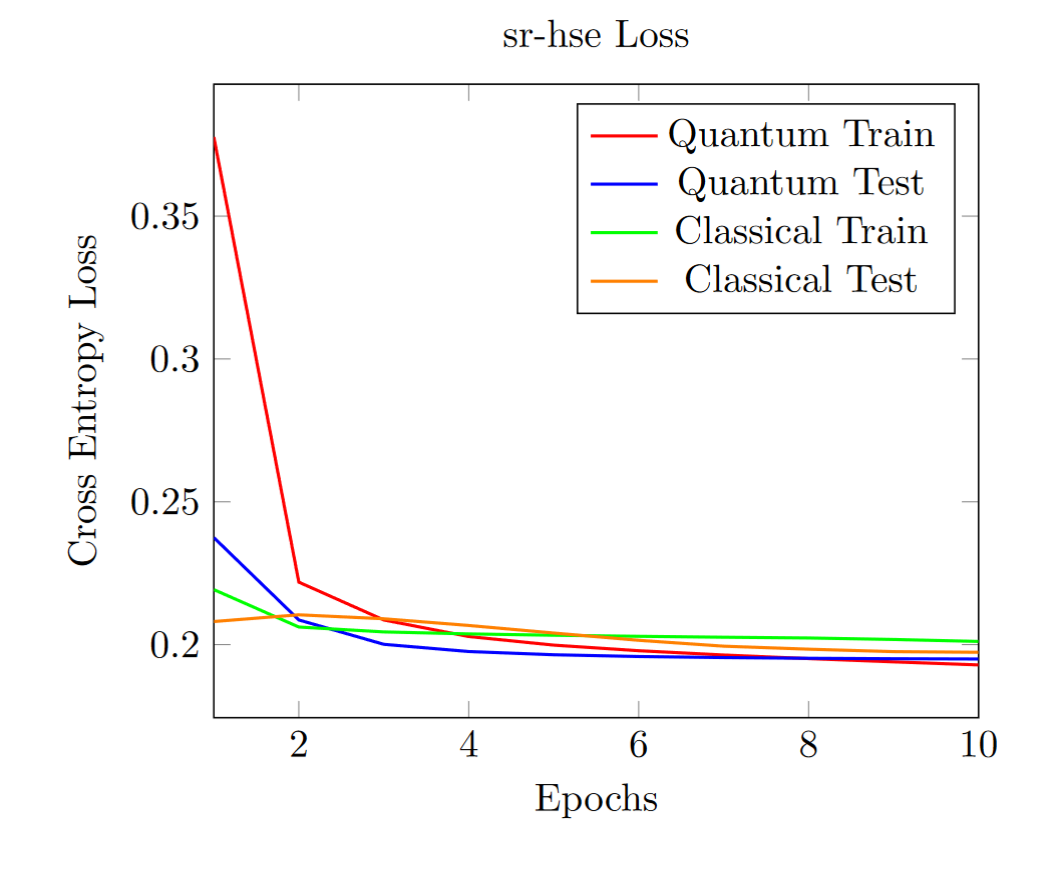}
        \caption{}
    \end{subfigure}
    \hfill
    \begin{subfigure}[b]{0.47\textwidth}
        \includegraphics[width=\textwidth]{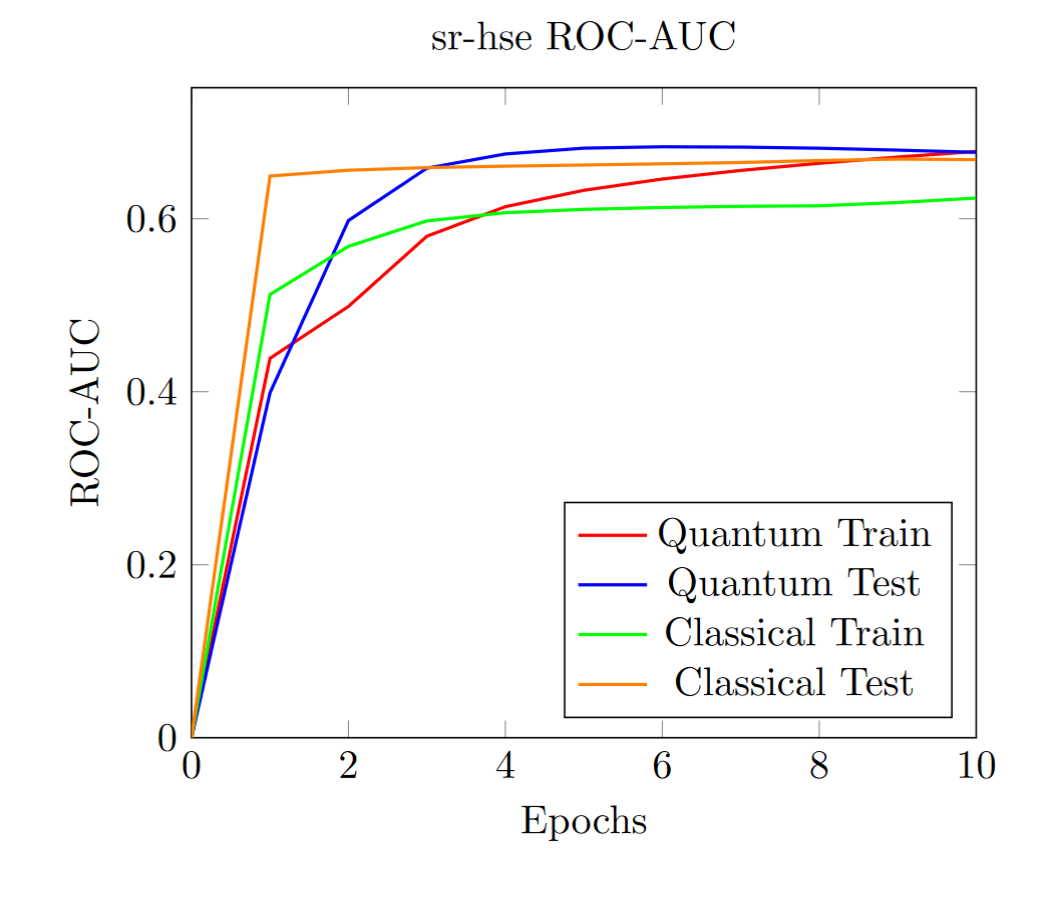}
        \caption{}
    \end{subfigure}
    \begin{subfigure}[b]{0.48\textwidth}
        \includegraphics[width=\textwidth]{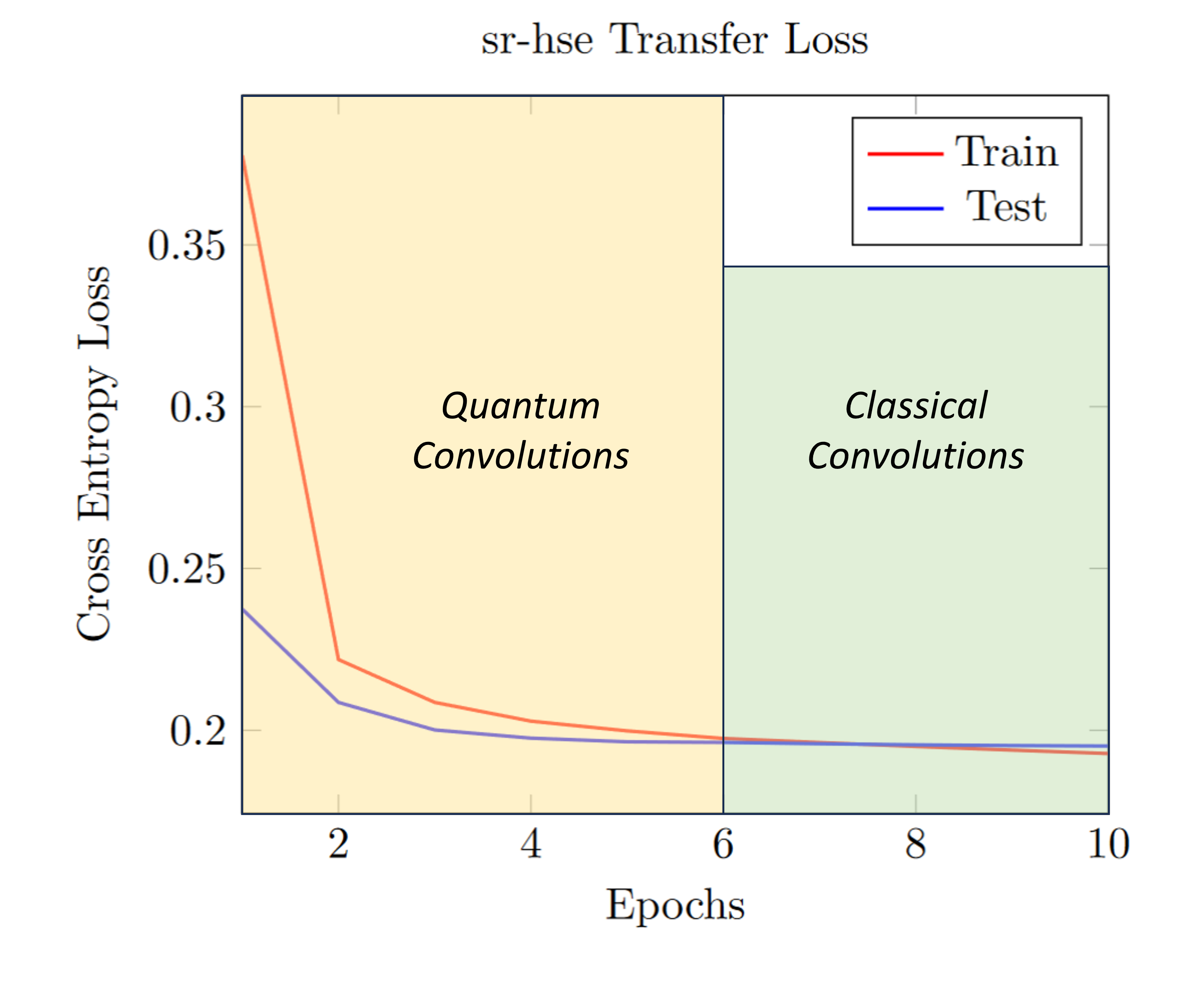}
        \caption{}
    \end{subfigure}
    \hfill
    \begin{subfigure}[b]{0.47\textwidth}
        \includegraphics[width=\textwidth]{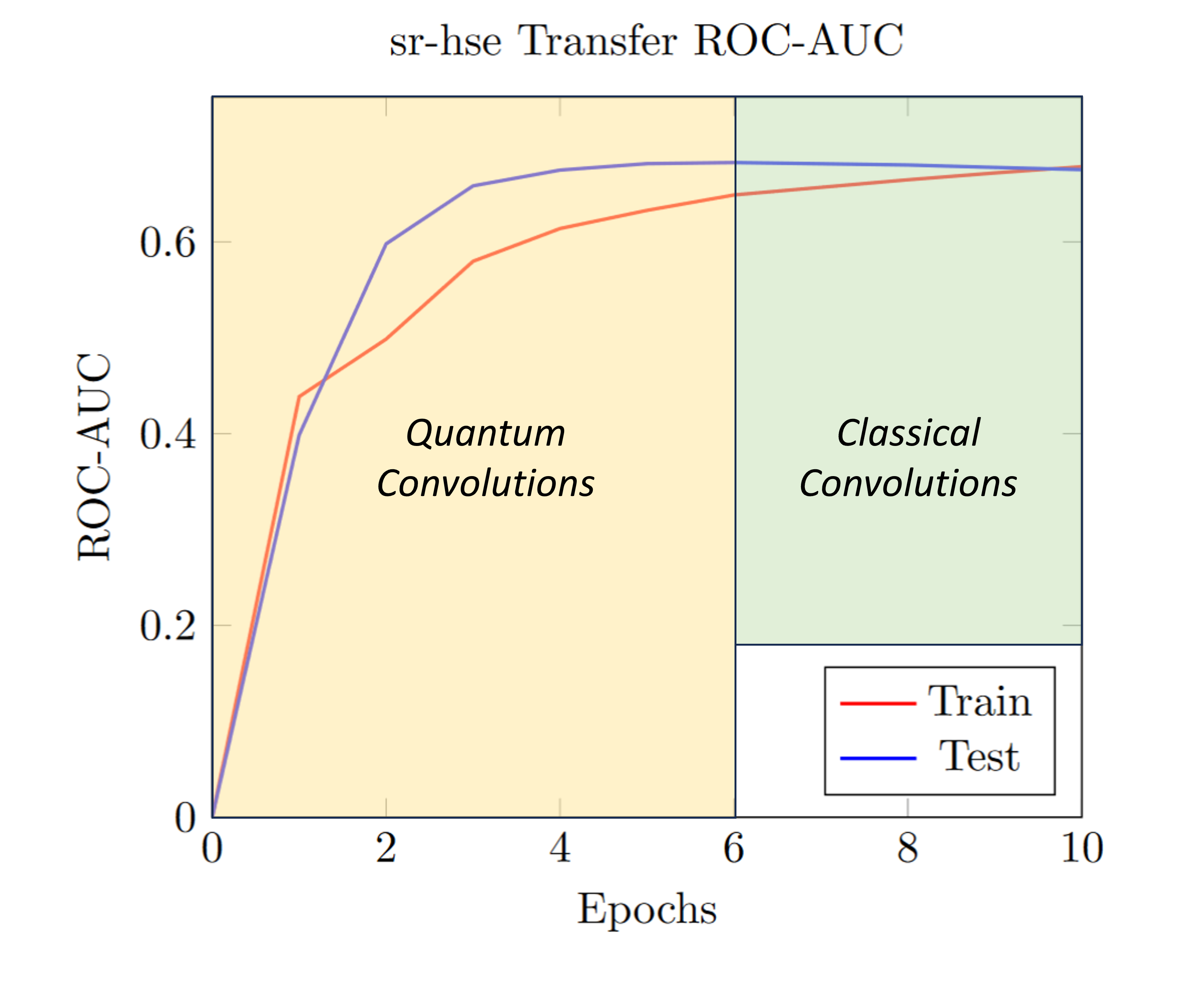}
        \caption{}
    \end{subfigure}
    \caption{Training curves for the sr-hse assay}
\end{figure}

\begin{figure}[!h]
    \centering
    \begin{subfigure}[b]{0.47\textwidth}
        \includegraphics[width=\textwidth]{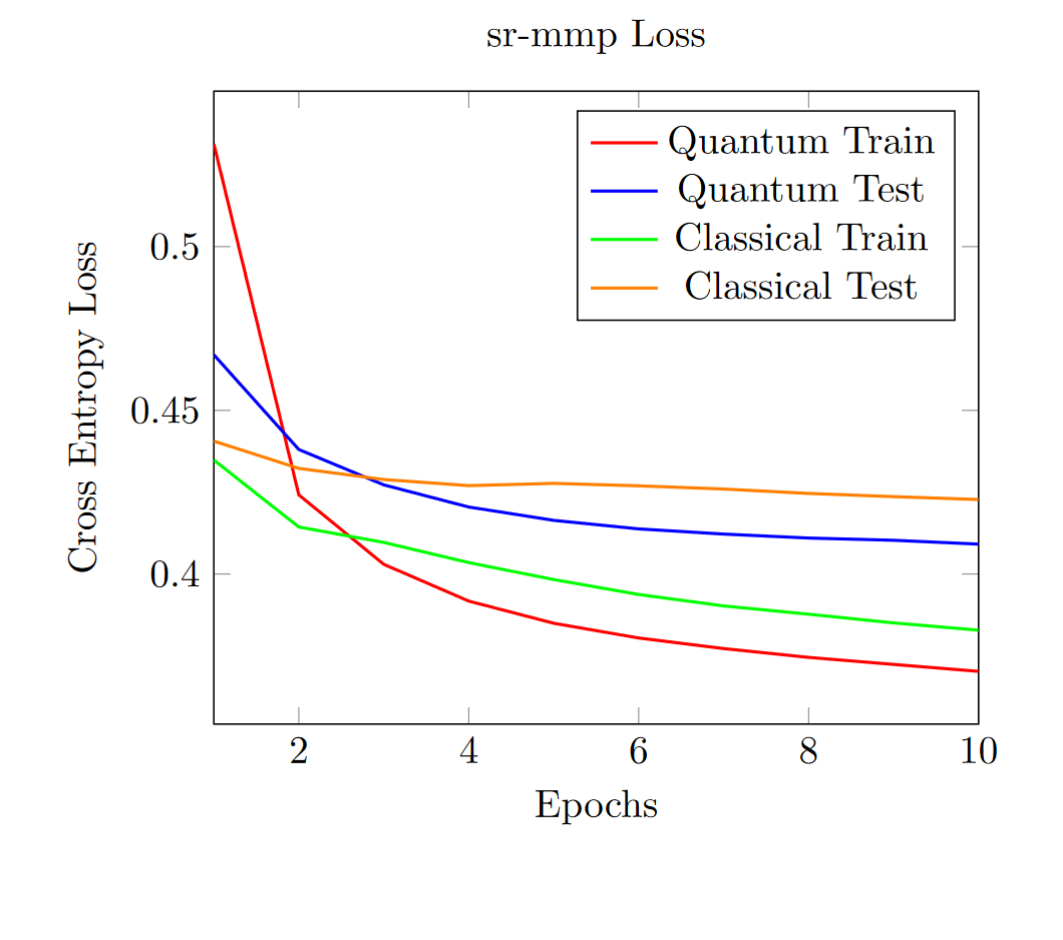}
        \caption{}
    \end{subfigure}
    \hfill
    \begin{subfigure}[b]{0.47\textwidth}
        \includegraphics[width=\textwidth]{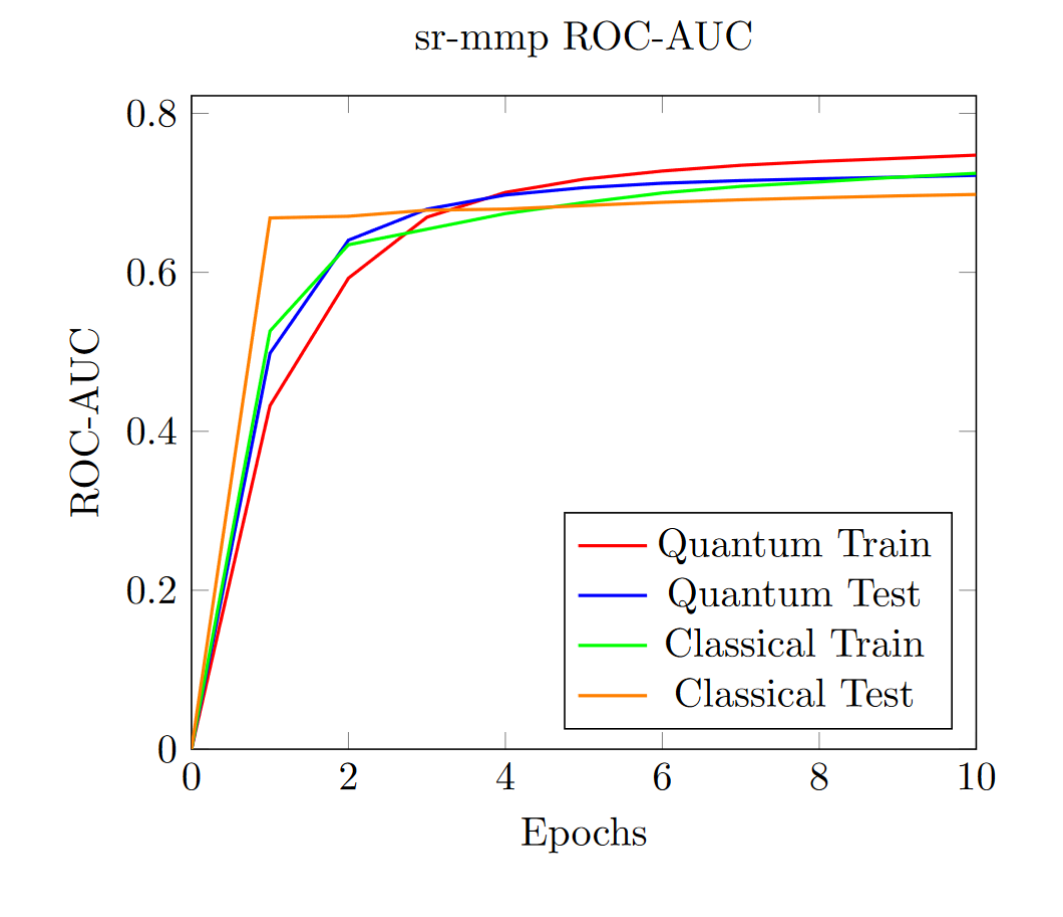}
        \caption{}
    \end{subfigure}
    \begin{subfigure}[b]{0.47\textwidth}
        \includegraphics[width=\textwidth]{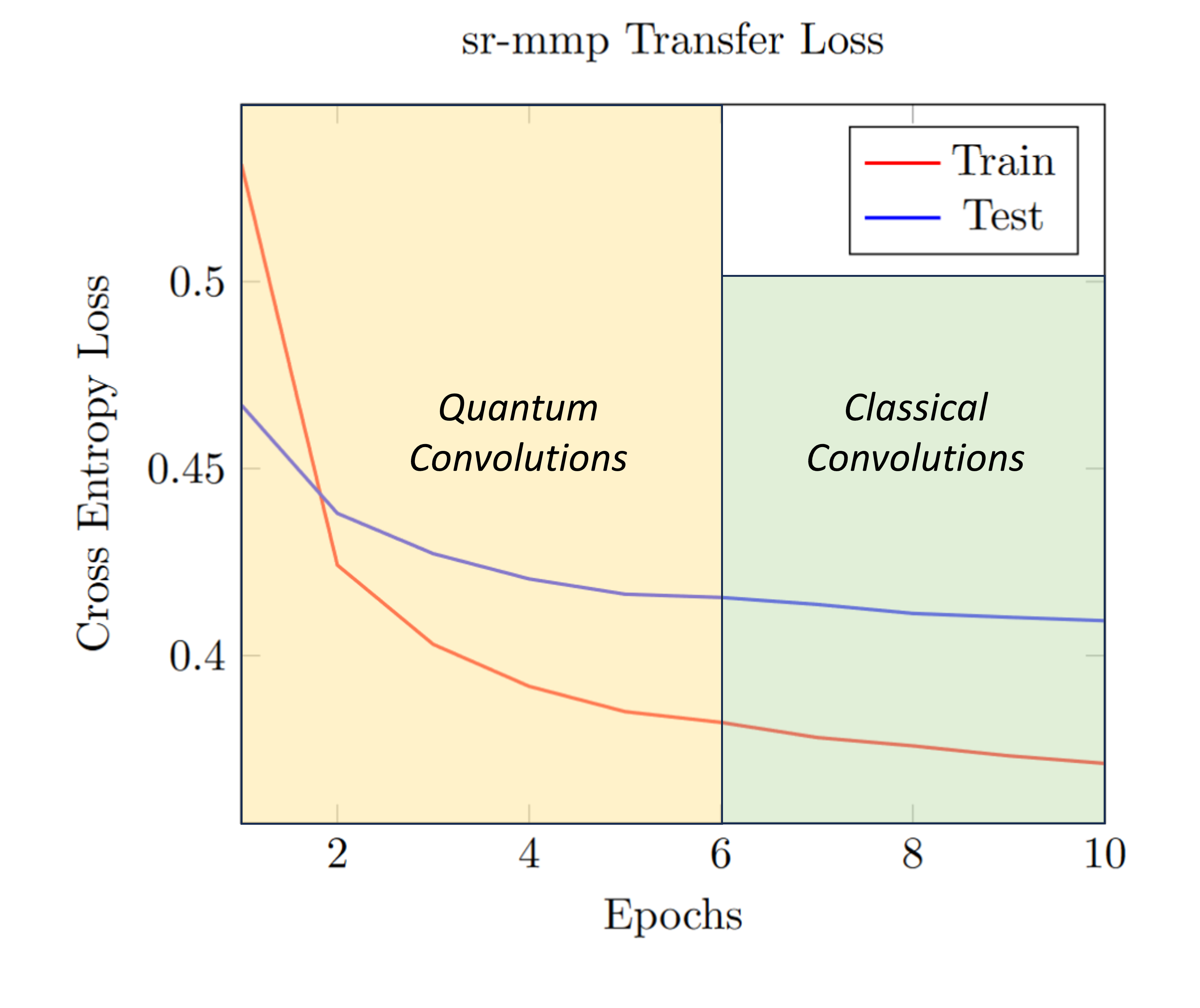}
        \caption{}
    \end{subfigure}
    \hfill
    \begin{subfigure}[b]{0.47\textwidth}
        \includegraphics[width=\textwidth]{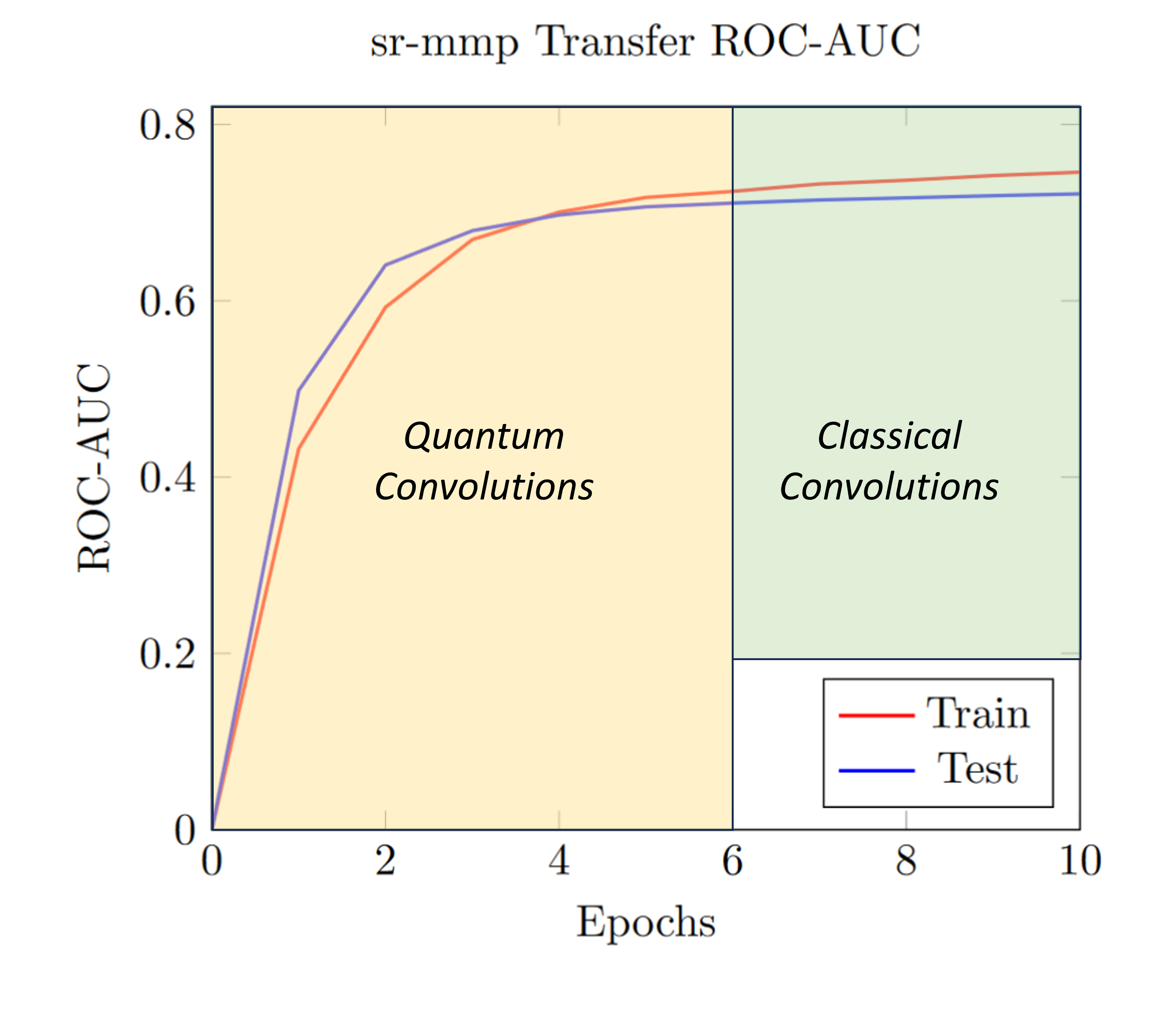}
        \caption{}
    \end{subfigure}
    \caption{Training curves for the sr-mmp assay}
\end{figure}

\begin{figure}[!h]
    \centering
    \begin{subfigure}[b]{0.47\textwidth}
        \includegraphics[width=\textwidth]{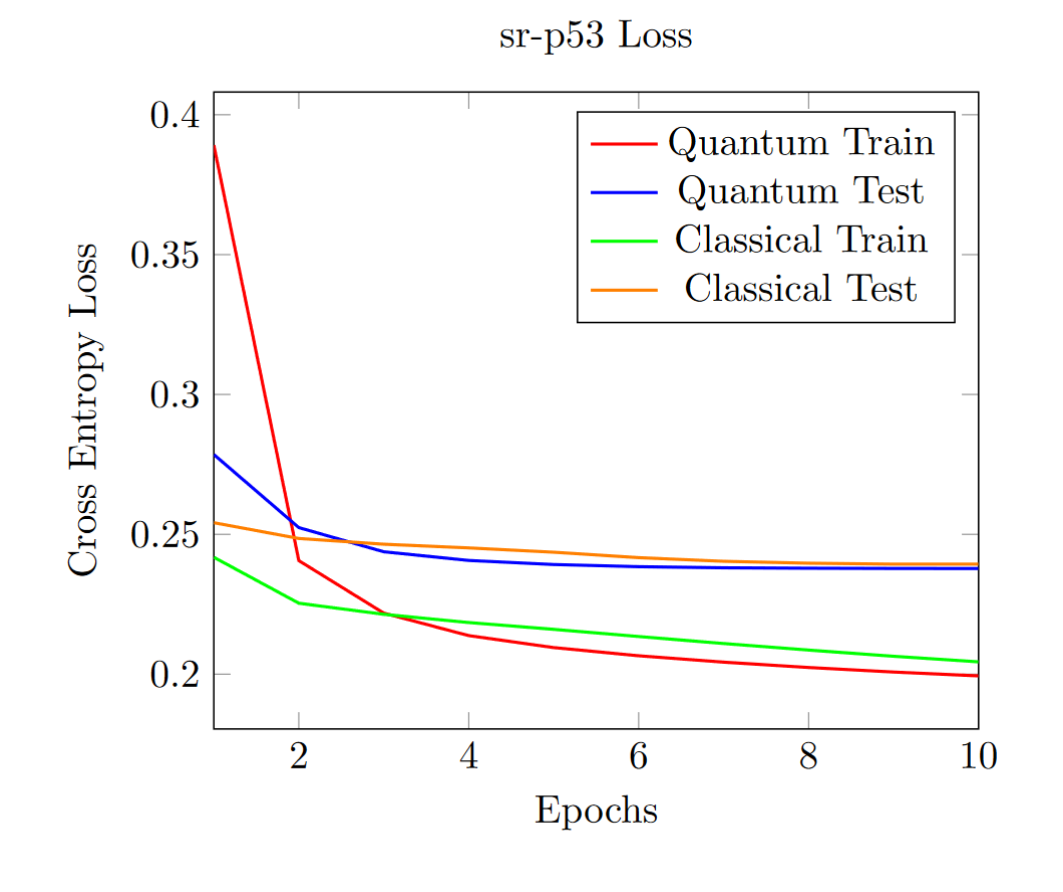}
        \caption{}
    \end{subfigure}
    \hfill
    \begin{subfigure}[b]{0.47\textwidth}
        \includegraphics[width=\textwidth]{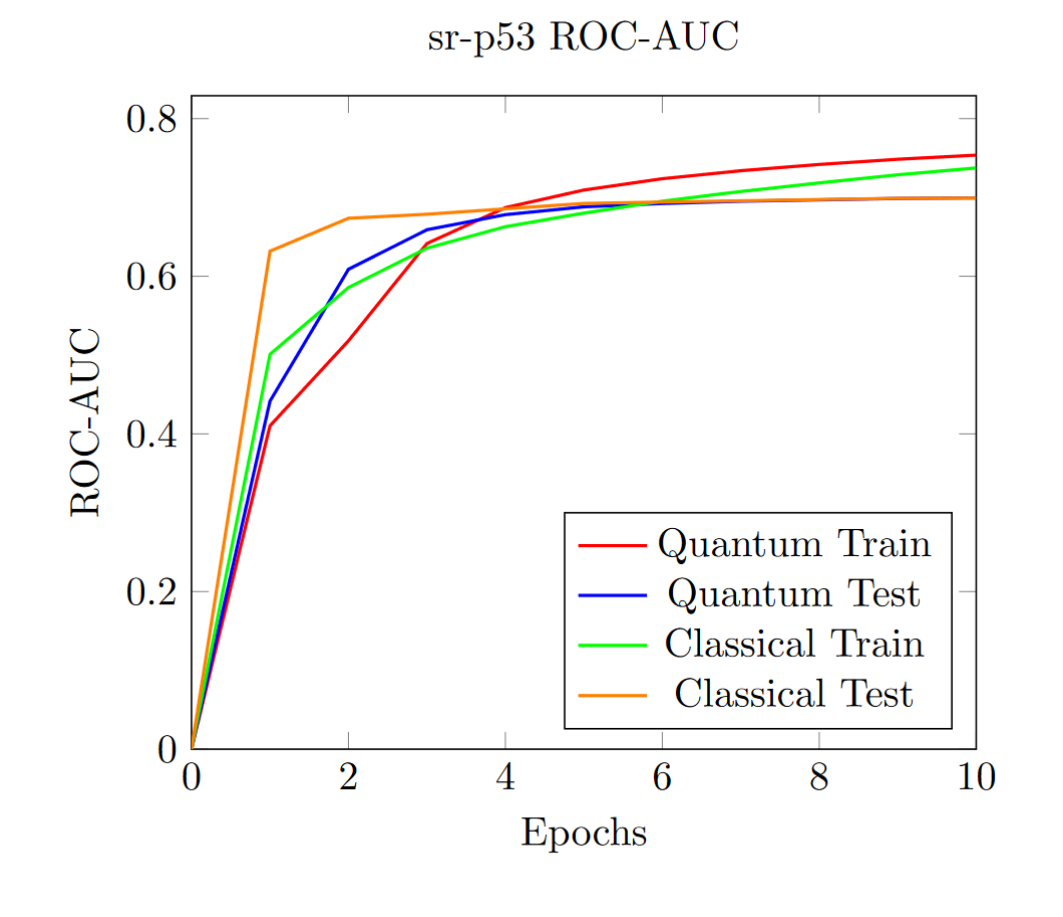}
        \caption{}
    \end{subfigure}
    \begin{subfigure}[b]{0.47\textwidth}
        \includegraphics[width=\textwidth]{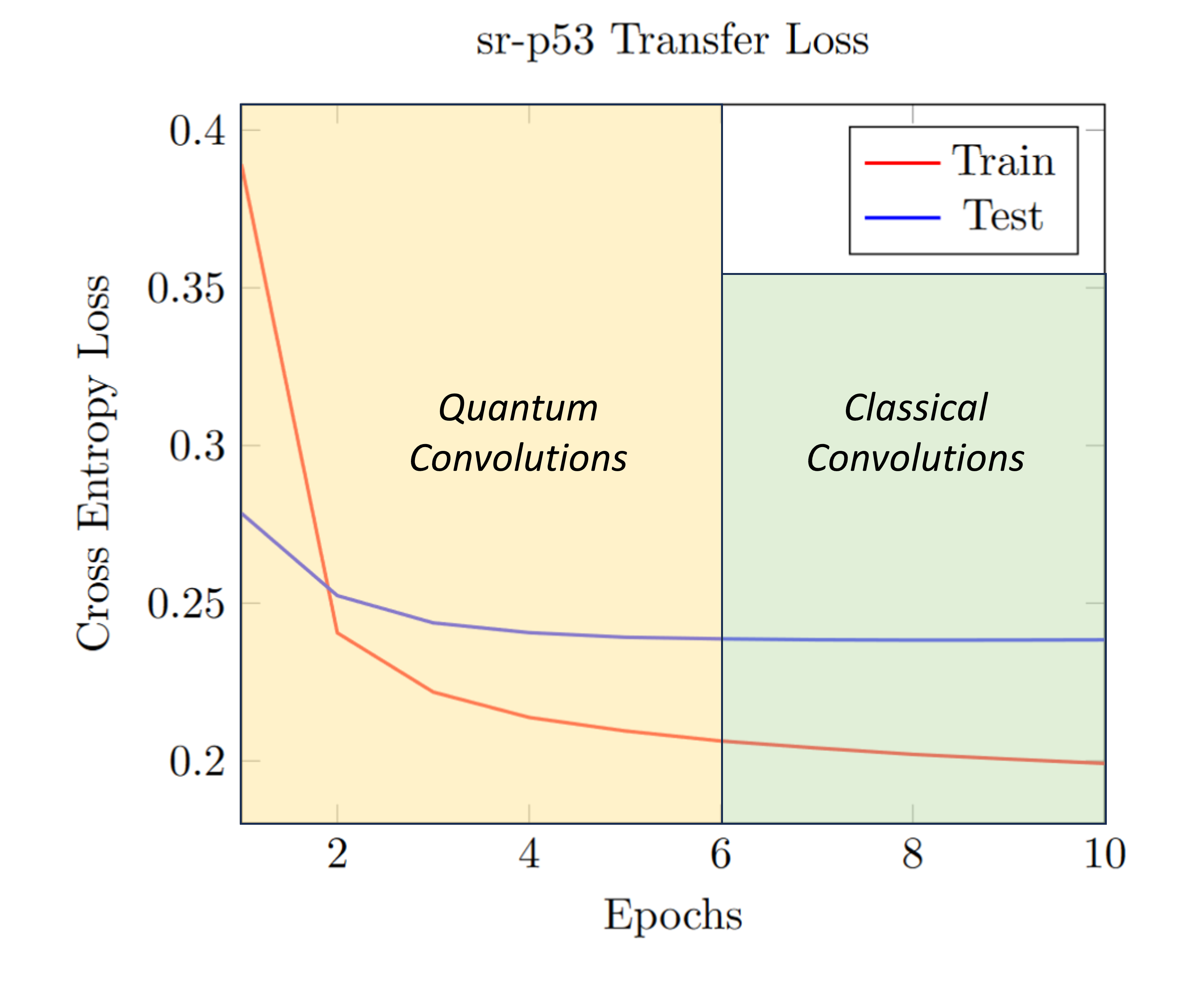}
        \caption{}
    \end{subfigure}
    \hfill
    \begin{subfigure}[b]{0.47\textwidth}
        \includegraphics[width=\textwidth]{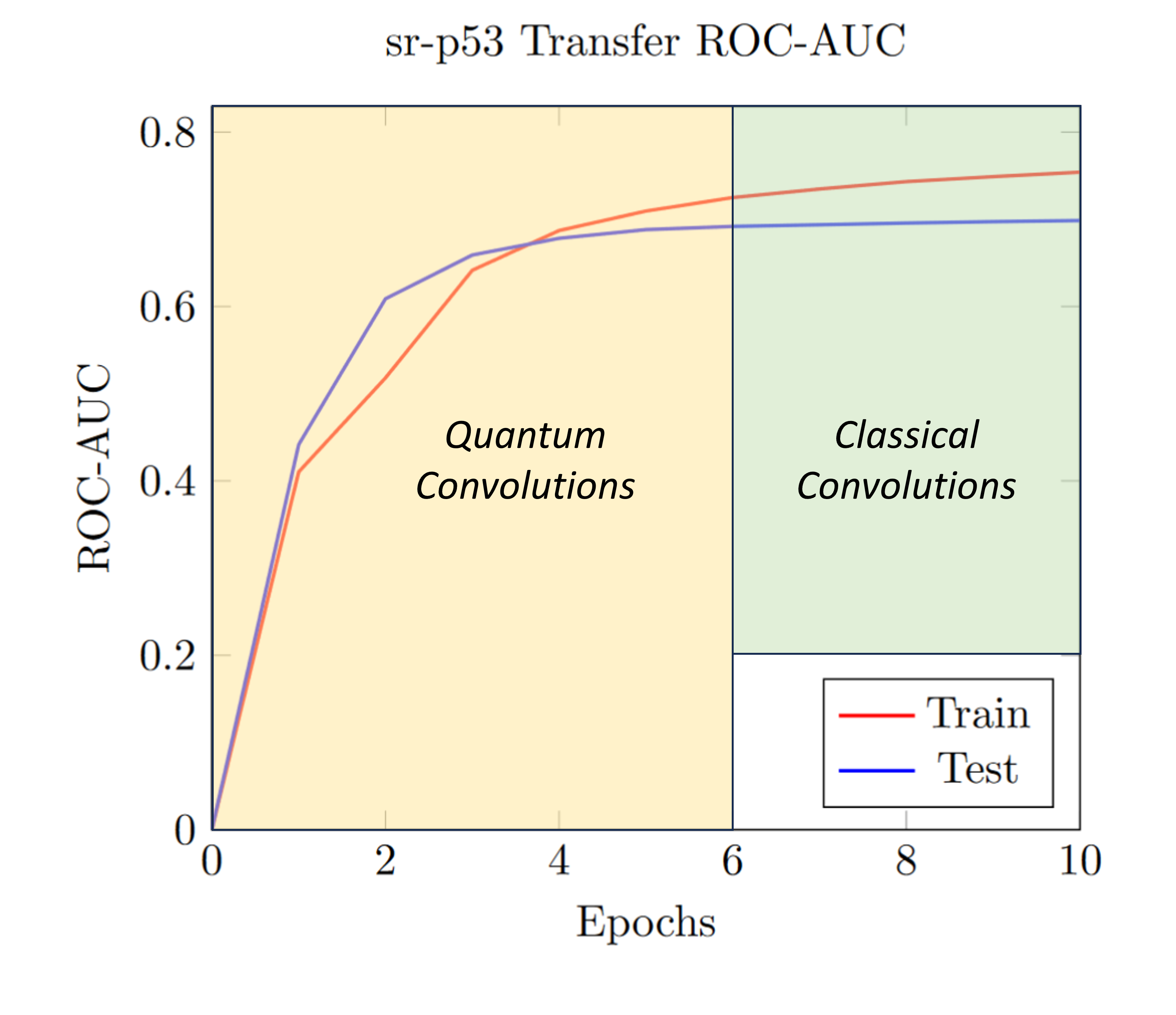}
        \caption{}
    \end{subfigure}
    \caption{Training curves for the sr-p53 assay}
\end{figure}

\end{document}